%% file: main.tex
\documentclass[a4paper,10pt,accepted=2025-08-28]{quantumarticle} 

\input{_preamb_quantum.sty}
\input{_tikz}

\begin{document}

\title{Systematic construction of stabilizer codes via gauging abelian boundary symmetries}

\author{Bram Vancraeynest-De Cuiper}\email{Bram.VancraeynestDeCuiper{@}UGent.be}
\affiliation{Department of Physics and Astronomy, Ghent University, Krijgslaan 281, 9000 Gent, Belgium}
\orcid{0000-0003-0080-2030}
\author{Jos\'e Garre-Rubio}
\affiliation{University of Vienna, Faculty of Mathematics, Oskar-Morgenstern-Platz 1, 1090 Vienna, Austria}
\affiliation{Instituto de F\'isica Te\'orica, UAM/CSIC, C. Nicol\'as Cabrera 13-15, Cantoblanco, 28049 Madrid, Spain}
\orcid{0000-0003-1845-7554}

\maketitle

\begin{abstract}
We propose a systematic framework to construct a (d+1)-dimensional stabilizer model from an initial generic d-dimensional abelian symmetry. Our approach builds upon the \emph{iterative gauging procedure}, developed by one of the authors in \href{https://doi.org/10.1038/s41467-024-52320-7}{[J. Garre-Rubio, Nature Commun. {\bf 15}, 7986 (2024)]}, in which an initial symmetric state is repeatedly gauged to obtain an emergent model in one dimension higher that supports the initial symmetry at its boundary. This method not only enables the construction of emergent states and corresponding commuting stabilizer Hamiltonians of which they are ground states, but it also provides a way to construct gapped boundary conditions for these models that amount to spontaneously breaking part of the boundary symmetry.\par

In a detailed introductory example, we showcase our paradigm by constructing three-dimensional Clifford-deformed surface codes from iteratively gauging a global 0-form symmetry that lives in two dimensions. We then provide a proof of our main result, hereby drawing upon a slight extension of the gauging procedure of Williamson. We additionally provide two more examples in d=2 in which different type-I fracton orders emerge from gauging initial linear subsystem and Sierpinski fractal symmetries. En passant, we provide explicit tensor network representations of all of the involved gauging maps and the emergent states.
\end{abstract}

\input{_intro}
\input{_topo_order}
\input{_general_framework}
\input{_fracton_order_LSS}
\input{_fracton_order_fractal}
\input{_conclusions}

\section*{Acknowledgments}
The authors thank Jacob Bridgeman, Clement Delcamp, Jutho Haegeman, Anasuya Lyons and Dominic Williamson for discussions and comments. This work has received funding from the Research Foundation Flanders (FWO) through Ph.D. fellowship No.~11O2423N awarded to BVDC. This work has been partially supported by the European Research Council (ERC) under the European Union’s Horizon 2020 research and innovation programme through the ERC-CoG SEQUAM (Grant Agreement No. 863476) and by the FWF Erwin Schrödinger Program (Grant DOI 10.55776/J4796).

\bibliographystyle{quantum}
\bibliography{refs}

\end{document}

%% file: _tikz.tex
\tikzset{
	every picture/.append style={
		line join=round,
		line width = 0.8pt,
            line cap=round
	}
}

\tikzstyle{lat}=[black, line width = 0.8pt, line join=round, line cap=round]
\tikzstyle{latdual}=[black!50, line width = 0.9pt, line join=round, line cap=round]
\tikzstyle{latb}=[\colorb, line width = 0.9pt, line join=round, line cap=round]
\tikzstyle{lata}=[\colora, line width = 0.8pt, line join=round, line cap=round]
\tikzstyle{GrayBlob}=[circle,fill=black!20, inner sep=0pt,minimum size=8pt]
\tikzstyle{op}=[black,rounded corners=2pt,fill=white,inner sep=.8pt] 
\tikzstyle{dir}=[line width = 0.6pt, arrows = {-Latex[width=2.5pt, length=2pt]}] 
\tikzstyle{dirdual}=[black!30, line width = 0.6pt, arrows = {-Latex[width=2.5pt, length=2pt]}] 
\tikzstyle{BlackNode}=[circle,draw,fill=black, inner sep=0pt,minimum size=2.5pt]
\tikzstyle{PlaquetteNode}=[circle,fill=\colora, inner sep=0pt,minimum size=4pt]
\tikzstyle{VertexNode}=[circle,fill=\colorb, inner sep=0pt,minimum size=4pt]

\pgfmathsetseed{\number\pdfrandomseed}

\tikzset{int/.style={transform shape, 
minimum width=2pt, minimum height=3pt, inner sep=0pt}}

\tikzset{->-/.style={
    decoration={
        markings, mark=at position #1 with {\arrow[line width=.25pt]{Latex}}
        },
    postaction={decorate}}
}

\tikzset{
    clip even odd rule/.code={\pgfseteorule},
    invclip/.style={clip,insert path=[clip even odd rule]{
    [reset cm](-\maxdimen,-\maxdimen)rectangle(\maxdimen,\maxdimen)
    }}} 

\newcommand\clipIntersection[1]{
    (#1.north west) -- (#1.north east) -- (#1.south east) -- (#1.south west) -- (#1.north west)
}

\newcommand\clipAroundNode[2]{
    \begin{pgfscope}
        \begin{scope}[overlay]
            \path[invclip] \clipIntersection{#1};#2
        \end{scope}
    \end{pgfscope}
}

\newcommand{\centerarc}[5]{
	\draw[#1] ($(#2)+({#5*cos(#3)},{#5*sin(#3)})$) arc (#3:#4:#5);
}
\newcommand{\SketchBulkBdry}[0]{
    \begin{tikzpicture}[line width=.4pt, baseline={([yshift=-.5ex]current bounding box.center)},,z={(0,1)},y={(0.4,.5)}]
        \def\l{1.1};
        \def\h{1.};
        \def\r{.1};

        \draw[decorate,decoration={brace,amplitude=3pt,mirror}] (2.8*\l,0,-\h/8) -- (2.8*\l,0,13*\h/8) node[anchor=west,xshift=4pt,midway,align=left,node font=\scriptsize]{d+1-dim.\\ stabilizer\\ system};

        \draw[decorate,decoration={brace,amplitude=3pt,mirror}] (2.5*\l,-.3*\l,-2*\h) -- (2.45*\l,1.1*\l,-2*\h) node[anchor=west,xshift=12pt,midway,align=left,node font=\scriptsize]{d-dim.\\ symmetric\\ system};

        \node[draw, circle, fill=BlueViolet!20, inner sep=3.pt] at (0,0,-2*\h) {};
        \node[draw, circle, fill=BlueViolet!20, inner sep=3.pt] at (\l,0,-2*\h) {};
        \node[draw, circle, fill=BlueViolet!20, inner sep=3.pt] at (0,\l,-2*\h) {};
        \node[draw, circle, fill=BlueViolet!20, inner sep=3.pt] at (\l,\l,-2*\h) {};
        \node[draw, circle, fill=BlueViolet!20, inner sep=3.pt] at (2*\l,0,-2*\h) {};
        \node[draw, circle, fill=BlueViolet!20, inner sep=3.pt] at (2*\l,\l,-2*\h) {};

        \draw[-Latex] (\l,\l/2,-3*\h/2) --+ (0,0,\h) node[fill=white, pos=.5] {\scriptsize Iterative gauging};
        
        \draw[dotted] (\l,0,0) -- (\l/2,\l/2,\h/2);
        \draw[dotted] (0,\l,0) -- (\l/2,\l/2,\h/2);
        \draw[dotted] (\l,\l,0) -- (\l/2,\l/2,\h/2);

        \draw[dotted] (2*\l,0,0) -- (3*\l/2,\l/2,\h/2);
        \draw[dotted] (\l,\l,0) -- (3*\l/2,\l/2,\h/2);
        \draw[dotted] (2*\l,\l,0) -- (3*\l/2,\l/2,\h/2);

        \node[draw, circle, fill=BlueViolet!20, inner sep=3.pt] at (\l,\l,0) {};
         
        \draw[dotted] (\l,0,0) -- (3*\l/2,\l/2,\h/2);
        
        \node[draw, circle, fill=BlueViolet!20, inner sep=3.pt] at (\l,0,0) {};
        \node[draw, circle, fill=BlueViolet!20, inner sep=3.pt] at (0,\l,0) {};
       
        \node[draw, circle, fill=BlueViolet!20, inner sep=3.pt] at (2*\l,0,0) {};
        \node[draw, circle, fill=BlueViolet!20, inner sep=3.pt] at (2*\l,\l,0) {};

        \draw[dotted] (0,0,0) -- (\l/2,\l/2,\h/2);
        \node[draw, circle, fill=BlueViolet!20, inner sep=3.pt] at (0,0,0) {};

        \draw[dotted] (\l,0,\h) -- (\l/2,\l/2,\h/2);
        \draw[dotted] (0,\l,\h) -- (\l/2,\l/2,\h/2);
        \draw[dotted] (\l,\l,\h) -- (\l/2,\l/2,\h/2);
        \draw[dotted] (0,0,\h) -- (\l/2,\l/2,\h/2);

        \draw[dotted] (2*\l,0,\h) -- (3*\l/2,\l/2,\h/2);
        \draw[dotted] (\l,\l,\h) -- (3*\l/2,\l/2,\h/2);
        \draw[dotted] (2*\l,\l,\h) -- (3*\l/2,\l/2,\h/2);
        \draw[dotted] (\l,0,\h) -- (3*\l/2,\l/2,\h/2);

        \node[draw, circle, fill=BrickRed!20, inner sep=3.pt] at (\l/2,\l/2,\h/2) {};
        \node[draw, circle, fill=BrickRed!20, inner sep=3.pt] at (3*\l/2,\l/2,\h/2) {};

        \node[draw, circle, fill=BlueViolet!20, inner sep=3.pt] at (0,0,\h) {};
        \node[draw, circle, fill=BlueViolet!20, inner sep=3.pt] at (\l,0,\h) {};
        \node[draw, circle, fill=BlueViolet!20, inner sep=3.pt] at (0,\l,\h) {};
        \node[draw, circle, fill=BlueViolet!20, inner sep=3.pt] at (\l,\l,\h) {};
        \node[draw, circle, fill=BlueViolet!20, inner sep=3.pt] at (2*\l,0,\h) {};
        \node[draw, circle, fill=BlueViolet!20, inner sep=3.pt] at (2*\l,\l,\h) {};
    \end{tikzpicture}
}
\newcommand{\smallcube}[0]{
    \begin{tikzpicture}[line width=.4pt, baseline={([yshift=-.5ex]current bounding box.center)},z={(0,1)},y={(0.5,0.5)}]
        \def\l{.15};
        \draw (0,\l,\l) -- (\l,\l,\l);
        \draw (\l,\l,0) -- (\l,\l,\l);
        \draw (0,0,\l) -- (0,\l,\l);
        \draw (\l,0,0) -- (\l,\l,0);
        \draw (\l,0,\l) -- (\l,\l,\l);
        \draw (0,0,0) rectangle (\l,0,\l);
    \end{tikzpicture}
}
\newcommand{\CubicLattice}{
    \begin{tikzpicture}[baseline={([yshift=-.5ex]current bounding box.center)},z={(0,1)},y={(0.55,0.5)}]
        \def\lx{1.1};
        \def\ly{1.1};
        \def\lz{1.1};
        \def\off{.25};
        
        \def\o{(3.25*\lx,\ly/2,\lz/4)};    
        \draw[dir] \o --+ (-\lx,0,0) node[anchor=east] {$\sss\hat x$};
        \draw[dir] \o --+ (0,-\ly,0) node[anchor=north east] {$\sss\hat y$};
        \draw[dir] \o --+ (0,0,\lz) node[anchor=south] {$\sss\hat z$};

        \begin{scope}[shorten <= -3pt, shorten >= -3pt]
            \node[int](mk) at (.55*\lx,0,\lz) {};
            
            \draw[latb] (0,0,0) -- (0,0,\lz);
            \draw[latb, dotted] (0,0,0) --++ (0,0,-\off);
            \draw[latb, dotted] (0,0,\lz) --++ (0,0,\off);
            \draw[latb] (\lx,0,0) -- (\lx,0,\lz);
            \draw[latb, dotted] (\lx,0,0) --++ (0,0,-\off);
            \draw[latb, dotted] (\lx,0,\lz) --++ (0,0,\off);
            \clipAroundNode{mk}{
                \draw[latb] (0,\ly,0) -- (0,\ly,\lz);
                \draw[latb, dotted] (0,\ly,0) --++ (0,0,-\off);
                \draw[latb, dotted] (0,\ly,\lz) --++ (0,0,\off);
            }
            \draw[latb] (\lx,\ly,0) -- (\lx,\ly,\lz);
            \draw[latb, dotted] (\lx,\lx,0) --++ (0,0,-\off);
            \draw[latb, dotted] (\lx,\lx,\lz) --++ (0,0,\off);
            
            \draw[lata] (0,0,0) -- (\lx,0,0);
            \draw[lata, dotted] (0,0,0) --++ (-\off,0,0);
            \draw[lata, dotted] (\lx,0,0) --++ (\off,0,0);
            \draw[lata] (0,0,0) -- (0,\ly,0);
            \draw[lata, dotted] (0,0,0) --++ (0,-\off,0);
            \draw[lata, dotted] (0,\ly,0) --++ (0,\off,0);
            \draw[lata] (\lx,0,0) -- (\lx,\ly,0);
            \draw[lata, dotted] (\lx,0,0) --++ (0,-\off,0);
            \draw[lata, dotted] (\lx,\ly,0) --++ (0,\off,0);
            \draw[lata] (0,\ly,0) -- (\lx,\ly,0);
            \draw[lata, dotted] (0,\ly,0) --++ (-\off,0,0);
            \draw[lata, dotted] (\lx,\ly,0) --++ (\off,0,0);

            \draw[lata] (0,0,\lz) -- (\lx,0,\lz);
            \draw[lata, dotted] (0,0,\lz) --++ (-\off,0,0);
            \draw[lata, dotted] (\lx,0,\lz) --++ (\off,0,0);
            \draw[lata] (0,0,\lz) -- (0,\ly,\lz);
            \draw[lata, dotted] (0,0,\lz) --++ (0,-\off,0);
            \draw[lata, dotted] (0,\ly,\lz) --++ (0,\off,0);
            \draw[lata] (\lx,0,\lz) -- (\lx,\ly,\lz);
            \draw[lata, dotted] (\lx,0,\lz) --++ (0,-\off,0);
            \draw[lata, dotted] (\lx,\ly,\lz) --++ (0,\off,0);
            \draw[lata] (0,\ly,\lz) -- (\lx,\ly,\lz);
            \draw[lata, dotted] (0,\ly,\lz) --++ (-\off,0,0);
            \draw[lata, dotted] (\lx,\ly,\lz) --++ (\off,0,0);
        \end{scope}
    \end{tikzpicture}
}
\newcommand{\TOPOPEPOState}{
    \begin{tikzpicture}[baseline={([yshift=1.25ex]current bounding box.center)},z={(0,1)},y={(0.6,0.5)}]
        \def\l{1.1};
        \def\h{1.25};

        \clip (0,-3.2) rectangle (-1.7,4.5);

        \node[int](mk1) at (-.6*\l,0,0) {};
        \node[int](mk2) at (-.6*\l,0,\h) {};

        \draw[lata, dotted] (-\l,0,-\h/2) -- (-\l,0,-3*\h/4);
        \draw[lata, dotted] (0,-\l,-\h/2) -- (0,-\l,-3*\h/4);
        \draw[latb,->-=.55] (\l/2,0,0) -- (0,0,0);
        \clipAroundNode{mk1}{
            \draw[latb,->-=.44] (0,0,0) -- (-\l,0,0);
        }
        \draw[latb,->-=.8] (-\l,0,0) -- (-3*\l/2,0,0); 

        \draw[latb, ->-=.55] (0,\l/2,0) -- (0,0,0);
        \draw[latb, ->-=.64] (0,0,0) -- (0,-\l,0);
        \draw[latb, ->-=.90] (0,-\l,0) -- (0,-3*\l/2,0);
        
        \draw[lata, ->-=.26, ->-=.88] (-\l,0,-\h/2) -- (-\l,0,\h/2);
        \draw[lata, ->-=.26, ->-=.88] (0,-\l,-\h/2) -- (0,-\l,\h/2);

        \draw[latb, ->-=.60] (0,0,0) -- (0,0,\h);

        \draw[lata,->-=.55] (\l/2,0,\h) -- (0,0,\h);
        \clipAroundNode{mk2}{
            \draw[lata,->-=.44] (0,0,\h) -- (-\l,0,\h);
        }
        \draw[lata,->-=.8] (-\l,0,\h) -- (-3*\l/2,0,\h);

        \draw[lata, ->-=.55] (0,\l/2,\h) -- (0,0,\h);
        \draw[lata, ->-=.64] (0,0,\h) -- (0,-\l,\h);
        \draw[lata, ->-=.90] (0,-\l,\h) -- (0,-3*\l/2,\h);

        \draw[lata, ->-=.80] (-\l,0,\h) -- (-\l,0,3*\h/2);
        \draw[lata, ->-=.80] (0,-\l,\h) -- (0,-\l,3*\h/2);
        \draw[latb, ->-=.80] (0,0,\h) -- (0,0,3*\h/2);

        \draw[lata, dotted] (-\l,0,3*\h/2) -- (-\l,0,7*\h/4);
        \draw[lata, dotted] (0,-\l,3*\h/2) -- (0,-\l,7*\h/4);

        \node[line width=.7pt, draw=black, fill=BlueViolet!20, circle, inner sep=3.2pt] at (0,0,0) {};
        \node at (0,0,0) {${\sss +}$};
        \node[line width=.7pt, draw=black, fill=BrickRed!20, circle, inner sep=3.2pt] at (-\l,0,0) {};
        \node at (-\l,0,0) {${\sss Z}$};
        \node[line width=.7pt, draw=black, fill=BrickRed!20, circle, inner sep=3.2pt] at (0,-\l,0) {};
        \node at (0,-\l,0) {${\sss Z}$};

        \node[line width=.7pt, draw=black, fill=BlueViolet!20, circle, inner sep=3.2pt] at (0,0,\h) {};
        \node at (0,0,\h) {${\sss Z}$};
        \node[line width=.7pt, draw=black, fill=BrickRed!20, circle, inner sep=3.2pt] at (-\l,0,\h) {};
        \node at (-\l,0,\h) {${\sss +}$};
        \node[line width=.7pt, draw=black, fill=BrickRed!20, circle, inner sep=3.2pt] at (0,-\l,\h) {};
        \node at (0,-\l,\h) {${\sss +}$};
    \end{tikzpicture}
}
\newcommand{\XZXpullingthrough}[1]{
    \begin{tikzpicture}[baseline={([yshift=-.5ex]current bounding box.center)},z={(0,1)},y={(0.6,0.5)}]
        \def\l{1.1};
        \def\h{0.65};

        \clip (-1.55,-2.25) rectangle (-.55,2.2);
        
        \draw[lata, ->-=.55] (\l/2,0,0) -- (0,0,0);
        \draw[lata, ->-=.6] (0,0,0) -- (-\l,0,0);
        \draw[lata, ->-=.6] (-\l,0,0) -- (-2*\l,0,0);
        \draw[lata, ->-=.80] (-2*\l,0,0) -- (-5*\l/2,0,0);
        
        \draw[lata,->-=.55] (0,\l/2,0) -- (0,0,0);
        \draw[lata, ->-=.90] (0,0,0) -- (0,-\l/2,0);
        \draw[lata,->-=.55] (-2*\l,\l/2,0) -- (-2*\l,0,0);
        \draw[lata, ->-=.90] (-2*\l,0,0) -- (-2*\l,-\l/2,0);
        \draw[lata, ->-=.80] (-\l,0,0) -- (-\l,0,\h);
        \draw[latb, ->-=.58] (0,0,-\h) -- (0,0,0);
        \draw[latb, ->-=.80] (0,0,0) -- (0,0,\h);
        \draw[latb, ->-=.58] (-2*\l,0,-\h) -- (-2*\l,0,0);
        \draw[latb, ->-=.80] (-2*\l,0,0) -- (-2*\l,0,\h);

        \node[line width=.7pt, draw=black, fill=BlueViolet!20, circle, inner sep=3.2pt] at (0,0,0) {};
        \node at (0,0,0) {${\sss Z}$};
        \node[line width=.7pt, draw=black, fill=BrickRed!20, circle, inner sep=3.2pt] at (-\l,0,0) {};
        \node at (-\l,0,0) {${\sss +}$};
        \node[line width=.7pt, draw=black, fill=BlueViolet!20, circle, inner sep=3.2pt] at (-2*\l,0,0) {};
        \node at (-2*\l,0,0) {${\sss Z}$};

        \ifthenelse{#1=1}{
            \node[anchor=south] at (-2*\l,0,\h) {$\sss X_\chi$};
            \node[anchor=south] at (-\l,0,\h) {$\sss Z_\chi^\dagger$};
            \node[anchor=south] at (0,0,\h) {$\sss X_\chi^\dagger$};
        }{}
        \ifthenelse{#1=2}{
            \node[anchor=north] at (-2*\l,0,-\h) {$\sss X_\chi$};
            \node[anchor=north] at (0,0,-\h) {$\sss X_\chi^\dagger$};
        }{}
    \end{tikzpicture}
}
\newcommand{\StabPullingThrough}[1]{
    \begin{tikzpicture}[baseline={([yshift=-.5ex]current bounding box.center)},z={(0,1)},y={(0.6,0.5)}]
        \def\l{1.1};
        \def\h{1.25};

        \ifthenelse{#1=1}{
            \draw[latb, ->-=.55] (-2*\l,0,0) -- (-2*\l,0,\h);
            \draw[latb, ->-=.55] (0,0,0) -- (0,0,\h);
            
            \node[anchor=south, yshift=-3.5pt] at (0,0,3*\h/2) {$\sss X_\chi^\dagger$};
            \node[anchor=south, yshift=-3.5pt] at (-\l,0,\h/2) {$\sss Z_\chi$};
            \node[anchor=south, yshift=-3.5pt] at (-\l,0,5*\h/2) {$\sss Z_\chi^\dagger$};
            \node[anchor=south, yshift=-3.5pt] at (-2*\l,0,3*\h/2) {$\sss X_\chi$};
        }{}
        \ifthenelse{#1=2}{
            \draw[latb, ->-=.35] (-2*\l,0,0) -- (-2*\l,0,\h);
            \draw[latb, ->-=.35] (0,0,0) -- (0,0,\h);
            
            \node[op] at (0,0,\h/2) {$\sss X_\chi^\dagger$};
            \node[anchor=south, yshift=-3pt] at (-\l,0,\h/2) {$\sss Z_\chi$};
            \node[op] at (-2*\l,0,\h/2) {$\sss X_\chi$};        
        }{}

        \draw[lata, dotted] (-\l,0,-3*\h/4) -- (-\l,0,-\h/2);
        \draw[latb, ->-=.09, ->-=.36, ->-=.68, ->-=.97] (\l/2,0,0) -- (-5*\l/2,0,0);
        \draw[latb, ->-=.28, ->-=.96] (-2*\l,\l/2,0) -- (-2*\l,-\l/2,0);
        \draw[latb, ->-=.28, ->-=.96] (0,\l/2,0) -- (0,-\l/2,0);
        \draw[lata, ->-=.6] (-\l,0,-\h/2) -- (-\l,0,0);
        \draw[lata, ->-=.74] (-\l,0,0) -- (-\l,0,\h/2);

        \draw[lata, ->-=.09, ->-=.36, ->-=.68, ->-=.97] (\l/2,0,\h) -- (-5*\l/2,0,\h);
        \draw[lata, ->-=.28, ->-=.96] (-2*\l,\l/2,\h) -- (-2*\l,-\l/2,\h);
        \draw[lata, ->-=.28, ->-=.96] (0,\l/2,\h) -- (0,-\l/2,\h);
        \draw[latb, ->-=.74] (-2*\l,0,\h) -- (-2*\l,0,3*\h/2);
        \draw[latb, ->-=.74] (0,0,\h) -- (0,0,3*\h/2);
        \draw[lata, ->-=.54] (-\l,0,\h) -- (-\l,0,2*\h);

        \draw[latb, ->-=.09, ->-=.36, ->-=.68, ->-=.97] (\l/2,0,2*\h) -- (-5*\l/2,0,2*\h);
        \draw[latb, ->-=.28, ->-=.96] (-2*\l,\l/2,2*\h) -- (-2*\l,-\l/2,2*\h);
        \draw[latb, ->-=.28, ->-=.96] (0,\l/2,2*\h) -- (0,-\l/2,2*\h);
        \draw[latb, ->-=.80] (-2*\l,0,2*\h) -- (-2*\l,0,5*\h/2);
        \draw[lata, ->-=.80] (-\l,0,2*\h) -- (-\l,0,5*\h/2);
        \draw[latb, ->-=.80] (0,0,2*\h) -- (0,0,5*\h/2);
        \draw[latb, dotted] (-2*\l,0,5*\h/2) -- (-2*\l,0,11*\h/4);
        \draw[latb, dotted] (0,0,5*\h/2) -- (0,0,11*\h/4);

        \node[line width=.7pt, draw=black, fill=BlueViolet!20, circle, inner sep=3.2pt] at (0,0,0) {};
        \node at (0,0,0) {${\sss +}$};
        \node[line width=.7pt, draw=black, fill=BrickRed!20, circle, inner sep=3.2pt] at (-\l,0,0) {};
        \node at (-\l,0,0) {${\sss Z}$};
        \node[line width=.7pt, draw=black, fill=BlueViolet!20, circle, inner sep=3.2pt] at (-2*\l,0,0) {};
        \node at (-2*\l,0,0) {${\sss +}$};

        \node[line width=.7pt, draw=black, fill=BlueViolet!20, circle, inner sep=3.2pt] at (0,0,\h) {};
        \node at (0,0,\h) {${\sss Z}$};
        \node[line width=.7pt, draw=black, fill=BrickRed!20, circle, inner sep=3.2pt] at (-\l,0,\h) {};
        \node at (-\l,0,\h) {${\sss +}$};
        \node[line width=.7pt, draw=black, fill=BlueViolet!20, circle, inner sep=3.2pt] at (-2*\l,0,\h) {};
        \node at (-2*\l,0,\h) {${\sss Z}$};

        \node[line width=.7pt, draw=black, fill=BlueViolet!20, circle, inner sep=3.2pt] at (0,0,2*\h) {};
        \node at (0,0,2*\h) {${\sss +}$};
        \node[line width=.7pt, draw=black, fill=BrickRed!20, circle, inner sep=3.2pt] at (-\l,0,2*\h) {};
        \node at (-\l,0,2*\h) {${\sss Z}$};
        \node[line width=.7pt, draw=black, fill=BlueViolet!20, circle, inner sep=3.2pt] at (-2*\l,0,2*\h) {};
        \node at (-2*\l,0,2*\h) {${\sss +}$};
    \end{tikzpicture}
}
\newcommand{\CSSHam}[1]{
    \begin{tikzpicture}[baseline={([yshift=-.5ex]current bounding box.center)},z={(0,1)},y={(0.6,0.5)}]
        \def\l{1.2};

        \ifthenelse{\equal{#1}{v}}{
            \draw[lata] (-\l,0,0) -- (\l,0,0);
            \draw[lata] (0,-\l,0) -- (0,\l,0);
            \draw[latb] (0,0,-\l) -- (0,0,\l);
    
            \node[op] at (0,0,.5*\l) {$\sss Z_{\!g}^\dagger$};
            \node[op] at (0,0,-.5*\l) {$\sss Z_{\!g}$};
            \node[op] at (0,.55*\l,0) {$\sss X_{\!g}^\dagger$};
            \node[op] at (0,-.5*\l,0) {$\sss X_{\!g}$};
            \node[op] at (-.5*\l,0,0) {$\sss X_{\!g}$};
            \node[op] at (.55*\l,0,0) {$\sss X_{\!g}^\dagger$};
        }{}
        \ifthenelse{\equal{#1}{vconst}}{
            \draw[lata] (-\l,0,0) -- (\l,0,0);
            \draw[lata] (0,-\l,0) -- (0,\l,0);
    
            \node[op] at (0,.5*\l,0) {$\sss X_{\!\chi}$};
            \node[op] at (0,-.5*\l,0) {$\sss X_{\!\chi}^\dagger$};
            \node[op] at (-.5*\l,0,0) {$\sss X_{\! \chi}^\dagger$};
            \node[op] at (.5*\l,0,0) {$\sss X_{\! \chi}$};
        }{}
        \ifthenelse{\equal{#1}{facefront}}{
            \draw[lata] (0,0,0) -- (\l,0,0);
            \draw[lata] (0,0,\l) -- (\l,0,\l);
            \draw[latb] (0,0,0) -- (0,0,\l);
            \draw[latb] (\l,0,0) -- (\l,0,\l);
            
            \node[op] at (0,0,.5*\l) {$\sss X_\chi$};
            \node[op] at (\l,0,.5*\l) {$\sss X_\chi^\dagger$};
            \node[op] at (.5*\l,0,0) {$\sss Z_\chi$};
            \node[op] at (.5*\l,0,\l) {$\sss Z_\chi^\dagger$};
        }{}
        \ifthenelse{\equal{#1}{faceright}}{
            \draw[lata] (\l,0,0) -- (\l,\l,0);
            \draw[lata] (\l,0,\l) -- (\l,\l,\l);
            \draw[latb] (\l,0,0) -- (\l,0,\l);
            \draw[latb] (\l,\l,0) -- (\l,\l,\l);
            
            \node[op] at (\l,0,.6*\l) {$\sss X_\chi$};
            \node[op] at (\l,\l,.4*\l) {$\sss X_\chi^\dagger$};
            \node[op] at (\l,.5*\l,0) {$\sss Z_\chi$};
            \node[op] at (\l,.5*\l,\l) {$\sss Z_\chi^\dagger$};
        }{}
        \ifthenelse{\equal{#1}{facebottom}}{
            \draw[lata] (0,0,0) -- (\l,0,0);
            \draw[lata] (\l,0,0) -- (\l,\l,0);
            \draw[lata] (0,0,0) -- (0,\l,0);
            \draw[lata] (0,\l,0) -- (\l,\l,0);
            
            \node[op] at (0,.5*\l,0) {$\sss Z_\chi$};
            \node[op] at (.4*\l,\l,0) {$\sss Z_\chi$};
            \node[op] at (\l,.5*\l,0) {$\sss Z_\chi^\dagger$};
            \node[op] at (.6*\l,0,0) {$\sss Z_\chi^\dagger$};
        }{}
        \ifthenelse{\equal{#1}{bdryex}}{
            \draw[lata] (0,0,0) -- (\l,0,0);
            \node[op] at (.5*\l,0,0) {$\sss Z_\chi$};
        }{}
        \ifthenelse{\equal{#1}{bdryey}}{
            \draw[lata] (0,0,0) -- (0,\l,0);
            \node[op] at (0,.5*\l,0) {$\sss Z_\chi$};
        }{}
        \ifthenelse{\equal{#1}{bdryfacefront}}{
            \draw[lata] (0,0,\l) -- (\l,0,\l);
            \draw[latb] (0,0,0) -- (0,0,\l);
            \draw[latb] (\l,0,0) -- (\l,0,\l);
            
            \node[op] at (0,0,.5*\l) {$\sss X_\chi^\dagger$};
            \node[op] at (\l,0,.5*\l) {$\sss X_\chi$};
            \node[op] at (.5*\l,0,\l) {$\sss Z_\chi^\dagger$};
        }{}
        \ifthenelse{\equal{#1}{bdrysinglez}}{
            \draw[latb] (0,0,0) -- (0,0,\l);
            \node[op] at (0,0,.5*\l) {$\sss Z_g$};
        }{}
        \ifthenelse{\equal{#1}{bdryv}}{
            \draw[lata] (-\l,0,0) -- (\l,0,0);
            \draw[lata] (0,-\l,0) -- (0,\l,0);
            \draw[latb] (0,0,0) -- (0,0,\l);
    
            \node[op] at (0,0,.5*\l) {$\sss Z_{\!g}^\dagger$};
            \node[op] at (0,.55*\l,0) {$\sss X_{\!g}^\dagger$};
            \node[op] at (0,-.5*\l,0) {$\sss X_{\!g}$};
            \node[op] at (-.5*\l,0,0) {$\sss X_{\!g}$};
            \node[op] at (.55*\l,0,0) {$\sss X_{\!g}^\dagger$};
        }{}
        \ifthenelse{\equal{#1}{bdryv2}}{
            \draw[lata] (-\l,0,0) -- (\l,0,0);
            \draw[lata] (0,-\l,0) -- (0,\l,0);

            \node[op] at (0,.55*\l,0) {$\sss X_{\!g}^\dagger$};
            \node[op] at (0,-.5*\l,0) {$\sss X_{\!g}$};
            \node[op] at (-.5*\l,0,0) {$\sss X_{\!g}$};
            \node[op] at (.55*\l,0,0) {$\sss X_{\!g}^\dagger$};
        }{}
        \ifthenelse{\equal{#1}{bdryfaceright}}{
            \draw[lata] (\l,0,\l) -- (\l,\l,\l);
            \draw[latb] (\l,0,0) -- (\l,0,\l);
            \draw[latb] (\l,\l,0) -- (\l,\l,\l);
            
            \node[op] at (\l,0,.6*\l) {$\sss X_\chi$};
            \node[op] at (\l,\l,.4*\l) {$\sss X_\chi^\dagger$};
            \node[op] at (\l,.5*\l,\l) {$\sss Z_\chi^\dagger$};
        }{}
    \end{tikzpicture}
}
\newcommand{\PEPOTensor}[7]{
    \begin{tikzpicture}[baseline={(0,0,.2)},z={(0,1)},y={(0.6,0.5)}]
        \def\l{.75};
        \def\h{.65}

        \ifthenelse{\equal{#1}{0}}{
            \draw[lata, ->-=.70] (0,0,0) -- (-\l,0,0) node[anchor=east, op, pos=1] {$\sss #5$};
            \draw[lata, ->-=.55] (0,\l,0) -- (0,0,0) node[anchor=west, op, sloped, pos=0] {$\sss #3$};
            \draw[lata, ->-=.55] (\l,0,0) -- (0,0,0) node[anchor=west, op, pos=0] {$\sss #4$};
            \draw[lata, ->-=.8] (0,0,0) -- (0,-\l,0) node[anchor=east, op, sloped, pos=1] {$\sss #6$};
            \draw[latb, ->-=.58] (0,0,-\h) -- (0,0,0) node[anchor=north, black, pos=0] {$\sss #2$};
            \draw[latb, ->-=.80] (0,0,0) -- (0,0,\h) node[anchor=south, black, pos=1] {$\sss #7$};
            
            \centerarc{dash pattern= on 0pt off 2.5pt, line width=1pt}{0,0}{25}{75}{.75*\l};
            \centerarc{dash pattern= on 0pt off 2.5pt, line width=1pt}{0,0}{-105}{-155}{.75*\l};
            
            \node[line width=.7pt, draw=black, fill=BlueViolet!20, circle, inner sep=3.2pt] at (0,0) {};
            \node at (0,0) {${\sss Z}$};
        }{}
        \ifthenelse{\equal{#1}{1}}
        {
            \draw[lata, ->-=.70] (0,0,0) -- (-\l,0,0) node[anchor=east, op] {$\sss #2$};
            \draw[lata, ->-=.80] (0,0,0) -- (0,0,\h) node[anchor=south, op] {$\sss #3$};
            \draw[lata, ->-=.55] (\l,0,0) node[anchor=west, op] {$\sss #4$} -- (0,0,0);
            
            \node[line width=.7pt, draw=black, fill=BrickRed!20, circle, inner sep=3.2pt] at (0,0) {};
            \node at (0,0) {${\sss +}$};
        }{}
    	\ifthenelse{\equal{#1}{-1}}
    	{
    		\draw[lata, ->-=.70] (0,0,0) -- (-\l,0,0) node[anchor=east, op] {$\sss #2$};
    		\draw[lata, ->-=.55] (0,0,\h) -- (0,0,0) node[anchor=south, op, pos=0] {$\sss #3$};
    		\draw[lata, ->-=.55] (\l,0,0) -- (0,0,0) node[anchor=west, op, pos=0] {$\sss #4$};
    		
    		\node[line width=.7pt, draw=black, fill=BrickRed!20, circle, inner sep=3.2pt] at (0,0) {};
    		\node at (0,0) {${\sss +}$};
    	}{}
        \ifthenelse{\equal{#1}{2}}
        {
            \draw[lata, ->-=.58] (0,0,-\h) -- (0,0,0) node[anchor=north, op, pos=0] {$\sss #2$};
            \draw[lata, ->-=.80] (0,0,0) -- (0,0,\h) node[anchor=south, op] {$\sss #3$};

            \draw[latb, ->-=.70] (0,0,0) -- (-\l,0,0) node[anchor=east, op] {$\sss #4$};
            \draw[latb, ->-=.55] (\l,0,0) -- (0,0,0) node[anchor=west, op, pos=0] {$\sss #5$};
            
            \node[line width=.7pt, draw=black, fill=BrickRed!20, circle, inner sep=3.2pt] at (0,0) {};
            \node at (0,0) {${\sss Z}$};
        }{}
        \ifthenelse{\equal{#1}{3}}{
            \draw[latb, ->-=.70] (0,0,0) -- (-\l,0,0) node[anchor=east, op] {$\sss #4$};
            \draw[latb, ->-=.55] (0,\l,0) -- (0,0,0) node[anchor=west, sloped, op, pos=0] {$\sss #3$};
            \draw[latb, ->-=.55] (\l,0,0) -- (0,0,0) node[anchor=west, op, pos=0] {$\sss #2$};
            \draw[latb, ->-=.80] (0,0,0) -- (0,-\l,0) node[anchor=east, op, sloped, pos=1] {$\sss #5$};
            \draw[latb, ->-=.80] (0,0,0) -- (0,0,\h) node[anchor=south, op] {$\sss #6$};
            \draw[latb, ->-=.58] (0,0,-\h) -- (0,0,0) node[anchor=north, op, pos=0] {$\sss #7$};
            
            \node[line width=.7pt, draw=black, fill=BlueViolet!20, circle, inner sep=3.2pt] at (0,0) {};
            \node at (0,0) {${\sss +}$};
            
            \centerarc{dash pattern= on 0pt off 2.5pt, line width=1pt}{0,0}{142}{155}{1.2*\l};
            \centerarc{dash pattern= on 0pt off 2.5pt, line width=1pt}{0,0}{-37}{-50}{1.2*\l};
        }{}
    \end{tikzpicture}
}
\newcommand{\GaussLaw}[1]{
    \begin{tikzpicture}[baseline={(0,0,.2)},z={(0,1)},y={(0.6,0.5)}]
        \def\l{1.1};
        \def\h{.65};

        \ifthenelse{#1=0}{
            \draw[lata, ->-=.64] (0,\l,0) -- (0,0,0);
            \draw[lata, ->-=.7] (\l,0,0) -- (0,0,0);
            \draw[lata, ->-=.64] (0,0,0) -- (0,-\l,0);        
            
            \node[int](mk1) at (-.6*\l,0,0) {};
            \node[int](mk2) at (.58*\l,0,0) {};
            
            \clipAroundNode{mk1}{
                \draw[lata, ->-=.44] (0,0,0) -- (-\l,0,0);
            }
    
            \draw[latb, ->-=.80] (0,0,0) -- (0,0,\h);
            \draw[lata, ->-=.80] (-\l,0,0) -- (-\l,0,\h);
            \draw[lata, ->-=.80] (\l,0,0) -- (\l,0,\h);
            \draw[lata, ->-=.80] (0,-\l,0) -- (0,-\l,\h);
            \draw[lata, ->-=.80] (0,\l,0) -- (0,\l,\h);
    
            \draw[latb, ->-=.58] (0,0,-\h) -- (0,0,0);
            \draw[lata, ->-=.58] (-\l,0,-\h) -- (-\l,0,0);
            \draw[lata, ->-=.58] (\l,0,-\h) -- (\l,0,0);
            \draw[lata, ->-=.58] (0,-\l,-\h) -- (0,-\l,0);
            \clipAroundNode{mk2}{
                \draw[lata, ->-=.62] (0,\l,-\h) -- (0,\l,0);
            }

            \node[line width=.7pt, draw=black, fill=BlueViolet!20, circle, inner sep=3.2pt] at (0,0) {};
            \node at (0,0) {${\sss Z}$};
            
            \node[line width=.7pt, draw=black, fill=BrickRed!20, circle, inner sep=3.2pt] at (-\l,0,0) {};
            \node at (-\l,0,0) {${\sss +}$};
            \node[line width=.7pt, draw=black, fill=BrickRed!20, circle, inner sep=3.2pt] at (\l,0,0) {};
            \node at (\l,0,0) {${\sss +}$};
            \node[line width=.7pt, draw=black, fill=BrickRed!20, circle, inner sep=3.2pt] at (0,-\l,0) {};
            \node at (0,-\l,0) {${\sss +}$};
            \node[line width=.7pt, draw=black, fill=BrickRed!20, circle, inner sep=3.2pt] at (0,\l,0) {};
            \node at (0,\l,0) {${\sss +}$};
            
            \centerarc{dash pattern= on 0pt off 2.5pt, line width=1pt}{0,0}{43}{70}{.79*\l};
            \centerarc{dash pattern= on 0pt off 2.5pt, line width=1pt}{0,0}{-115}{-145}{.81*\l};
        }{}
        \ifthenelse{#1=1}{
            \draw[lata, ->-=.58] (0,0,-\h) -- (0,0,0);
            \draw[lata, ->-=.80] (0,0,0) -- (0,0,\h);
    
            \draw[latb, ->-=.58] (-\l,0,-\h) -- (-\l,0,0);
            \draw[latb, ->-=.80] (-\l,0,0) -- (-\l,0,\h);
    
            \draw[latb, ->-=.55] (\l,0,-\h) -- (\l,0,0);
            \draw[latb, ->-=.80] (\l,0,0) -- (\l,0,\h);
    
            \draw[latb, ->-=.6] (0,0,0) -- (-\l,0,0);
            \draw[latb, ->-=.6] (\l,0,0) -- (0,0,0);
    
            \node[line width=.7pt, draw=black, fill=BlueViolet!20, circle, inner sep=3.2pt] at (-\l,0) {};
            \node at (-\l,0) {${\sss +}$};
            \node[line width=.7pt, draw=black, fill=BrickRed!20, circle, inner sep=3.2pt] at (0,0) {};
            \node at (0,0) {${\sss Z}$};
            \node[line width=.7pt, draw=black, fill=BlueViolet!20, circle, inner sep=3.2pt] at (\l,0) {};
            \node at (\l,0) {${\sss +}$};
        }{}
    \end{tikzpicture}
}
\newcommand{\qformPEPOUnit}[1]{
    \begin{tikzpicture}[baseline={([yshift=-.5ex]current bounding box.center)},z={(0,1)},y={(0.6,0.5)}]
        \def\l{1.1};
        \def\h{.63};

        \ifthenelse{\equal{#1}{0}}{
	        \draw[lata, ->-=.55] (0,\l/2,0) -- (0,0,0);
	        \draw[lata, ->-=.55] (\l/2,0,0) -- (0,0,0);
	        \draw[lata, ->-=.64] (0,0,0) -- (0,-\l,0);
	        \draw[latb, ->-=.58] (0,0,-\h) -- (0,0,0);
	        \draw[latb, ->-=.80] (0,0,0) -- (0,0,\h);
	    
	        \draw[lata, ->-=.80] (-\l,0,0) -- (-3*\l/2,0,0);
	        \draw[lata, ->-=.80] (-\l,0,0) -- (-\l,0,\h);
	        
	        \draw[lata, ->-=.90] (0,-\l,0) -- (0,-3*\l/2,0);
	        \draw[lata, ->-=.80] (0,-\l,0) --node[int, pos=.80](mp){} (0,-\l,\h);
	        
	        \clipAroundNode{mp}{
	            \draw[lata, ->-=.44] (0,0,0) -- (-\l,0,0);
	        }
	        
	        \node[line width=.7pt, draw=black, fill=BlueViolet!20, circle, inner sep=3.pt] at (0,0) {};
	        \node at (0,0) {${\sss Z}$};
	        \node[line width=.7pt, draw=black, fill=BrickRed!20, circle, inner sep=3.pt] at (-\l,0) {};
	        \node at (-\l,0) {${\sss +}$};        
	        \node[line width=.7pt, draw=black, fill=BrickRed!20, circle, inner sep=3.pt] at (0,-\l) {};
	        \node at (0,-\l) {${\sss +}$};        
        }{}
        \ifthenelse{\equal{#1}{1}}{
        
            \draw[latb, ->-=.55] (\l/2,0,0) -- (0,0,0);
            \draw[latb, ->-=.55] (0,\l/2,0) -- (0,0,0);
            \draw[latb, ->-=.64] (0,0,0) -- (0,-\l,0);
            \draw[latb, ->-=.80] (0,0,0) -- (0,0,\h);
            
            \draw[latb, ->-=.80] (-\l,0,0) -- (-3*\l/2,0,0);
            
            \draw[lata, ->-=.58] (0,-\l,-\h) -- (0,-\l,0);
            \draw[lata, ->-=.80] (0,-\l,0) --node[int, pos=.8](mk){} (0,-\l,\h);
            \clipAroundNode{mk}{
                \draw[latb, ->-=.44] (0,0,0) -- (-\l,0,0);
            }
            \draw[lata, ->-=.80] (-\l,0,0) -- (-\l,0,\h);
            \draw[latb, ->-=.90] (0,-\l,0) -- (0,-3*\l/2,0);
            \draw[lata, ->-=.58] (-\l,0,-\h) -- (-\l,0,0);

            \node[line width=.7pt, draw=black, fill=BlueViolet!20, circle, inner sep=3.2pt] at (0,0) {};
            \node at (0,0) {${\sss +}$};

            \node[line width=.7pt, draw=black, fill=BrickRed!20, circle, inner sep=3.2pt] at (-\l,0) {};
            \node at (-\l,0) {${\sss Z}$};
            
            \node[line width=.7pt, draw=black, fill=BrickRed!20, circle, inner sep=3.2pt] at (0,-\l) {};
            \node at (0,-\l) {${\sss Z}$};
        }{} 
    
    	\centerarc{dash pattern= on 0pt off 2.5pt, line width=1pt}{0,0}{35}{75}{.5*\l};
    	\centerarc{dash pattern= on 0pt off 2.5pt, line width=1pt}{0,0}{-115}{-155}{1.2*\l};
    \end{tikzpicture}
}
\newcommand{\LSS}[1]{
    \begin{tikzpicture}[baseline={([yshift=-.5ex]current bounding box.center)},z={(0,1)},y={(0.5,0.5)}]
        \def\l{1.3};

        \ifthenelse{\equal{#1}{0}}{
            \fill[rounded corners,Tan!40] (-\l/2,\l+\l/10,0) -- (2*\l+\l/2,\l+\l/10,0) -- (2*\l+\l/2,\l-\l/10,0) -- (-\l/2,\l-\l/10,0) -- cycle;
            \fill[rounded corners,black!10] (\l-\l/10,-\l/4,0) -- (\l-\l/10,2*\l+\l/4,0) -- (\l+\l/10,2*\l+\l/4,0) -- (\l+\l/10,-\l/4,0) -- cycle;

            \foreach \x in {0,...,2}{
                \draw[lat] (\l*\x,-\l/2,0) -- (\l*\x,2*\l+\l/2,0);
                \draw[lat, dotted] (\l*\x,-\l/2,0) --+ (0,-\l/2,0);
                \draw[lat, dotted] (\l*\x,2*\l+\l/2,0) --+ (0,\l/2,0);
            }
            \foreach \y in {0,...,2}{
                \draw[lat] (-\l/2,\l*\y,0) -- (\l*2+\l/2,\l*\y,0);
                \draw[lat, dotted] (-\l/2,\l*\y,0) --+ (-\l/2,0,0);
                \draw[lat, dotted] (\l*2+\l/2,\l*\y,0) --+ (\l/2,0,0);
            }
            \foreach \x in {0,...,2}{
                \foreach \y in {0,...,2}{
                    \node[VertexNode] at (\x*\l,\y*\l) {};
                }
            }
            \foreach \x in {0,...,2}{
                \node[color=Tan] at (\x*\l-\l/4,\l+\l/4,\l/12) {$\sss Z_{\!g}$};
            }
            \foreach \y in {0,...,2}{
                \node[color=black!40] at (\l+\l/3,\l*\y-\l/3,0) {$\sss Z_{\!g'}$};
            }

            \def\o{(3.2*\l,3*\l,0)};
            \draw[dir] \o --+ (-.7*\l,0,0) node[anchor=east] {$\sss\hat x$};
            \draw[dir] \o --+ (0,-.7*\l,0) node[anchor=north east, pos=.9] {$\sss\hat y$};
        }{}

        \ifthenelse{\equal{#1}{1}}{
            \def\l{1.6};
        
            \fill[rounded corners,Tan!40] (-\l/2,\l/2+\l/10,0) -- (2*\l+\l/2,\l/2+\l/10,0) -- (2*\l+\l/2,\l/2-\l/10,0) -- (-\l/2,\l/2-\l/10,0) -- cycle;
            \fill[rounded corners,black!10] (\l/2-\l/10,-\l/4,0) -- (\l/2-\l/10,2*\l+\l/4,0) -- (\l/2+\l/10,2*\l+\l/4,0) -- (\l/2+\l/10,-\l/4,0) -- cycle;

            \foreach \x in {0,...,2}{
                \draw[lat] (\l*\x,-\l/2,0) -- (\l*\x,2*\l+\l/2,0);
                \draw[lat, dotted] (\l*\x,-\l/2,0) --+ (0,-\l/2,0);
                \draw[lat, dotted] (\l*\x,2*\l+\l/2,0) --+ (0,\l/2,0);
            }
            \foreach \y in {0,...,2}{
                \draw[lat] (-\l/2,\l*\y,0) -- (\l*2+\l/2,\l*\y,0);
                \draw[lat, dotted] (-\l/2,\l*\y,0) --+ (-\l/2,0,0);
                \draw[lat, dotted] (\l*2+\l/2,\l*\y,0) --+ (\l/2,0,0);
            }
            \foreach \x in {0,...,2}{
                \foreach \y in {0,...,2}{
                    \node[VertexNode] at (\x*\l,\y*\l) {};
                }
            }

            \foreach \x in {0,...,1}{
                \node[Tan] at (\x*\l+\l/4,\l/2+\l/4,0) {$\sss Z_{\!\chi}$};
            }
            \foreach \y in {0,...,1}{
                \node[black!50] at (\l/2+\l/4,\l*\y+\l/4.5,0) {$\sss Z_{\!\chi'}$};
            }
            
            \foreach \x in {0,1}{
                \draw[latdual] (\l*\x+\l/2,-\l/2,0) -- (\l*\x+\l/2,2*\l+\l/2,0);
                \draw[dotted, latdual] (\l*\x+\l/2,-\l/2,0) --+ (0,-\l/2,0);
                \draw[dotted, latdual] (\l*\x+\l/2,2*\l+\l/2,0) --+ (0,\l/2,0);
            }
            \foreach \y in {0,1}{
                \draw[latdual] (-\l/2,\l*\y+\l/2,0) -- (\l*2+\l/2,\l*\y+\l/2,0);
                \draw[dotted, latdual] (-\l/2,\l*\y+\l/2,0) --+ (-\l/2,0,0);
                \draw[dotted, latdual] (\l*2+\l/2,\l*\y+\l/2,0) --+ (\l/2,0,0);
            }

            \foreach \x in {0,...,1}{
                \foreach \y in {0,...,1}{
                    \node[PlaquetteNode] at (\x*\l+\l/2,\y*\l+\l/2) {};
                }
            }

            \def\o{(3.5*\l,-1*\l,0)};
            \draw[dirdual] \o --+ (-.7*\l,0,0) node[black, fill=white, left] {$\sss\hat y^\vee$};
            \draw[dirdual] \o --+ (0,.7*\l,0) node[black, fill=white, inner sep=2pt, pos=1.4] {$\sss\hat x^\vee$};
        }{}
    \end{tikzpicture}
}
\newcommand{\LSSGauss}[1]{
    \begin{tikzpicture}[baseline={([yshift=-.5ex]current bounding box.center)},z={(0,1)},y={(0.6,0.5)}]
        \def\l{1.4};

        \foreach \x in {0,1}{
            \draw (\l*\x,-\l/2,0) -- (\l*\x,\l+\l/2,0);
            \draw[dotted] (\l*\x,-\l/2,0) --+ (0,-\l/2,0);
            \draw[dotted] (\l*\x,\l+\l/2,0) --+ (0,\l/2,0);
        }
        \foreach \y in {0,1}{
            \draw (-\l/2,\l*\y,0) -- (\l+\l/2,\l*\y,0);
            \draw[dotted] (-\l/2,\l*\y,0) --+ (-\l/2,0,0);
            \draw[dotted] (\l+\l/2,\l*\y,0) --+ (\l/2,0,0);
        }

        \foreach \x in {0,1}{
            \foreach \y in {0,1}{
                \node[VertexNode] at (\x*\l,\y*\l) {};
            }
        }
    \ifthenelse{\equal{#1}{0}}{
        \begin{scope}[thin, decoration={markings, mark=at position 0.65 with {\arrow[line width=.25pt]{Latex}}}]
            \def\va{(0,\l,0)};
            \def\pa{(\l/2,\l/2,0)};
            \def\pb{(-\l/2,\l/2,0)};
            \def\pc{(\l/2,3*\l/2,0)};
            \def\pd{(-\l/2,3*\l/2,0)};

            \draw[postaction={decorate}] \pa -- \va;
            \draw[postaction={decorate}] \pd -- \va;
            \draw[postaction={decorate}] \va -- \pb;
            \draw[postaction={decorate}] \va -- \pc;

            \node[PlaquetteNode] at \pa {};
            \node[right] at \pa {$\sss  X_{\!g}^\dagger$};
            \node[PlaquetteNode] at \pb {};
            \node[left] at \pb {$\sss  X_{\!g}$};
            \node[PlaquetteNode] at \pc {};
            \node[above] at \pc {$\sss  X_{\!g}$};
            \node[PlaquetteNode] at \pd {};
            \node[above] at \pd {$\sss  X_{\!g}^\dagger$};

            \node[VertexNode] at \va {};
            \node[xshift=1pt] at (0,\l,\l/5) {$\sss  Z_{\!g}$};

            \node[draw] at (-3*\l/2,2*\l,0) {$\mc G_g^\msf v$};
            
        \end{scope}
    }{
    \ifthenelse{\equal{#1}{1}}{
        \begin{scope}[thin,decoration={markings, mark=at position 0.65 with {\arrow[line width=.25pt]{Latex}}}]
                
                \def\va{(\l/2,\l/2,0)};
                \def\pa{(\l,0,0)};
                \def\pb{(\l,\l,0)};
                \def\pc{(0,0,0)};
                \def\pd{(0,\l,0)};
    
                \draw[postaction={decorate}] \va -- \pa;
                \draw[postaction={decorate}] \va -- \pd;
                \draw[postaction={decorate}] \pb -- \va;
                \draw[postaction={decorate}] \pc -- \va;
    
                \node[VertexNode] at \pa {};
                \node[below right] at \pa {$\sss  X_{\!\chi}$};
                \node[VertexNode] at \pb {};
                \node[above left] at \pb {$\sss  X_{\!\chi}^\dagger$};
                \node[VertexNode] at \pc {};
                \node[above left] at \pc {$\sss  X_{\!\chi}^\dagger$};
                \node[VertexNode] at \pd {};
                \node[above left] at \pd {$\sss  X_{\!\chi}$};

                \node[PlaquetteNode] at \va {};
                \node[above] at (\l/2+0.16,\l/3+.12,0) {$\sss  Z_{\!\chi}$};

                \node[draw] at (-3*\l/2,2*\l,0) {$\mc G_\chi^\msf p$};
                
            \end{scope}
        }{}
    }
    \end{tikzpicture}
}
\newcommand{\FractonHam}[1]{
    \begin{tikzpicture}[baseline={([yshift=-.5ex]current bounding box.center)},z={(0,1)},y={(0.4,0.5)}]
        \def\l{1.2};

        \ifthenelse{\equal{#1}{v}}{
            
            \draw[latb] (0,0,-\l) -- (0,0,0);
            \node[op] at (0,0,-.5*\l) {$\sss Z_{\!g}$};

            \fill[BrickRed, opacity=.3] (-\l,-\l,0) -- (-\l,\l,0) -- (\l,\l,0) -- (\l,-\l,0) -- cycle;
            
            \draw[latb] (0,0,0) -- (0,0,\l);
    
            \node[op] at (0,0,.65*\l) {$\sss Z_{\!g}^\dagger$};
            \node[] at (.5*\l,.5*\l,0) {$\sss X_{\!g}$};
            \node[] at (-.5*\l,.5*\l,0) {$\sss X_{\!g}^\dagger$};
            \node[] at (.5*\l,-.5*\l,0) {$\sss X_{\!g}^\dagger$};
            \node[] at (-.5*\l,-.5*\l,0) {$\sss X_{\!g}$};

            \draw[white] (0,-\l,0) -- (0,\l,0);
            \draw[white] (-\l,0,0) -- (\l,0,0);
        }{
            \ifthenelse{\equal{#1}{STU}}{
           
            \def\va{(0,0,0)};
            \def\vb{(-\l,0,0)};
            \def\vc{(0,-\l,0)};
            \def\vd{(-\l/3,-\l/3,.75*\l)};
            \def\ve{(-\l/3,-\l/3,-.75*\l)};

            \draw[latdual] \ve -- \va -- \vd;
            \draw[latdual] \ve -- \vb -- \vd;
            \draw[latdual] \ve -- \vc --node[pos=.45](mp){} \vd;

            \clipAroundNode{mp}{
                \draw[latdual] \va -- \vb -- \vc -- cycle;
            }
            
            \node[op] at \va {$\sss X$};
            \node[op] at \vb {$\sss X$};
            \node[op] at \vc {$\sss X$};

            \node[op] at \vd {$\sss Z$};
            \node[op] at \ve {$\sss Z$};
        }{
        \ifthenelse{\equal{#1}{c}}{

            \node[op] at (0,0,0) {$\ssss Z_{\!\chi}$};
            
            \def\va{(-.5*\l,.5*\l,0)};
            \def\vb{(.5*\l,.5*\l,0)};
            \def\vc{(.5*\l,-.5*\l,0)};
            \def\vd{(-.5*\l,-.5*\l,0)};

            \draw[latb] \va --+ (0,0,\l);
            \draw[latb] \vb --+ (0,0,\l);
            \draw[latb] \vc --+ (0,0,\l);
            \draw[latb] \vd --+ (0,0,\l);
    
            \node[op] at (-.5*\l,.5*\l,\l/2) {$\sss X_{\!\chi}$};
            \node[op] at (.5*\l,.5*\l,\l/2) {$\sss X_{\!\chi}^\dagger$};
            \node[op] at (.5*\l,-.5*\l,\l/2) {$\sss X_{\!\chi}$};
            \node[op] at (-.5*\l,-.5*\l,\l/2) {$\sss X_{\!\chi}^\dagger$};

            \fill[BrickRed, opacity=.3] (-\l/2,-\l/2,0) -- (-\l/2,\l/2,0) -- (\l/2,\l/2,0) -- (\l/2,-\l/2,0) -- cycle;

            \fill[BrickRed, opacity=.3] (-\l/2,-\l/2,\l) -- (-\l/2,\l/2,\l) -- (\l/2,\l/2,\l) -- (\l/2,-\l/2,\l) -- cycle;

            \node[] at (0,0,\l) {$\ssss Z_{\!\chi}^\dagger$};
        }{
        \ifthenelse{\equal{#1}{d}}{
            
            \def\va{(-.5*\l,.5*\l,0)};
            \def\vb{(.5*\l,.5*\l,0)};
            \def\vc{(.5*\l,-.5*\l,0)};
            \def\vd{(-.5*\l,-.5*\l,0)};

            \draw[latb] \va --+ (0,0,\l);
            \draw[latb] \vb --+ (0,0,\l);
            \draw[latb] \vc --+ (0,0,\l);
            \draw[latb] \vd --+ (0,0,\l);
    
            \node[op] at (-.5*\l,.5*\l,\l/2) {$\sss X_{\!\chi}$};
            \node[op] at (.5*\l,.5*\l,\l/2) {$\sss X_{\!\chi}^\dagger$};
            \node[op] at (.5*\l,-.5*\l,\l/2) {$\sss X_{\!\chi}$};
            \node[op] at (-.5*\l,-.5*\l,\l/2) {$\sss X_{\!\chi}^\dagger$};

            \fill[BrickRed, opacity=.3] (-\l/2,-\l/2,\l) -- (-\l/2,\l/2,\l) -- (\l/2,\l/2,\l) -- (\l/2,-\l/2,\l) -- cycle;

            \node[] at (0,0,\l) {$\sss Z_{\!\chi}^\dagger$};
        }{}}}}        
    \end{tikzpicture}
}
\newcommand{\LSSCheck}[1]{
    \begin{tikzpicture}[baseline={([yshift=-.5ex]current bounding box.center)},z={(0,1)},y={(0.6,0.5)}]
        \def\l{1.4};

        \foreach \x in {0,1}{
            \draw (\l*\x,-\l/2,0) -- (\l*\x,\l+\l/2,0);
            \draw[dotted] (\l*\x,-\l/2,0) --+ (0,-\l/2,0);
            \draw[dotted] (\l*\x,\l+\l/2,0) --+ (0,\l/2,0);
        }
        \foreach \y in {0,1}{
            \draw (-\l/2,\l*\y,0) -- (\l+\l/2,\l*\y,0);
            \draw[dotted] (-\l/2,\l*\y,0) --+ (-\l/2,0,0);
            \draw[dotted] (\l+\l/2,\l*\y,0) --+ (\l/2,0,0);
        }

        \foreach \x in {0,1}{
            \foreach \y in {0,1}{
                \node[VertexNode] at (\x*\l,\y*\l) {};
            }
        }

        \ifthenelse{\equal{#1}{0}}{
            \node[above left] at (\l,0,0) {$\sss  X_{\!\chi}^\dagger$};
            \node[above left] at (\l,\l,0) {$\sss  X_{\!\chi}$};
            \node[above left] at (0,0,0) {$\sss  X_{\!\chi}$};
            \node[above left] at (0,\l,0) {$\sss  X_{\!\chi}^\dagger$};
    
            \node[draw] at (-1.75*\l,2.5*\l,0) {$X^c_\chi$};
        }{}
        \ifthenelse{\equal{#1}{1}}{
            \node[above left] at (\l,0,0) {$\sss  X_{\!\chi}$};
            \node[above left] at (\l,\l,0) {$\sss  X_{\!\chi}^\dagger$};
            \node[above left] at (0,0,0) {$\sss  X_{\!\chi}^\dagger$};
            \node[above left] at (0,\l,0) {$\sss  X_{\!\chi}$};
    
            \node[draw] at (-1.75*\l,2.5*\l,0) {$X^{c\dagger}_\chi$};
        }{}
        
    \end{tikzpicture}
}
\newcommand{\LSSPEPOUnit}[1]{
    \begin{tikzpicture}[baseline={([yshift=-.5ex]current bounding box.center)},z={(0,1)},y={(0.4,.52)}]
        \def\l{1.1};
        \def\h{0.65};
        
        \ifthenelse{\equal{#1}{0}}{
            \draw[lata, ->-=.80] (0,0,0) -- (0,0,\h);
        
            \draw[lata, ->-=.60] (-\l/2,-\l/2,0) -- (0,0,0);
            \draw[lata, ->-=.52] (\l,\l,0) --+ (-\l,-\l,0);
            \draw[lata, ->-=.87] (0,0,0) -- (-\l/2,\l/2,0);
            \draw[lata, ->-=.87] (0,0,0) --+ (\l/2,-\l/2,0);
    
            \draw[lata, ->-=.60] (3*\l/2,\l/2,0) --+ (-\l/2,\l/2,0);
            \draw[lata, ->-=.52] (\l/2,3*\l/2,0) --+ (\l/2,-\l/2,0);
            \draw[lata, ->-=.70] (\l,\l,0) --+ (\l/2,\l/2,0);

            \draw[latb, ->-=.58] (\l,\l,-\h) -- (\l,\l,0);
            \draw[latb, ->-=.80] (\l,\l,0) -- (\l,\l,\h);
            
            \node[line width=.7pt, draw=black, fill=BrickRed!20, circle, inner sep=3.pt] at (0,0,0) {};
            \node at (0,0,0) {${\sss +}$};
            \node[line width=.7pt, draw=black, fill=BlueViolet!20, circle, inner sep=3.pt] at (\l,\l,0) {};
            \node at (\l,\l,0) {${\sss Z}$}; 
        }{}
        
        \ifthenelse{\equal{#1}{1}}{
            \draw[latb, ->-=.60] (-\l/2,-\l/2,0) --+ (\l/2,\l/2,0);
            \draw[latb, ->-=.52] (\l,\l,0) --+ (-\l,-\l,0);
            \draw[latb, ->-=.87] (0,0,0) --+ (-\l/2,\l/2,0);
            \draw[latb, ->-=.87] (0,0,0) --+ (\l/2,-\l/2,0);
    
            \draw[latb, ->-=.60] (3*\l/2,\l/2,0) --+ (-\l/2,\l/2,0);
            \draw[latb, ->-=.52] (\l/2,3*\l/2,0) --+ (\l/2,-\l/2,0);
            \draw[latb, ->-=.70] (\l,\l,0) --+ (\l/2,\l/2,0);

            \draw[lata, ->-=.58] (0,0,-\h) -- (0,0,0);
            \draw[lata, ->-=.80] (0,0,0) -- (0,0,\h);
            \draw[latb, ->-=.80] (\l,\l,0) -- (\l,\l,\h);
    
            \node[line width=.7pt, draw=black, fill=BrickRed!20, circle, inner sep=3.2pt] at (0,0,0) {};
            \node at (0,0,0) {${\sss Z}$};
            \node[line width=.7pt, draw=black, fill=BlueViolet!20, circle, inner sep=3.2pt] at (\l,\l,0) {};
            \node at (\l,\l,0) {${\sss +}$};
        }{}
    \end{tikzpicture}
}
\newcommand{\LSSPEPOState}{
    \begin{tikzpicture}[baseline={([yshift=-.5ex]current bounding box.center)}, z={(0,1)},y={(.4,.52)}]
        \def\l{1.1};
        \def\h{1.30};
        
        \draw[lata, ->-=.55] (0,0,0) -- (0,0,\h);
        
        \draw[lata, ->-=.60] (-\l/2,-\l/2,0) -- (0,0,0);
        \draw[lata, ->-=.52] (\l,\l,0) --+ (-\l,-\l,0);
        \draw[lata, ->-=.87] (0,0,0) -- (-\l/2,\l/2,0);
        \draw[lata, ->-=.87] (0,0,0) --+ (\l/2,-\l/2,0);

        \draw[lata, ->-=.60] (3*\l/2,\l/2,0) --+ (-\l/2,\l/2,0);
        \draw[lata, ->-=.52] (\l/2,3*\l/2,0) --+ (\l/2,-\l/2,0);
        \draw[lata, ->-=.70] (\l,\l,0) --+ (\l/2,\l/2,0);

        \draw[latb, dotted] (\l,\l,-3*\h/4) -- (\l,\l,-\h/2);
        \draw[latb, ->-=.60] (\l,\l,-\h/2) -- (\l,\l,0);
        \draw[latb, ->-=.80] (\l,\l,0) -- (\l,\l,\h/2);

        \draw[latb, ->-=.60] (-\l/2,-\l/2,\h) --+ (\l/2,\l/2,0);
        \draw[latb, ->-=.52] (\l,\l,\h) --+ (-\l,-\l,0);
        \draw[latb, ->-=.87] (0,0,\h) --+ (-\l/2,\l/2,0);
        \draw[latb, ->-=.87] (0,0,\h) --+ (\l/2,-\l/2,0);

        \draw[latb, ->-=.60] (3*\l/2,\l/2,\h) --+ (-\l/2,\l/2,0);
        \draw[latb, ->-=.52] (\l/2,3*\l/2,\h) --+ (\l/2,-\l/2,0);
        \draw[latb, ->-=.70] (\l,\l,\h) --+ (\l/2,\l/2,0);
        
        \draw[lata, ->-=.80] (0,0,\h) -- (0,0,3*\h/2);
        \draw[latb, ->-=.80] (\l,\l,\h) -- (\l,\l,3*\h/2);
        \draw[latb, dotted] (\l,\l,3*\h/2) -- (\l,\l,7*\h/4);

        \node[line width=.7pt, draw=black, fill=BrickRed!20, circle, inner sep=3.2pt] at (0,0,0) {};
        \node at (0,0,0) {${\sss +}$};
        \node[line width=.7pt, draw=black, fill=BlueViolet!20, circle, inner sep=3.2pt] at (\l,\l,0) {};
        \node at (\l,\l,0) {${\sss Z}$};
        \node[line width=.7pt, draw=black, fill=BrickRed!20, circle, inner sep=3.2pt] at (0,0,\h) {};
        \node at (0,0,\h) {${\sss Z}$};
        \node[line width=.7pt, draw=black, fill=BlueViolet!20, circle, inner sep=3.2pt] at (\l,\l,\h) {};
        \node at (\l,\l,\h) {${\sss +}$};
        
    \end{tikzpicture}
}
\newcommand{\SierpinskiLattice}[1]{
\begin{tikzpicture}[ baseline={([yshift=-.5ex]current bounding box.center)},z={(0,1)},y={(0.5,0.5)}]
        \def\l{1.2};
        
        \foreach \x in {1,...,2}{
            \draw[lat] (\l*\x,0,0) -- (\l*\x,3*\l,0);
        }
        
        \draw[lat, ->-=.55] (0,3*\l,0) -- (0,0,0);
        \draw[lat, ->-=.55] (3*\l,3*\l,0) -- (3*\l,0,0);
        
        \foreach \y in {0,...,3}{
            \draw[lat] (0,\l*\y,0) -- (\l*3,\l*\y,0);
        }
        
        \begin{scope}[decoration={markings,
                mark=at position 0.5 with {\arrow[line width=.25pt]{Latex}},
                mark=at position 0.54 with {\arrow[line width=.25pt]{Latex}}}
            ]
            \draw[lat, postaction={decorate}] (3*\l,0,0) -- (0,0,0);
            \draw[lat, postaction={decorate}] (3*\l,3*\l,0) -- (0,3*\l,0);
        \end{scope}
        
        \foreach \x in {0,...,2}{
            \foreach \y in {0,...,2}{
                \draw[lat] (\x*\l+\l,\y*\l) --++ (-\l,\l) {};
            }
        }
            
        \def\o{(3.75*\l,3.5*\l,0)};
        \draw[dir] \o --+ (-.7*\l,0,0) node[left] {$\sss\hat x$};
        \draw[dir] \o --+ (0,-.7*\l,0) node[pos=1.4] {$\sss\hat y$};

        \ifthenelse{\equal{#1}{0}}{
            \foreach \x in {1,...,3}{
                \foreach \y in {1,...,3}{
                    \node[fill, circle, inner sep=.75pt] at ({(\x-.35)*\l},{(\y-.35)*\l},0) {};
                    \draw[thin, dotted] ({(\x-.35)*\l},{(\y-.35)*\l},0) -- (\x*\l,\y*\l,0);
                }
            }
        }{}
        
        \ifthenelse{\equal{#1}{1}}{
            \node[op, above] at (3*\l/4,5*\l/4,0) {$\sss Z$};
            \node[op, above] at (11*\l/4,5*\l/4,0) {$\sss Z$};
            \node[op, above] at (3*\l/4,9*\l/4,0) {$\sss Z$};
            \node[op, above] at (7*\l/4,9*\l/4,0) {$\sss Z$};
            \node[op, above] at (7*\l/4,13*\l/4,0) {$\sss Z$};
            \node[op, above] at (11*\l/4,13*\l/4,0) {$\sss Z$};
        }{};

        \foreach \x in {0,...,2}{
            \foreach \y in {0,...,2}{
                    \fill[BrickRed, opacity=.3] (\x*\l+\l,\y*\l,0) -- (\x*\l+\l,\y*\l+\l,0) -- (\x*\l,\y*\l+\l,0) -- cycle;
            }
        }
        
        \foreach \x in {0,...,3}{
            \foreach \y in {0,...,3}{
                \node[VertexNode] at (\x*\l,\y*\l) {};
            }
        }
    \end{tikzpicture}
}
\newcommand{\SierpinskiGauss}[1]{
\begin{tikzpicture}[ baseline={([yshift=-.5ex]current bounding box.center)},z={(0,1)},y={(0.5,0.5)}]
        \def\l{1.2};
        
        \foreach \x in {0,...,1}{
                \foreach \y in {0,...,1}{
                    \fill[BrickRed, opacity=.3] (\x*\l+\l,\y*\l,0) -- (\x*\l+\l,\y*\l+\l,0) -- (\x*\l,\y*\l+\l,0) -- cycle;
            }
        }
        \foreach \x in {0,...,2}{
            \draw[lat] (\l*\x,-\l/2,0) -- (\l*\x,5*\l/2,0);
            \draw[lat, dotted] (\l*\x,-\l/2,0) --++ (0,-\l/2,0);
            \draw[lat, dotted] (\l*\x,5*\l/2,0) --++ (0,\l/2,0);
        }
        \foreach \y in {0,...,2}{
            \draw[lat] (-\l/2,\l*\y,0) -- (5*\l/2,\l*\y,0);
            \draw[lat, dotted] (-\l/2,\l*\y,0) --++ (-\l/2,0,0);
            \draw[lat, dotted] (5*\l/2,\l*\y,0) --++ (\l/2,0,0);
        }
        \ifthenelse{\equal{#1}{0}}{
            \node[above] at (\l,\l,0) {$\sss Z$};
            \node[above] at (3*\l/4,4*\l/3,0) {$\sss X$};
            \node[above] at (3*\l/4,\l/3,0) {$\sss X$};
            \node[above] at (7*\l/4,\l/3,0) {$\sss X$};

            \node[draw] at (-3*\l/2,3*\l,0) {$\mc G_1^\msf v$};
        }{
            \node[above] at (\l,\l,0) {$\sss X$};
            \node[above] at (2*\l,\l,0) {$\sss X$};
            \node[above] at (2*\l,0,0) {$\sss X$};
            \node[above] at (7*\l/4,\l/3,0) {$\sss Z$};

            \node[draw] at (-3*\l/2,3*\l,0) {$\mc G_1^{\msf p^+}$};
        };
        \foreach \x in {0,...,1}{
            \foreach \y in {0,...,1}{
                \draw[lat] (\x*\l+\l,\y*\l) --++ (-\l,\l) {};
            }
        };
        \foreach \x in {0,...,2}{
            \foreach \y in {0,...,2}{
                \node[VertexNode] at (\x*\l,\y*\l) {};
            }
        };
    \end{tikzpicture}
}
\newcommand{\SierpinskiCheck}[0]{
\begin{tikzpicture}[ baseline={([yshift=-.5ex]current bounding box.center)},z={(0,1)},y={(0.5,0.5)}]
        \def\l{1.2};

        \foreach \x in {0,...,1}{
            \draw[lat] (\l*\x,-\l/2,0) -- (\l*\x,3*\l/2,0);
            \draw[lat, dotted] (\l*\x,-\l/2,0) --++ (0,-\l/2,0);
            \draw[lat, dotted] (\l*\x,3*\l/2,0) --++ (0,\l/2,0);
        }
        \foreach \y in {0,...,1}{
            \draw[lat] (-\l/2,\l*\y,0) -- (3*\l/2,\l*\y,0);
            \draw[lat, dotted] (-\l/2,\l*\y,0) --++ (-\l/2,0,0);
            \draw[lat, dotted] (3*\l/2,\l*\y,0) --++ (\l/2,0,0);
        }
        \node[above] at (0,\l,0) {$\sss X$};
        \node[above] at (\l,0,0) {$\sss X$};
        \node[above] at (\l,\l,0) {$\sss X$};

        \foreach \x in {0,...,1}{
            \foreach \y in {0,...,1}{
                \node[VertexNode] at (\x*\l,\y*\l,0) {};
            }
        }
        
    \end{tikzpicture}
}
\newcommand{\SierpinskiHam}[1]{
    \begin{tikzpicture}[baseline={([yshift=-.5ex]current bounding box.center)},z={(0,1)},y={(0.28,0.25)}]
        \def\l{1.4};
        \ifthenelse{\equal{#1}{0}}{
           
            \def\va{(0,0,0)};
            \def\vb{(-\l,0,0)};
            \def\vc{(-\l,-\l,0)};
            \def\vd{(-3*\l/4,-2*\l/3,.75*\l)};
            \def\ve{(-3*\l/4,-2*\l/3,-.75*\l)};

            \draw[latdual] \ve -- \va -- \vd;
            \draw[latdual] \ve -- \vc -- \vd;
            \draw[latdual] \va -- \vb -- \vc -- node[pos=.25](mp){}  cycle;

            \clipAroundNode{mp}{
                \draw[latdual] \ve -- \vb -- \vd;
            }
            
            \node[op] at \va {$\color{\colora}\sss X$};
            \node[op] at \vb {$\color{\colora}\sss X$};
            \node[op] at \vc {$\color{\colora}\sss X$};

            \node[op] at \vd {$\color{\colorb}\sss Z$};
            \node[op] at \ve {$\color{\colorb}\sss Z$};
        }{}
        \ifthenelse{\equal{#1}{1}}{
           
            \def\va{(0,0,0)};
            \def\vb{(-\l,0,0)};
            \def\vc{(0,-\l,0)};
            \def\vd{(-\l/3,-\l/3,.75*\l)};
            \def\ve{(-\l/3,-\l/3,-.75*\l)};

            \draw[latdual] \ve -- \va -- \vd;
            \draw[latdual] \ve -- \vb -- \vd;
            \draw[latdual] \ve -- \vc --node[pos=.3](mp){} \vd;
            
            \clipAroundNode{mp}{
                \draw[latdual] \va -- \vb -- \vc -- cycle;
            }
            
            \node[op] at \va {$\color{\colorb}\sss X$};
            \node[op] at \vb {$\color{\colorb}\sss X$};
            \node[op] at \vc {$\color{\colorb}\sss X$};

            \node[op] at \vd {$\color{\colora}\sss Z$};
            \node[op] at \ve {$\color{\colora}\sss Z$};
        }{}
        \ifthenelse{\equal{#1}{3}}{
           
            \def\va{(0,0,0)};
            \def\vb{(-\l,0,0)};
            \def\vc{(0,-\l,0)};
            \def\vd{(-\l/3,-\l/3,.75*\l)};

            \draw[latdual] \vd -- \va ;
            \draw[latdual] \vd -- \vb ;
            \draw[latdual] \vd -- \vc ;

            \clipAroundNode{mp}{
                \draw[latdual] \va -- \vb -- \vc -- cycle;
            }
            
            \node[op] at \va {$\color{\colorb}\sss X$};
            \node[op] at \vb {$\color{\colorb}\sss X$};
            \node[op] at \vc {$\color{\colorb}\sss X$};

            \node[op] at \vd {$\color{\colora}\sss Z$};
        }{}
    \end{tikzpicture}
}
\newcommand{\ZtwoPEPOTensor}[7]{
    \begin{tikzpicture}[baseline={(0,0,.2)},z={(0,1)},y={(0,.5)}]
        \def\l{1.5};
        \def\h{.65};

        \ifthenelse{\equal{#1}{0}}{
        
            \draw[lata] (\l,0,0) -- (2*\l/3,2*\l/3,0) node[op, pos=1.3] {$\sss #2$};
            \draw[lata] (\l,0,0) -- (2*\l/3,-\l/3,0) node[op, pos=1.3] {$\sss #3$};
            \draw[lata] (\l,0,0) -- (5*\l/3,-\l/3,0) node[op, pos=1.3] {$\sss #4$};

            \draw[lata] (\l,0,-\h) -- (\l,0,0) node[op, pos=-.3] {$\sss #5$};
            \draw[lata] (\l,0,0) -- (\l,0,\h) node[op, pos=1.3] {$\sss #6$};
            
            \node[line width=.7pt, draw=black, fill=BrickRed!20, circle, inner sep=3.2pt] at (\l,0,0) {};
            \node at (\l,0,0) {${\sss Z}$};
        }{}
        \ifthenelse{\equal{#1}{1}}
        {
            \draw[lata] (2*\l/3,-\l/3,0) -- (\l,0,0) node[black, pos=1.3] {$\sss #2$};
            \draw[lata] (2*\l/3,-\l/3,0) -- (\l,-\l,0) node[black, pos=1.3] {$\sss #3$};
            \draw[lata] (2*\l/3,-\l/3,0) -- (0,0,0) node[black, pos=1.3] {$\sss #4$};
            \draw[lata] (2*\l/3,-\l/3,0) -- (2*\l/3,-\l/3,\h) node[black, pos=1.3] {$\sss #4$};
            
            \node[line width=.7pt, draw=black, fill=BrickRed!20, circle, inner sep=3.2pt] at (2*\l/3,-\l/3,0) {};
            \node at (2*\l/3,-\l/3,0) {${\sss +}$};
        }{}
    \end{tikzpicture}
}

\newcommand{\ZtwoPEPOUnit}[2]{
    \begin{tikzpicture}[baseline={(0,0,.2)},z={(0,1)},y={(0,.5)}]
        \def\l{1.5};
        \def\h{0.65};

        \ifthenelse{\equal{#2}{Z}}{
            \draw[lata] (\l,0,-\h) -- (\l,0,0);
        }{}
        \ifthenelse{\equal{#1}{Z}}{
            \draw[lata] (2*\l/3,-\l/3,-\h) -- (2*\l/3,-\l/3,0);
        }{}
        
        \draw[lata] (\l,0,0) -- (2*\l/3,2*\l/3,0);
        \draw[lata] (\l,0,0) -- (2*\l/3,-\l/3,0);
        \draw[lata] (\l,0,0) -- (5*\l/3,-\l/3,0);
        
        \draw[lata] (\l,0,0) -- (\l,0,\h);

        \draw[lata] (2*\l/3,-\l/3,0) -- (\l,0,0);
        \draw[lata] (2*\l/3,-\l/3,0) -- (\l,-\l,0);
        \draw[lata] (2*\l/3,-\l/3,0) -- (0,0,0);
        \draw[lata] (2*\l/3,-\l/3,0) -- (2*\l/3,-\l/3,\h);
        
        \node[line width=.7pt, draw=black, fill=BrickRed!20, circle, inner sep=3.2pt] at (2*\l/3,-\l/3,0) {};
        \node at (2*\l/3,-\l/3,0) {${\sss #1}$};
        
        \node[line width=.7pt, draw=black, fill=BrickRed!20, circle, inner sep=3.2pt] at (\l,0,0) {};
        \node at (\l,0,0) {${\sss #2}$};

    \end{tikzpicture}
}
\newcommand{\FractonPEPOState}{
    \begin{tikzpicture}[baseline={([yshift=-.5ex]current bounding box.center)},z={(0,1)},y={(.1,.52)}]
        \def\l{1.35};
        \def\h{1.1};

        \draw[lata, dotted] (\l,0,-3*\h/4) --+ (0,0,\h/4);
        \draw[lata] (\l,0,-\h/2) --+ (0,0,\h/2);
        \draw[lata] (\l,0,0) --+ (0,0,\h/2);
        
        \draw[lata] (\l,0,0) --+ (-\l/3,2*\l/3,0);
        \draw[lata] (\l,0,0) --+ (-\l/3,-\l/3,0);
        \draw[lata] (\l,0,0) --+ (2*\l/3,-\l/3,0);

        \draw[lata] (2*\l/3,-\l/3,0) --+ (\l/3,\l/3,0);
        \draw[lata] (2*\l/3,-\l/3,0) --+ (\l/3,-2*\l/3,0);
        \draw[lata] (2*\l/3,-\l/3,0) --+ (-2*\l/3,\l/3,0);
        \draw[lata] (2*\l/3,-\l/3,0) --+ (0,0,\h);

        \draw[lata] (\l,0,\h) --+ (-\l/3,2*\l/3,0);
        \draw[lata] (\l,0,\h) --+ (-\l/3,-\l/3,0);
        \draw[lata] (\l,0,\h) --+ (2*\l/3,-\l/3,0);
        \draw[lata] (\l,0,\h) --+ (0,0,\h/2);
        \draw[lata, dotted] (\l,0,3*\h/2) --+ (0,0,\h/4);

        \draw[lata] (2*\l/3,-\l/3,\h) --+ (\l/3,\l/3,0);
        \draw[lata] (2*\l/3,-\l/3,\h) --+ (\l/3,-2*\l/3,0);
        \draw[lata] (2*\l/3,-\l/3,\h) --+ (-2*\l/3,\l/3,0);
        \draw[lata] (2*\l/3,-\l/3,\h) --+ (0,0,\h/2);
        
        \node[line width=.7pt, draw=black, fill=BrickRed!20, circle, inner sep=3.2pt] at (2*\l/3,-\l/3,0) {};
        \node at (2*\l/3,-\l/3,0) {${\sss +}$};
        \node[line width=.7pt, draw=black, fill=BrickRed!20, circle, inner sep=3.2pt] at (\l,0,0) {};
        \node at (\l,0,0) {${\sss Z}$};
        \node[line width=.7pt, draw=black, fill=BrickRed!20, circle, inner sep=3.2pt] at (2*\l/3,-\l/3,\h) {};
        \node at (2*\l/3,-\l/3,\h) {${\sss Z}$};
        \node[line width=.7pt, draw=black, fill=BrickRed!20, circle, inner sep=3.2pt] at (\l,0,\h) {};
        \node at (\l,0,\h) {${\sss +}$};

    \end{tikzpicture}
}

%% file: _intro.tex
\section*{Introduction}
Different incarnations of \emph{bulk-boundary correspondence} are ubiquitous in condensed matter physics. Generically, a bulk-boundary correspondence establishes a relation between the non-trivial physics in the bulk of a certain physical system and the physics constrained to its boundary. One example of such a relation is the conjecture that the bulk of certain \emph{anomaly-free} topological orders is fully determined by their gapped boundary theory by means of an appropriate notion of the categorical \emph{center}~\cite{Kitaev11,Kong14,Kong17}. A bulk-boundary correspondence has also proven to be the key for finding cohomological invariants that differentiate certain symmetry-protected topological phases (SPTs) protected by an invertible symmetry, whereby these invariants pertain to the transformation properties of the edge modes under the symmetry group. For the particular case of (1+1)d, this approach has led to a complete classification of gapped one-dimensional phases with group-like symmetry \cite{Chen11,Schuch11}. More recently, it has also been argued how the boundaries of particular fracton phases host \emph{emergent} subsystem symmetries whose explicit realization can be traced back to the characteristics of the bulk excitation spectrum~\cite{Schuster23}. Contrary to the case of intrinsic topological orders, it was demonstrated that the \emph{bulk reconstruction} of the fracton orders under scrutiny depends in a non-trivial way on the relative orientation of the boundary with respect to the bulk. In a certain sense all of the above examples of bulk-boundary correspondence can thus be brought together under the umbrella term of a \emph{holographic principle}.

In another recent series of developments an opposite avenue has been pursued in which a given quantum field theory (QFT) with a (higher) \emph{categorical symmetry} is regarded as an interval compactification of a topological field theory in one dimension higher, coined `SymTFT', with a \emph{topological} gapped boundary condition determining the symmetry of the QFT and a not necessarily gapped \emph{physical} boundary condition that dictates the dynamics of the QFT~\cite{Apruzzi21,Kaidi22}. In this picture the focus is very much on the understanding of certain features of the symmetric QFT, such as (dis)order parameters, gapped phases and anomalies, from the higher-dimensional SymTFT. Aspects of this paradigm have previously appeared under the name of `strange correlator', which admits an explicit lattice realization in terms of tensor network representations of the higher-dimensional topological order~\cite{You13,Vanhove18,Vanhove21}.

\bigskip\noindent In this manuscript we realize an explicit bulk-boundary correspondence for generic abelian \emph{boundary symmetries} in any dimension by making use of the \emph{iterative gauging procedure} proposed by one of the authors in~\cite{Garre24} for \emph{one-dimensional abelian boundary symmetries}. That work builds on the observation that gauging an \emph{initial} abelian (anomaly-free global 0-form) symmetry gives rise to \emph{dual} on-site global symmetry generators labeled by the one-dimensional representations of the underlying symmetry group. Since this emergent symmetry is on-site, a tensor product of local unitaries, it is free of any 't Hooft anomaly, and so it can be gauged, thereby recovering the initial $G$ symmetry. The crux of the paradigm put forward in~\cite{Garre24} is that these gauging procedures can be repeated indefinitely and that in this way a state emerges which naturally lives in one dimension higher than the initial symmetry. Such an \emph{emergent state} was shown in~\cite{Garre24} to be stabilized by certain local operators whose origin can be traced back to the Gauss constraints of the involved gauging procedures. In this picture, the initial symmetry which is being gauged can be thought of as a \emph{boundary symmetry} for the emergent state. This idea is made very tangible by making use of the explicit \emph{gauging map} for quantum states proposed in~\cite{Haegeman14} and generalized later in \cite{Williamson16}. One of the main advantages of using this approach is that a tensor network description of the gauging maps is readily available, which can in turn be used to write down an explicit tensor network representation of the emergent state, facilitating the analytical study of its properties and its numerical simulation.

In the current manuscript, Ref.~\cite{Garre24} is extended to arbitrary dimensions and generic abelian symmetries. Specifically, we propose a slight generalization of the gauging procedure of~\cite{Williamson16} to arbitrary abelian groups, and demonstrate that the iterative gauging procedure of~\cite{Garre24} in \emph{any dimension} gives rise to emergent states with corresponding \emph{local stabilizer} Hamiltonians. The abelian symmetries which are being gauged in the framework of~\cite{Williamson16} are completely specified by a \emph{complete collection} of `checks' which commute with all symmetry generators. As such, that framework includes the gauging of \emph{fractal} and \emph{subsystem} symmetries as specific cases. In our general framework we are also able to relate inequivalent choices of \emph{gapped boundary stabilizers} to different spontaneous symmetry breaking patters of the boundary symmetry.

To showcase our framework we provide three detailed worked examples where the initial boundary symmetry is two-dimensional. Gauging a global on-site \emph{0-form} symmetry in (2+1)d gives rise to emergent \emph{1-form} -- or \emph{Wilson loop}-- symmetries. In this case the iterative gauging framework gives rise to three-dimensional \emph{Clifford-deformed surface codes}~\cite{Huang23}, which are up to a local basis change equivalent to the three-dimensional abelian quantum doubles~\cite{Kitaev97}. Two-dimensional linear subsystem symmetries map to dual linear subsystem symmetries on the dual lattice. Applying the framework in this case gives rise to foliated anisotropic type-I fracton orders~\cite{Shirley19}. Finally, we leverage the full potential of our framework and the gauging procedure of~\cite{Williamson16} to gauge fractal boundary symmetry, specifically of Sierpinski type, and obtain the fracton model proposed by Caselnovo and Chamon in~\cite{Castelnovo12} for the study of quantum glassiness. For each of these examples we also provide a tensor network description of the involved gauging maps, and hence the emergent state, which is based on the tensor network description of the gauging map for global on-site 0-form symmetries of~\cite{Haegeman14}.

Our manuscript is organized as follows. In \cref{sec:TopoOrder} we provide an introductory example to explicitly demonstrate the iterative gauging procedure applied to an initial two-dimensional global on-site symmetry. All the details of our framework including the study of symmetry-breaking boundary conditions and the tensor network representation of the emergent Clifford-deformed CSS codes are discussed in detail. This section also serves as an introduction to many of the conventions and notations used throughout this manuscript. We then proceed by putting these observations on systematic footing in \cref{sec:GeneralFramework}. First, we slightly generalize the gauging procedure of \cite{Williamson16} to generic abelian groups and then prove that the iterative gauging procedure in any dimension gives rise to emergent states with a corresponding commuting stabilizer Hamiltonian of which it is a ground state. We then provide a way to add boundary stabilizers to the emergent Hamiltonian which commute with all terms in the bulk and among each other, thereby preserving the gap. We argue that different gapped boundary stabilizers correspond to certain symmetry breaking patterns of the boundary symmetry. We then conclude with two more examples in \cref{sec:FractonLSS} and \cref{sec:FractonFractal} that pertain to the derivation of different type-I fracton models from the iterative gauging of linear subsystem boundary symmetry and Sierpinski fractal symmetry respectively.
\begin{figure}
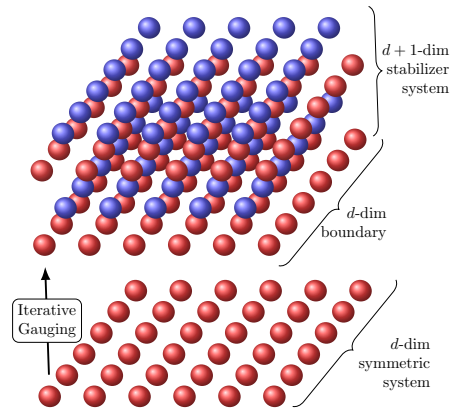

    \centering
    \SketchBulkBdry
    \caption{Schematic overview of the iterative gauging procedure. Given a d-dimensional symmetric state, iterative gauging of its symmetry and the emerging dual symmetries give rise to a (d+1)-dimensional stabilizer model with the initial state as boundary.}
    \label{fig}
\end{figure}

%% file: _topo_order.tex
\section{Introductory example:\texorpdfstring{\\}{ }Clifford-deformed surface codes from abelian higher-form symmetries}\label{sec:TopoOrder}
{\it In this section we review the gauging procedure of abelian on-site global 0-form symmetries proposed in~\cite{Haegeman14} for two-dimensional symmetric states. We argue that the gauged state is in the even sector of a dual global 1-form $\widehat G$ symmetry acting non-trivially solely on the introduced gauge degrees of freedom. We use an explicit gauging procedure for this 1-form symmetry, following refs.~\cite{Rayhaun23,Williamson16}, that yields in turn again the original $G$ symmetry. For both gauging maps we construct a tensor network representation of bond dimension $|G|$. We show that the concatenation of these 2D gauging maps as in~\cite{Garre24} produces 3D Clifford-deformed surface codes \cite{Huang23}, which are unitarily equivalent to abelian $G$ quantum doubles~\cite{Kitaev97}.}
\Sep

\noindent Let us consider the two-dimensional torus endowed with a lattice $\Lambda$, whose sets of vertices, oriented edges and plaquettes will be denoted by $\msf V(\Lambda)$, $\msf E(\Lambda)$ and $\msf P(\Lambda)$ respectively. The starting point of the gauging procedure laid out in~\cite{Haegeman14} is then a two-dimensional state $\ket{\psi}$ living in a tensor product Hilbert space of the so-called \emph{matter} degrees of freedom supported on the vertices of $\Lambda$, $\ket{\psi}\in\bigotimes_{\msf v\in\msf V(\Lambda)}\mc H^\msf v$. The local Hilbert space $\mc H^\msf v$ forms a unitary representation of $G$, which is assumed to be abelian throughout the manuscript, and whose matrix representation is written as $\{u_g | g\in G\}$. We adapt the notation $\overline g$ to denote the inverse group element of $g\in G$ and $1$ to denote the identity element. The state $\ket{\psi}$ is then required to transform trivially under the tensor product representation $\{\bigotimes_{\msf v\in\msf V(\Lambda)} u_g^\msf v|g\in G\}$\footnote{Indeed, if $\ket{\psi}$ transforms non-trivially under $G$, the gauging map defined below annihilates $\ket{\psi}$. However, a different gauging map can be defined for each symmetry sector.}, i.e.:
\begin{equation}
	\bigg(\bigotimes_{\msf v\in\msf V(\Lambda)} u_g^\msf v\bigg)\ket{\psi}=\ket{\psi},\quad \forall g\in G .
\end{equation}
Notice that the symmetry operators considered here act on all matter degrees of freedom $\mc H^\msf v$, $\msf v\in\msf V(\Lambda)$, simultaneously. Such a global $G$ symmetry is in modern jargon referred to as an \emph{invertible global 0-form} symmetry~\cite{Gaiotto14}.

Here and below, for $q=0,1$, a \emph{q-form} operator is a \emph{topological} operator that acts non-trivially only on a codimension q submanifold of the spatial manifold, i.e. the two-dimensional torus. The case of the $\bigotimes_\msf v u_g^\msf v$ symmetry operators then naturally corresponds to $q=0$.

For the sake of being concrete we will henceforth make an explicit choice of local Hilbert space and its representation.\footnote{It should be noted that the gauging procedure which follows can be carried out for any on-site realization of the symmetry~\cite{Haegeman14}. However, in the construction of the emergent state below, the choice of the representation of the G symmetry on the boundary does not influence the bulk stabilizers of the emergent state.} To this end we consider the character group $\widehat G := \Hom(G,\rm U(1))$ of $G$, which is isomorphic to $G$ here since $G$ is abelian. We then choose the local Hilbert space as $\mc H^\msf v=\mbb C[\widehat G]$ where $\mbb C[\widehat G]:={\rm Span}_\mbb C\{\ket{\chi}|\chi\in\widehat G\}$ is endowed with the inner product $\braket{\chi_1}{\chi_2}=\delta_{\chi_1,\chi_2}$. This local Hilbert space is then assumed to carry the $G$-representation $u_g=Z_g$ defined as $Z_g := \sum_{\chi\in\widehat G} \chi(g) \ketbra{\chi}$, $\forall g\in G$. Let us also introduce the shorthand $\mbb U_g := \bigotimes_{\msf v\in\msf V(\Lambda)} Z^\msf v_g$.

\subsection{Review: gauging two-dimensional abelian 0-form symmetries}
Gauging a global on-site 0-form $G$ symmetry \`a la~\cite{Haegeman14}\footnote{Notice that the framework of~\cite{Haegeman14} applies more generally to non-abelian groups as well. The gauged theory in that case enjoys Wilson loop symmetries labeled by representations of the symmetry group. Wilson loops corresponding to higher-dimensional representations of the group can be realized on the lattice by non-trivial \emph{matrix product operators}~\cite{Lootens20,Delcamp21,Delcamp24}.} begins with supplementing the Hilbert space with gauge degrees of freedom supported on the edges $\msf E(\Lambda)$ and valued in $\mbb C[G]:={\rm Span}_\mbb C\{\ket{g}|g\in G\}$, $\braket{g_1}{g_2}=\delta_{g_1,g_2}$. The group algebra $\mbb C[G]$ is naturally endowed with the left regular representation given by:
\begin{equation}
	X_g := \sum_{h\in G} \ketbra{gh}{h},\quad \forall g\in G.
\end{equation}
On this enlarged Hilbert space
\begin{equation}
	\left(\bigotimes_{\msf v\in\msf V(\Lambda)} \mbb C[\widehat G]\right) \otimes\left(\bigotimes_{\msf e\in\msf E(\Lambda)} \mbb C[G]\right)
\end{equation}
we then define a \emph{gauge transformation} for every vertex $\msf v\in\msf V(\Lambda)$ and every $g\in G$:
\begin{equation}\label{eq:0formGaugeTrans}
    \mc G_g^\msf v := Z_g^\msf v \,\bigotimes_{\msf e(\rightarrow\msf v)}\, X_{\bar g}^\msf e \,\bigotimes_{\msf e(\leftarrow\msf v)}\, X_g^\msf e\ .
\end{equation}
Here, $\msf e(\rightarrow\!\msf v)$, respectively $\msf e(\leftarrow\!\msf v)$, stands for the set of edges of $\Lambda$ that point into, respectively point away from, the fixed vertex $\msf v$.

Notice that the operators $\mc G_g^\msf v$ in \cref{eq:0formGaugeTrans} form a \emph{local} representation of $G$ in the sense that they are supported on a small neighborhood around $\msf v$ and obey $\mc G_{g_1}^\msf v\cdot\mc G_{g_2}^\msf v=\mc G_{g_1g_2}^\msf v$ for every $\msf v\in\msf V(\Lambda)$ and for every $g_1,g_2\in G$. Moreover, notice that the global symmetry operators are recovered by taking the product of the local symmetry operators on all vertices, namely
\begin{equation}\label{eq:ProdGauss0form}
    \prod_{\msf v\in\msf V(\Lambda)} \mc G^\msf v_g = \mbb U_g,\quad \forall g\in G.
\end{equation}
Next, a \emph{local} projector is obtained by averaging $\mc G_g^\msf v$ over $G$:
\begin{equation}\label{eq:LocalProj}
    \mc P^\msf v := \frac{1}{|G|}\sum_{g\in G} \mc G_g^\msf v.
\end{equation}
Indeed it is straightforward to verify that $\mc P^\msf v\cdot\mc P^\msf v = \mc P^\msf v$, $\forall\msf v\in\msf V(\Lambda)$ by a relabeling of summation variables.

The commutation of the local projectors $\big[\mc P^\msf v, \mc P^{\msf v'}\big]=0$ allows us to unambiguously define the \emph{global} projector onto the gauge invariant subspace of the enlarged Hilbert space:
\begin{equation}
    \mc P := \prod_{\msf v\in\msf V(\Lambda)} \mc P^\msf v.
\end{equation}
The state $\ket{\psi}$ is then gauged by applying the projector $\mc P$ to the initial state $\ket{\psi}$ coupled to a trivial flat gauge field $\bigotimes_{\msf e\in\msf E(\Lambda)} \ket{1}^{\msf e}$. Hence, we define the \emph{gauging map} $\mc G^0$ for the global 0-form $G$ symmetry as:
\begin{gather}
    \G^{0} :=  \mc P  \bigg(\bigotimes_{\msf e\in\msf E(\Lambda)} \ket{1}^{\msf e}\bigg)\,: \label{eq:G0}\\
    \,\bigg(\bigotimes_{\msf v\in\msf V(\Lambda)} \mbb C[\widehat G]\bigg) \rightarrow \bigg(\bigotimes_{\msf v\in\msf V(\Lambda)} \mbb C[\widehat G]\bigg) \otimes\bigg(\bigotimes_{\msf e\in\msf E(\Lambda)} \mbb C[G]\bigg), \nonumber
\end{gather}
and the \emph{gauged state} is by definition $\G^{0}\ket{\psi}$.

In order to elucidate the above procedure and facilitate the construction of the emergent state below, let us present a tensor network representation of the gauging map $\mc G^0$. To this end we introduce two types of tensors. For every vertex of the lattice with $m$ incoming edges and $n$ outgoing edges, we introduce the rank-$(m+n)$ \emph{vertex tensor}, or \emph{Z-type tensor}, defined as follows:
\begin{align}
    &\PEPOTensor{0}{}{}{}{}{}{}\nonumber\\[-20pt]
    &:= \sum_{\substack{\chi_1,\chi_2\\ \{g_k\}_{\! k}}}\frac{1}{|G|} \,\, \PEPOTensor{0}{\chi_1}{g_1}{g_m}{g_{m+n}}{g_{m+1}}{\chi_2} \,\,\,\ketbra{\chi_2}{\chi_1} \nonumber\\[-20pt]
    &\hspace{110pt} \otimes \ketbra{g_{m+1},...,g_{m+n}}{g_1,...,g_m} \nonumber\\[10pt]
    &:= \sum_{g\in G} \frac{1}{|G|} Z_g \otimes \ketbra{g,...,g}{g,...,g}. \label{eq:ZTensor}
\end{align}
In this and subsequent figures, the dots are a placeholder for tensor legs which are omitted for clarity.

In order to implement the flatness of the gauge field supported on the edges of the lattice we consider the rank-3 \emph{+-type} tensors given by:
\begin{equation}
\begin{split}
    \PEPOTensor{1}{}{}{}{}{}{}
    &:= \sum_{\substack{g,h,k \\ \in G}} \PEPOTensor{1}{g}{h}{k}{}{}{} \ketbra{g,h}{k}\\
    &:=\sum_{g,h\in G} \ketbra{g,h}{gh},
\end{split}
\end{equation}
and
\begin{equation}\label{eq:+Tensor}
\begin{split}
    \PEPOTensor{-1}{}{}{}{}{}{}
    &:= \sum_{\substack{g,h,k \\ \in G}} \PEPOTensor{-1}{k}{h}{g}{}{}{} \ketbra{k}{g,h}\\
    &:=\sum_{g,h\in G} \ketbra{gh}{g,h}.
\end{split}
\end{equation}
Using these tensors, the local projector $\mc P^\msf v$ introduced in \cref{eq:LocalProj} admits the simple tensor network representation
\begin{equation}
	\mc P^\msf v = \GaussLaw{0}\, ,
\end{equation}
where the orientation of the horizontal red legs is assumed to be in correspondence with the orientation of the edges of the lattice. Note that the commutation of two local projectors $\mc P^\msf v, \mc P^{\msf v'}$ on different vertices $\msf v\neq\msf v'$ follows immediately from the definition of the +-type tensors. Combining this observation with the fact that projecting one of the incoming legs of the +-type tensor \cref{eq:+Tensor} onto $\ket{1}$ effectively removes that leg, it follows that the gauging map $\mc G^0$ admits a representation as projected entangled-pair operator (PEPO) of bond dimension $|G|$, whose unit cell can be depicted as:
\begin{equation}\label{eq:G0PEPO}
	\qformPEPOUnit{0}\,\,\,.
\end{equation}
In this construction it should be understood that the full PEPO on the torus is obtained by placing copies of this unit cell on all vertices of the lattice and contracting along the virtual red legs according to the connectivity of the lattice. 

Let us proceed by dissecting the symmetry properties of the gauging map $\mc G^0$. One of the upshots of the PEPO representation is that these symmetry properties automatically follow from the symmetry properties of the local tensors from which the PEPO is made up. It follows readily from its definition that the vertex tensor satisfies
\begin{equation}\label{eq:ZTensorSym}
	\PEPOTensor{0}{Z_g}{}{}{}{}{} = \PEPOTensor{0}{}{X_{\!g}}{X_{\! g}}{X_{\! g}^\dagger}{X_{\! g}^\dagger}{} = \PEPOTensor{0}{}{}{}{}{}{Z_g}\ ,
\end{equation}
whereas the edge tensor obeys the relations
\begin{equation}\label{eq:+TensorSym}
	\PEPOTensor{1}{}{}{X_{\! g}}{}{}{} = \PEPOTensor{1}{}{X_{\! g}}{}{}{}{} = \PEPOTensor{1}{X_{\! g}}{}{}{}{}{}\,.
\end{equation}
Consider first gauge invariance of the gauging map, that is $\mc G^\msf v_g\cdot\mc G^0=\mc G^0$ for all $g\in G$ and $\msf v\in\msf V(\square)$. Note that it is exactly in this sense that this gauging procedure \emph{localizes} the global 0-form $G$ symmetry.

From the PEPO representation this property can be deduced by combining the second relation in \cref{eq:ZTensorSym} with both identities in \cref{eq:+TensorSym} and noticing that all $X_g^{(\dagger)}$'s on the virtual level exactly cancel with each other. Note that algebraically this follows directly from a renaming of summation variable in the definition of $\mc P^\msf v$ \cref{eq:LocalProj}. One also has that $\mc G^0\cdot \mbb U_g = \mc G^0$ for all $g\in G$. Physically this means that the gauging map projects out states which transform non-trivially under the global symmetry $G$. As such, $\mc G^{0}$ has a non-trivial kernel spanned by states in the original matter Hilbert space $\bigotimes_{\msf v\in\msf V(\Lambda)}\mbb C[\widehat G]$ which are charged under the $G$ symmetry. This can be seen from application of the first identity in \cref{eq:ZTensorSym} on every vertex together with the invariance of the edge tensors, namely
\begin{equation}\label{eq:+TensorInv}
	\PEPOTensor{1}{X_{\! g}}{}{X_{\! g}^\dagger}{}{}{} = \PEPOTensor{1}{}{}{}{}{}{}\,,
\end{equation}
which is deduced from \cref{eq:+TensorSym}.

As mentioned in the introduction, the gauged states exhibit at least a global 1-form $\widehat G$ symmetry as a consequence of the imposed Gauss laws. This 1-form symmetry is the topic of the next section.

\bigskip\noindent To conclude this paragraph, a comment which is of importance for what follows is in place. As argued in~\cite{Haegeman14} it is possible to disentangle the initial matter degrees of freedom from the newly introduced gauge degrees of freedom by means of a finite depth unitary quantum circuit, rendering the Gauss constraints $\mc G_g^\msf v\overset{!}{=}1$ trivial and effectively freezing the matter degrees of freedom to the $+1$ eigenstate of $Z_g$. As showcased below, we will construct a three-dimensional state by concatenating gauging operators, wherein the `matter' degrees of freedom in every layer will play the role of physical degrees of freedom of the three-dimensional state. We therefore refrain from applying this unitary circuit.

\subsection{The dual 1-form symmetry}\label{sec:dual1form}
Let us consider $\ell\in Z_1(\Lambda,\mbb Z)$ to be an oriented closed path supported on the edges of the lattice $\Lambda$, and introduce
\begin{equation}
	Z_\chi := \sum_{g\in G} \chi(g) \ketbra{g}{g} ,\quad\forall\chi\in\widehat G,
\end{equation}
which together with $X_g$ obeys the following commutation relation:
\begin{equation}\label{eq:ZchiXg}
	Z_\chi X_g = \chi(g)\,X_gZ_\chi.
\end{equation}
We define then following \emph{Wilson loop} operators acting exclusively on the gauge degrees of freedom:
\begin{equation}\label{eq:defUell}
\mbb U_\chi[\ell] := \bigotimes_{\msf e\in \ell} Z_{o_\ell^\msf e(\chi)}^{\msf e}.
\end{equation}
Here, $o_\ell^\msf e(\chi)$ is defined so that $o_\ell^\msf e(\chi)=\chi$ if the orientation of $\msf e$ and $\ell$ coincide and $\bar\chi$ otherwise. As we will momentarily show, it holds that
\begin{equation}\label{eq:UG=G}
    \mbb U_\chi[\ell] \cdot \G^{0} = \G^{0},
\end{equation}
$\forall \chi \in \widehat G,\ell\in Z_1(\Lambda,\mbb Z)$. This implies that the gauged state $\mc G^0\ket{\psi}$ transforms trivially under $\mbb U_\chi[\ell]$ for every valid choice of $\ell$ and every character $\chi\in\widehat G$. Making use of the fact that 
\begin{equation}
    \chi_1(g)\chi_2(g)=(\chi_1\otimes\chi_2)(g),
\end{equation}
$\forall \chi_1,\chi_2\in\widehat G, g\in G,$ it follows that the operators $\mbb U_\chi[\ell]$, with same $\ell$, multiply according to the product of characters in $\widehat G$, i.e.:
\begin{equation}
    \mbb U_{\chi_1}[\ell] \cdot \mbb U_{\chi_2}[\ell] = \mbb U_{\chi_1\otimes\chi_2}[\ell].
\end{equation}
Also note that the operators $\mbb U_\chi[\ell]$ are invariant under continuous deformations of $\ell$, rendering them \emph{topological}. These observations demonstrate that after gauging a global 0-form $G$ symmetry, a \emph{global 1-form $\widehat G$} symmetry emerges.\footnote{In fact, one should think of the emergent 1-form $\widehat G$ symmetry, or more precisely ${\rm Rep}(G)$ symmetry, as being the invertible component of the full symmetry structure of the gauge theory. The complete symmetry structure was shown to constitute the 1-form symmetry together with all its 0-form \emph{condensation descendants} and is organized in the \emph{fusion 2-category} $2{\rm Rep}(G)$ \cite{Roumpedakis22,Lin22,Bartsch22,Bhardwaj22, Delcamp24,BVDC25}.} Alternatively, one could think of (\ref{eq:UG=G}) as a \emph{kinematical}  constraint satisfied by all allowed states in the gauge invariant Hilbert space.

In order to verify \cref{eq:UG=G} we can again resort to the PEPO representation \cref{eq:G0PEPO}. The +-type tensor \cref{eq:+Tensor} satisfies following invariance property
\begin{equation}
    \PEPOTensor{1}{Z_\chi}{Z_\chi}{Z_\chi^\dagger}{}{}{} = \PEPOTensor{1}{}{}{}{}{}{}\,,
\end{equation}
whereas the Z-type tensor \cref{eq:ZTensor} obeys
\begin{equation}
	\PEPOTensor{0}{}{}{Z_\chi}{}{Z_\chi^\dagger}{} = \PEPOTensor{0}{}{}{}{}{}{}\, .
\end{equation}
Both these invariance properties once more follow directly from the definitions of these respective tensors. From them, the claim \cref{eq:UG=G} follows readily as again all occurrences of $Z_\chi^{(\dagger)}$ that appear on the virtual level after application of $\mbb U_\chi[\ell]$ cancel.

For the sake of completeness, let us also verify this more explicitly algebraically. To this end, one checks that for every such loop $\ell$ this line operator commutes with the local gauge transformations:
\begin{equation}\label{eq:comUP}
    \mbb U_\chi[\ell] \cdot \mc G_g^\msf v = \mc G_g^\msf v \cdot  \mbb U_\chi[\ell] , \quad \forall g \in G, \msf v\in \msf V(\Lambda).
\end{equation}

To prove this, first notice that if $\mbb U_\chi[\ell]$ and $\mc G_g^\msf v$ overlap, they overlap in an even number of edges. Let us consider just the case of two overlapping edges that we call $\msf e_1$, $\msf e_2$. Then for the given vertex $\msf v$, if both $\msf e_1$, $\msf e_2$ have the same orientation as $\ell$ or both have the opposite orientation as $\ell$, commuting $\mbb U_\chi[\ell]$ through $\mc G_g^\msf v$ yields two cancelling phases $\chi(g)$ and $\bar \chi(g)$ by virtue of the definition of $\mc G_g^\msf v$ and \cref{eq:ZchiXg}. If on the other hand $\ell$ has the same orientation as $\msf e_1$ but has the opposite orientation as $\msf e_2$, then we also obtain, due to the relative difference in orientation, phases $\chi(g)$ and $\bar\chi(g)$ that cancel. From \cref{eq:comUP} together with the fact that
\begin{equation}
	\mbb U_\chi[\ell] \big(\bigotimes_{\msf e\in \msf E(\Lambda)} \ket{1}^{\msf e} \big) = \bigotimes_{\msf e\in\msf E(\Lambda)} \ket{ 1}^{\msf e}
\end{equation}
we can conclude \cref{eq:UG=G}.

\subsection{Gauging the dual 1-form symmetry}\label{sec:gauge1form}
Since the emergent 1-form symmetry is realized on-site, it is anomaly-free in the sense of~\cite{Wen18} and can hence be gauged. To this end we use a similar procedure as in \cite{Haegeman14,Rayhaun23, Williamson16}. After gauging this 1-form symmetry one expects the original 0-form $G$ symmetry to re-emerge \cite{Gaiotto14, Rayhaun23}.

Following \cite{Rayhaun23, Williamson16, Moradi23}, gauging of the 1-form symmetry happens via introduction of gauge degrees of freedom on the \emph{vertices} $\msf V(\Lambda)$ of the lattice valued in the group algebra $\mbb C[\widehat G]$.\footnote{By invoking a form of Poincar\'e duality, one can identify the vertices of the primal lattice $\Lambda$ with the plaquettes of the dual lattice $\Lambda^\vee$, i.e. $\msf V(\Lambda)\equiv\msf P(\Lambda^\vee)$. In that sense the gauge field can indeed be thought of as a \emph{2-form gauge field} for the global 1-form symmetry~\cite{Moradi23}.} A local \emph{1-form gauge transformation} for every edge $\msf e\in\msf E(\Lambda)$ is then constructed as follows:
\begin{equation}\label{eq:1formGaugeTrans}
    \mc G^{\msf e}_\chi := X_\chi^{\msf v(\leftarrow \msf e)}\otimes Z_\chi^\msf e\otimes X_{\bar\chi}^{\msf v(\rightarrow\msf e)},
\end{equation}
where we used an analogue notation to the one in \cref{eq:0formGaugeTrans} denoting by $\msf v(\leftarrow\!\msf e)$ the vertex which is the target of $\msf e$ and by $\msf v(\rightarrow\!\msf e)$ the source vertex of the edge $\msf e$.  Naturally, $X_\chi$ is defined as
\begin{equation}
	X_\chi := \sum_{\eta\in\widehat G} \ketbra{\chi\eta}{\eta}.
\end{equation}
Akin to the case of the gauging of a global 0-form symmetry, these operators form a \emph{local} representation of $\widehat G$ supported near $\msf e$, $\mc G^\msf e_{\chi_1}\cdot\mc G^\msf e_{\chi_2}=\mc G^\msf e_{\chi_1\chi_2}$, $\forall \chi_1,\chi_2\in\widehat G$. Notice that the global 1-form symmetries can be recovered by taking the (oriented) product of the 1-form gauge transformations along the loop $\ell$, i.e.:
\begin{equation}
    \mbb U_\chi[\ell]=\prod_{\msf e\in\ell}\mc G^\msf e_{o_\ell^\msf e(\chi)},
\end{equation}
$\forall\chi\in\widehat G,\ \ell\in Z_1(\Lambda,\mbb Z)$, analogous to \cref{eq:ProdGauss0form}.

Group-averaging over $\widehat G$ allows us to construct the local projectors
\begin{equation}
    \mc P^\msf e := \frac{1}{|G|}\sum_{\chi\in\widehat G} \mc G_\chi^\msf e,
\end{equation}
where we made use of $|G|=|\widehat G|$. From these we can again unambiguously construct the global projector onto the gauge invariant subspace as:
\begin{equation}\label{eq:1formproj}
    \mc P := \prod_{\msf e\in\msf E(\Lambda)} \mc P^\msf e.
\end{equation}
Gauging the 1-form symmetry then ultimately happens in a similar fashion as in the 0-form case. We construct a gauging map by initializing all newly introduced gauge degrees of freedom in the trivial representation $\ul 1\in\widehat G$ followed by acting with the projector onto the gauge invariant subspace:
\begin{gather}
    \G^{1} := \mc P  \bigg(\bigotimes_{\msf v\in\msf V(\Lambda)} \ket{\ul 1}^{\msf e}\bigg)\,: \label{eq:G10}\\
    \bigg(\bigotimes_{\msf e\in\msf E(\Lambda)} \mbb C[G]\bigg) \rightarrow \bigg(\bigotimes_{\msf e\in\msf E(\Lambda)} \mbb C[G]\bigg) \otimes\bigg(\bigotimes_{\msf v\in\msf V(\Lambda)} \mbb C[\widehat G]\bigg). \nonumber
\end{gather}
Let us draw attention to the fact that without loss of generality the gauging map $\G^{1}$ has as its domain only the edge degrees of freedom. This can be traced back to the fact that the emergent 1-form symmetry generators $\mbb U_\chi[\ell]$ only act (non-trivially) on these degrees of freedom.

Let us now proceed by showcasing the tensor network representation of the gauging map $\mc G^1$, and its symmetry properties. In a similar vein as for the gauging map of the global 0-form $G$ symmetry, we introduce tensors that will be placed on the edges of the lattice that read:
\begin{align}
    &\PEPOTensor{2}{}{}{}{}{}{} \nonumber\\[-10pt]
    &:= \sum_{\substack{g_1,g_2 \nonumber\\
    \chi_1,\chi_2}} \frac{1}{|G|} \, \PEPOTensor{2}{g_1}{g_2}{\chi_2}{\chi_1}{}{} \, \ketbra{g_2}{g_1}\otimes \ketbra{\chi_2}{\chi_1} \nonumber\\
    &:= \sum_{\chi\in\widehat G} \frac{1}{|G|} Z_\chi\otimes \ketbra{\chi}.
\end{align}
Tensors incorporating the flatness of the (2-form) gauge field placed on the vertices read then
\begin{align}
&\PEPOTensor{3}{}{}{}{}{}{} \nonumber\\[-20pt]
    &:= \sum_{\{\chi_k\}_{\! k}} \PEPOTensor{3}{\chi_2}{\chi_1}{\chi_{m+2}}{\chi_{m+1}}{\chi_{m+n}}{\chi_m} \,\,\ket{\chi_{m+1},...,\chi_{m+n}} \nonumber\\[-20pt]
    &\hspace{130pt}\otimes\bra{\chi_1,...,\chi_m} \nonumber\\[5pt]
    &:= \sum_{\{\chi_k\}_{\! k}} \delta_{\prod_{k=1}^m\chi_k,\prod_{k=m+1}^{m+n}\chi_k} \ket{\chi_{m+1},...,\chi_{m+n}} \nonumber\\[-10pt]
    &\hspace{130pt} \otimes\bra{\chi_1,...,\chi_m}.
\end{align}
Their relevant symmetry properties are:
\begin{equation}\label{eq:ZDualTensorSym}
	\PEPOTensor{2}{Z_{\!\chi}}{}{}{}{}{} = \PEPOTensor{2}{}{}{X_{\!\chi}^\dagger}{X_{\!\chi}}{}{} = \PEPOTensor{2}{}{Z_{\!\chi}}{}{}{}{} \,,
\end{equation}
and
\begin{equation}\label{eq:+DualTensorSym}
	\PEPOTensor{3}{}{}{}{}{X_{\!\chi}}{}{} = \PEPOTensor{3}{}{}{}{}{}{X_{\!\chi}}\,.
\end{equation}
Making use of these tensors then results in following PEPO representation of the Gauss law $\mc P^\msf e$:
\begin{equation}
    \mc P^\msf e = \GaussLaw{1}\,.
\end{equation}
From this, the PEPO representation of the full gauging map $\mc G^1$ then readily follows. One unit cell is given by
\begin{equation}\label{eq:G1PEPO}
	\qformPEPOUnit{1}\,\,,
\end{equation}
where, as alluded to above, the blue +-type tensors are placed in alignment with the vertices of the primal lattice.

Gauge invariance which in this case reads $\mc G^\msf e_\chi\cdot \mc G^1 = \mc G^1$ for all $\chi\in\widehat G$ and $\msf e\in\msf E(\Lambda)$ is a direct consequence of the second equality in \cref{eq:ZDualTensorSym} combined with the identities of the vertex tensor in \cref{eq:+DualTensorSym}. Also the property $\mc G^{1}\cdot \mbb U_\chi[\ell] = \mc G^{1}$ holds for all closed paths $\ell$ and $\chi\in\widehat G$. In other words, the gauging map $\mc G^{1}$ annihilates states which are not in the even sector of the global 1-form $\widehat G$ symmetry. This follows from the first equation in \cref{eq:ZDualTensorSym} and the invariance properties of the vertex tensor, one of which reads:
\begin{equation}
	\PEPOTensor{3}{}{X_{\!\chi}^\dagger}{}{X_{\!\chi}}{}{}{} = \PEPOTensor{3}{}{}{}{}{}{} \,.
\end{equation}
This is not an independent property as it can be derived immediately from \cref{eq:+DualTensorSym}.

Finally and crucially, after gauging the 1-form symmetry the initial global 0-form $G$ symmetry reappears as anticipated. Indeed, it is found that
\begin{equation}\label{eq:UgG1=G1}
    \mbb U_g \cdot \mc G^{1} = \mc G^{1},
\end{equation}
for all $g\in G$. 

For future reference we demonstrate this here algebraically. To this end, first note that for two vertices $\msf v,\msf v'$ which share an edge $\msf e$, $Z_g^\msf v\otimes Z_g^{\msf v'}$ commutes with the gauge transformation $\mc G^\msf e_\chi$, $\forall g\in G$ by virtue of the identity
\begin{equation}\label{eq:ZgXchi}
    Z_gX_\chi = \chi(g) X_\chi Z_g.
\end{equation}
Combined with the observation that
\begin{equation}
    \mbb U_g \big(\bigotimes_{\msf v\in\msf V(\Lambda)}\ket{1}^\msf v\big)=\bigotimes_{\msf v\in\msf V(\Lambda)}\ket{1}^\msf v,
\end{equation}
we recover indeed \cref{eq:UgG1=G1}.

In order to verify this using the PEPO representation \cref{eq:G1PEPO} on the other hand, the relevant invariance properties this time read:
\begin{equation}
    \PEPOTensor{3}{Z_{\! g}}{Z_{\! g}}{Z_{\! g}^\dagger}{Z_{\! g}^\dagger}{Z_{\! g}^\dagger}{Z_{\! g}} = \PEPOTensor{3}{}{}{}{}{}{}\, ,
\end{equation}
and
\begin{equation}
    \PEPOTensor{2}{}{}{Z_{\! g}}{Z_{\! g}^\dagger}{}{} = \PEPOTensor{2}{}{}{}{}{}{}\, .
\end{equation}
From these properties, \cref{eq:UgG1=G1} follows readily since after application of $\mbb U_g$ on the gauging map, all occurrences of $Z_g^{(\dagger)}$ again cancel on the virtual bonds.

\subsection{3D local symmetries from iterative gauging}
Constructing a three-dimensional state from the above gauging maps now happens as follows. We start by choosing a two-dimensional state $\ket{\psi}$ that will serve as the boundary condition of the three-dimensional state and fixing a choice of abelian group $G$. We assume, as anticipated above, that this state lives in $\bigotimes_{\msf v\in\msf V(\Lambda)}\mbb C[\widehat G]$ and transforms trivially under the abelian 0-form $G$ symmetry $\mbb U_g$, i.e. $\mbb U_g\ket{\psi}=\ket{\psi}$, $\forall g\in G$. Throughout the above discussions $\Lambda$ was kept completely general, and as such the construction that follows also can be applied to this most general case.

On $\ket{\psi}$ we can thus first act with the gauging operator $\G^{0}$ defined explicitly in \cref{eq:G0}, thereby introducing degrees of freedom valued in $\mbb C[G]$ and supported on the edges of $\Lambda$. Indeed, recall that $\G^{0}$ acts according to $\mc G^0:\mbb C[\widehat G]^{\otimes\msf V}\rightarrow \mbb C[\widehat G]^{\otimes\msf V} \otimes \mbb C[G]^{\otimes\msf E}$, in which $\otimes\msf E$ and $\otimes\msf V$ stand for tensor products over the vertices and edges of the lattice. The gauged state $\G^{0}\ket{\psi}$ has then as argued in \cref{sec:dual1form} an emergent dual 1-form $\widehat G$ symmetry which is completely supported on the edges of $\Lambda$ and that in turn can be gauged using the procedure outlined in \cref{sec:gauge1form}. Notice that the gauging map for this 1-form symmetry, $\G^{1}$ defined in (\ref{eq:G10}), acts only on the $\mbb C[G]$-valued degrees of freedom supported on the edges, leaving the original `matter' degrees of freedom on the vertices $\msf v\in\msf V(\Lambda)$ valued in $\mbb C[\widehat G]$ untouched. With slight abuse of notation: $\mc G^1:\mbb C[G]^{\otimes\msf E}\rightarrow \mbb C[G]^{\otimes\msf E} \otimes \mbb C[\widehat G]^{\otimes\msf V}$. In turn we can then again gauge the emergent 0-form $G$ symmetry that arises after the application of $\G^{1}$ using the gauging map $\G^{0}$. In this step $\G^{0}$ does not act on the degrees of freedom in $\mbb C[\widehat G]$ living on $\msf E(\Lambda)$. By repeatedly concatenating the gauging maps $\G^{0}$ and $\G^{1}$ we thus in every step are left with degrees of freedom in $\mbb C[\widehat G]$ and $\mbb C[G]$ which are not acted upon by the next gauging map. These are the degrees of freedom which, following the paradigm of Ref.~\cite{Garre24}, are interpreted as \emph{physical} degrees of freedom of a \emph{foliated} three-dimensional state $\ket{\Psi[\psi]}$ whose 2D layers are made up of the `matter' degrees of freedom in every application of the gauging maps. Formally $\ket{\Psi[\psi]}$ reads:
\begin{equation}\label{eq:Topo3DStateDef}
    \ket{\Psi[\psi]} := \cdots \circ\G^{1}_{\msf 2}\circ\G^{0}_{\msf{\frac{3}{2}}}\circ\G^{1}_{\msf 1}\circ\G^{0}_{\msf{\frac{1}{2}}} \ket{\psi}.
\end{equation}
The exact meaning of the subscripts will be explained momentarily.

For concreteness we specialize to the square lattice $\square=\Lambda$ henceforth. It can then be argued that the physical degrees of freedom of $\ket{\Psi[\psi]}$ naturally organize themselves on the edges of a three-dimensional cubic lattice that we will denote by $\smallcube$, and where the initial two-dimensional state serves as a physical boundary of the lattice.

We therefore consider the cubic lattice as being made up of alternating layers of vertical edges followed by a layer of horizontal edges, that we label by ${\msf{\frac{1}{2},\frac{3}{2}},\dots}$ and $\msf{1,2},\dots$ respectively. The cubic lattice hence has a physical boundary made up of vertical edges.

For every layer of vertical edges we associate these edges with the vertices of a copy of $\square$ and the corresponding $\mbb C[\widehat G]$-valued `matter' degrees of freedom that are not acted upon by the gauging maps $\mc G^1$. Analogously, we associate with every horizontal edge the $\mbb C[G]$-valued degrees of freedom that are untouched by the $\mc G^0$'s.

Using a {\color{\colorb} blue} color for edges that carry a copy of the local Hilbert space $\mbb C[\widehat G]$ and {\color{\colora} red} for those with a copy of $\mbb C[G]$, one can depict the cubic lattice away from the boundary as follows:
\begin{equation}
    \CubicLattice\;\;\;.
\end{equation}
One unit cell of the state $\ket{\Psi[\psi]}$ away from the boundary then can be depicted by combining \cref{eq:G0PEPO} and \cref{eq:G1PEPO} as follows:
\begin{equation}\label{eq:StateTN}
    \TOPOPEPOState .
\end{equation}
In this figure it is understood that all horizontal legs are contracted with neighboring unit cells and the dots represent a contraction with unit cells in the vertical direction. The physical degrees of freedom of the state then correspond to the uncontracted vertical legs.

By virtue of the local symmetries of all the gauging maps appearing in \cref{eq:Topo3DStateDef}, the obtained state $\ket{\Psi[\psi]}$ also inherits those local symmetries in a slightly modified way. More precisely, the emergent state $\ket{\Psi[\psi]}$ is stabilized by plaquette operators for every plaquette of the cubic lattice, coming from the localization of the $1$-form symmetry, and vertex operators for every vertex, from gauging the global $0$-form symmetry. We denote the plaquette stabilizers which are labeled by $\chi\in\widehat G$ by $\mbb b^\msf p$:
\begin{align}
    \mbb b^{\msf p_{xz}}_\chi \ket{\Psi[\psi]} &= \ket{\Psi[\psi]}, \\
    \mbb b^{\msf p_{yz}}_\chi \ket{\Psi[\psi]} &= \ket{\Psi[\psi]}, \\
    \mbb b^{\msf p_{xy}}_\chi \ket{\Psi[\psi]} &= \ket{\Psi[\psi]},
\end{align}
and the vertex stabilizers $\mbb a^\msf v$ carry a group label $g\in G$:
\begin{equation}
    \mbb a^{\msf v}_g \ket{\Psi[\psi]} = \ket{\Psi[\psi]}.
\end{equation}
Explicitly, the plaquette and vertex terms are given by
\begin{align}
    \mbb b^{\msf p_{xz}}_\chi &:= \CSSHam{facefront}\,\,\,, \\
    \mbb b^{\msf p_{yz}}_\chi &:= \CSSHam{faceright}\,\,\,, \\
    \mbb b^{\msf p_{xy}}_\chi &:= \CSSHam{facebottom}\,\,\,, \label{eq:TopoBpxy}\\
    \mbb a^{\msf v}_g &:= \CSSHam{v}\,\,\,.
\end{align}

Let us first show algebraically that $\mbb b^{\msf p_{xz}}_\chi$ and $\mbb b^{\msf p_{yz}}_\chi$ stabilize $\ket{\Psi[\psi]}$ before residing to our graphical PEPO representation of the state. To this end we single out one of those plaquette term. From the commutation relation \cref{eq:ZgXchi} together with the definition of $\mc G^0$ in terms of the local gauge transformations (\ref{eq:0formGaugeTrans}), it follows that
\begin{equation}\label{eq:XZXG0}
\begin{split}
    &\big(X_\chi^{\msf v(\leftarrow\msf e)}\otimes Z_{\bar\chi}^\msf e\otimes X_{\bar\chi}^{\msf v(\rightarrow\msf e)}\big)\cdot\mc G^0 \\
    &\hspace{45pt}= \mc G^0\cdot\big(X_\chi^{\msf v(\leftarrow\msf e)}\otimes X_{\bar\chi}^{\msf v(\rightarrow\msf e)}\big).
\end{split}
\end{equation}
Recognizing then $X_\chi^{\msf v(\leftarrow\msf e)}\otimes Z_\chi\otimes X_{\bar\chi}^{\msf v(\rightarrow\msf e)} $ as the local 1-form gauge transformation (\ref{eq:1formGaugeTrans}), which leaves $\mc G^1$ invariant, together with the fact that $Z_{\bar\chi}\cdot\mc G^1=\mc G^1\cdot Z_{\bar\chi}$ proves that $\mbb b^{\msf p_{xz}}_\chi$ and $\mbb b^{\msf p_{yz}}_\chi$ are stabilizers of the state.

Let us now give an interpretation of these stabilizers in terms of our PEPO representation of the state \cref{eq:StateTN}. Note first that from the commutation relation \cref{eq:ZgXchi} it follows that
\begin{equation}
    \PEPOTensor{0}{}{}{}{}{}{X_\chi} = \PEPOTensor{0}{X_\chi}{}{}{}{Z_\chi^\dagger}{} = \PEPOTensor{0}{X_\chi}{}{Z_\chi^\dagger}{}{}{}\, ,
\end{equation}
from which, together with \cref{eq:ZDualTensorSym}, one finds that in the $xz$-plane \cref{eq:XZXG0} can be depicted as:
\begin{equation}\label{eq:XZXpullingthrough}
\begin{multlined}
    \XZXpullingthrough{1} \\
    =  \XZXpullingthrough{2},
\end{multlined}
\end{equation}
and similar in the $yz$-plane. Considering now a slice of the tensor network representation of our state \cref{eq:StateTN} along the $xz$-plane, we compute that:
{\allowdisplaybreaks
\begin{align}
    \mbb b^{\msf p_{xz}}_\chi \ket{\Psi[\psi]}
    \, &=\,  \StabPullingThrough{1} \\
    &=\, \StabPullingThrough{2} \, = \ket{\Psi[\psi]}, \nonumber
\end{align}
}
where in the second equality we made use of the fact that the operator $Z_\chi^\dagger$ on top can be freely \emph{pulled through} the Z-type tensor on which it acts, together with \cref{eq:XZXpullingthrough}. In the final equality we then recognized the gauge invariance of $\mc G^1$. The $yz$-plaquette term can be derived in similar fashion.

The final $xy$-plaquette terms can be recognized as the (minimal) global 1-form symmetries of the $\G^{0}$ gauging map. In a similar vein to \cref{eq:XZXG0}, one can check that $\mbb a^{\msf v}_g$ applied to $\mc G^1$ maps to $\mc G_g^\msf v$ which is exactly the local symmetry of $\mc G^0$, from which it follows that $\mbb a_g^\msf v$ is also a stabilizer.

These observations allow us to consider the state $\ket{\Psi[\psi]}$ a ground state of the commuting projector bulk Hamiltonian (a stabilizer code):
\begin{equation}\label{eq:BulkHam}
\begin{multlined}
    \mbb H^{\rm blk} := - \sum_{\msf v\in\,\msf V(\smallcube)}\mbb P^\msf v - \sum_{\msf p_{xz}\in\,\msf P(\smallcube)} \mbb P^{\msf p_{xz}} \\
    - \sum_{\msf p_{yz}\in\,\msf P(\smallcube)} \mbb P^{\msf p_{yz}} - \sum_{\msf p_{xy}\in\,\msf P(\smallcube)} \mbb P^{\msf p_{xy}}\,\,\,.
\end{multlined}
\end{equation}
In this expression the sum over $\msf V(\smallcube)$ is over all vertices of the cubic lattice, and the sums over $\msf P(\smallcube)$ are over plaquettes in the $xz$, $yz$ and $xy$ planes respectively. The projectors are explicitly given by
\begin{equation}
    \mbb P^\msf v := \frac{1}{|G|}\sum_{g\in G} \mbb a^{\msf v}_ g,
\end{equation}
and
\begin{align}
    \mbb P^{\msf p_{xz}} &:= \frac{1}{|G|}\sum_{\chi\in\widehat G} \,\,\, \mbb b^{\msf p_{xz}}_\chi, \\
    \mbb P^{\msf p_{yz}} &:= \frac{1}{|G|}\sum_{\chi\in\widehat G} \,\,\, \mbb b^{\msf p_{yz}}_\chi, \\
    \mbb P^{\msf p_{xy}} &:= \frac{1}{|G|}\sum_{\chi\in\widehat G} \,\,\, \mbb b^{\msf p_{xy}}_\chi.
\end{align}
It turns out that the Hamiltonian in \cref{eq:BulkHam} is unitarily equivalent to the $G$ abelian quantum double~\cite{Kitaev97}. In its displayed form (\ref{eq:BulkHam}) it appeared previously in \cite{Huang23} under the name of \emph{Clifford-deformed surface code}. It can be brought in its more familiar form~\cite{Kitaev97} defined on $\bigotimes_{\msf e\in\msf E(\smallcube)}\mbb C[G]$ by applying the generalized Hadamard transformation
\begin{equation}\label{eq:UBasisTrans}
    \mc U := \sum_{\substack{g\in G,\\\chi\in\widehat G}} \frac{\chi(g)}{\sqrt{|G|}}\ketbra{g}{\chi}
\end{equation}
to every vertical edge carrying a copy of $\mbb C[\widehat G]$, as it can be checked that
\begin{align}
    \mc U\,Z_g\,\mc U^\dagger &= X_g^\dagger \,, \\
    \mc U\,X_\chi\,\mc U^\dagger &= Z_\chi \,.
\end{align}

\subsection{Gapped boundary conditions}\label{sec:TopoOrderBdry}
The local stabilizers analyzed so far act on the bulk of the three-dimensional state $\ket{\Psi[\psi]}$. In this section we study the boundary stabilizers. These naturally are given by a subset of truncated bulk stabilizers commuting with the bulk stabilizers $\mbb b$ and $\mbb a$ defined above, where the subset depends on the choice of the boundary state $\ket{\psi}$, in particular the way in which $\ket{\psi}$ is invariant under the $G$ 0-form symmetry. Indeed, for every choice of subgroup $H\leq G$ one can construct explicitly a $G$-symmetric boundary state $\ket{\psi}_H$ for which $\ket{\Psi[\psi_H]}$ is a ground state of the bulk Hamiltonian $\mbb H^{\rm blk}$ (\ref{eq:BulkHam}) plus additional boundary terms which are projectors and which all commute with the terms in $\mbb H^{\rm blk}$ (\ref{eq:BulkHam}) and among themselves.

Let us proceed by constructing these boundary states and the corresponding boundary Hamiltonian terms explicitly. To this end we first define the following operators near the boundary $\dCube$(effectively truncating $\mbb b$) for any $\chi\in\widehat G$:
\begin{align}
    \mbb d^{\msf p_{xz}}_\chi &:= \CSSHam{bdryfacefront}\,\,\, , \label{eq:BdryStabxz}\\
    \mbb d^{\msf p_{yz}}_\chi &:= \CSSHam{bdryfaceright}\,\,\, . \label{eq:BdryStabyz}
\end{align}
These terms commute both with the bulk vertex operators $\mbb a^\msf v_g$ and plaquette operators $\mbb b^\msf p_\chi$. Naturally, they also commute all among each other. From the explicit form of the gauging maps $\mc G^{0}_{\frac{1}{2}}$ and $\mc G^1_1$ it follows that given an irrep $\chi \in \widehat{G}$ the operators (\ref{eq:BdryStabxz}-\ref{eq:BdryStabyz}) are stabilizers of $\ket{\Psi[\psi]}$ if and only if $(X^{\msf v}_{\bar \chi}\otimes X^{\msf v'}_\chi)\ket{\psi}=\ket{\psi}$, equivalently
\begin{equation}\label{symbdry}
\begin{multlined}
    \mbb d_\chi \ket{\Psi[\psi]}= \ket{\Psi[\psi]} \iff \\
\bra{\psi}(X^{\msf v}_{\bar \chi}\otimes X^{\msf v'}_\chi )\ket{\psi} = 1
\end{multlined}
\end{equation}
for neighboring vertices $\msf v,\msf v'\in \msf V(\square)$. Notice that $X_{\bar \chi}\otimes X_\chi$ is an order parameter for the (partial) breaking of the global 0-form $G$ symmetry. Its expectation value is a two-point symmetric correlation function.

In what follows we show that given a choice of subgroup $H\leq G$,  $\mbb d_\chi$ is a stabilizer of $\ket{\Psi[\psi_H]}$ for an appropriate boundary state $\ket{\psi}_H$ depending on that subgroup, provided that $\chi$ is trivial when restricted to $H$. It is a fact that characters in $\widehat G$ that are trivial when restricted to $H$ form a subgroup of $\widehat G$ isomorphic to $\widehat{G/H}$. From now on we will abuse notation and denote this subgroup simply by $\widehat{G/H}\leq\widehat G$. For the chosen subgroup $H\leq G$ we then propose following boundary state $\ket{\psi}_H$. First, we define the normalized states
\begin{equation}\label{eq:+Hstate}
    \ket{+}_H := \frac{1}{\sqrt{|G/H|}}\sum_{[\chi]\in\widehat{G/H}} \ket{\chi}.
\end{equation}
Notice that $\ket{+}_H$ is stabilized by $H$ in the sense that $Z_h\ket{+}_H = \ket{+}_H$ for all $h\in H$. Given $\ket{+}_H$ we define
\begin{equation}
    \ket{\psi}_H := \frac{1}{\sqrt{|G/H|}} \sum_{[g]\in G/H} \bigg(\bigotimes_{\msf v\in\msf V(\square)}Z_g^\msf v \ket{+}^\msf v_H\bigg).
\end{equation}
By construction $\ket{\psi}_H$ is symmetric under the 0-form $G$ symmetry and, as anticipated, is symmetric under $\{X^{\msf v}_{\bar \chi}\otimes X^{\msf v'}_\chi, \chi \in \widehat{G/H} \}$ since $X^{\msf v}_\chi$ leaves invariant $\ket{+}_H$ and $X^{\msf v}_{\bar \chi}\otimes X^{\msf v'}_\chi$ commutes with $Z_g^\msf v \otimes Z_g^{\msf v'}$. Notice one can think of $\ket{\psi}_H$ as an equal weight superposition of the $|G/H|$ ground states of a $G$ symmetric Hamiltonian whose symmetry is spontaneously broken down to $H$. It is therefore meaningful to refer to these boundaries as \emph{symmetry breaking} boundaries.
We also define the following operators acting on the boundary:
\begin{equation}
    \mbb c^{\msf e_{z}}_ g :=\CSSHam{bdrysinglez} ,
\end{equation}
which are symmetries of $\ket{\Psi[\psi_H]}$, and which commute with $\mbb d_\chi$ only if $g\in H$. Defining then
\begin{align}
    \mbb Q^{\msf p_{xz}}_H &:= \frac{1}{|G/H|} \sum_{\chi\in \widehat{G/H}} \mbb d^{\msf p_{xz}}_\chi \, ,\\
    \mbb Q^{\msf p_{yz}}_H &:= \frac{1}{|G/H|} \sum_{\chi\in \widehat{G/H}} \mbb d^{\msf p_{yz}}_\chi \, ,\\
    \mbb Q^{\msf e_{z}}_H &:= \frac{1}{|H|} \sum_{h\in H} \mbb c^{\msf e_{z}}_h \, ,
\end{align}
the Hamiltonian in the presence of such boundary terms reads
\begin{equation}
\begin{multlined}
    \mbb H_H = \mbb H^{\rm blk}
    + \sum_{\msf p_{xz}\in\msf P(\dCube\!)} \mbb Q^{\msf p_{xz}}_H \\
    + \sum_{\msf p_{yz}\in\msf P(\dCube\!)} \mbb Q^{\msf p_{yz}}_H
    + \sum_{\msf e_z\in\msf E(\dCube\!)} \mbb Q^{\msf e_{z}}_H,
\end{multlined}
\end{equation}
where the dependence on the choice of subgroup $H\leq G$ is explicit. As an example, one can consider the extreme cases of $H=G$ and $H=\{1\}$ that can be identified with \emph{Dirichlet} and \emph{Neumann} boundary conditions respectively. In the context of lattice gauge theory as of here it is also customary to refer to these cases as realizing the \emph{smooth} and \emph{rough} boundary conditions. \footnote{Notice that we don't recover \emph{twisted} boundary conditions that amount to pasting an SPT phase on top of the symmetry broken boundary phase~\cite{Wang17,Wang18,Ji22,Luo22,Zhao22}. It is expected that all such boundary conditions, which are in one-to-one correspondence with pairs $(H\leq G, [\psi]\in H^3(H, {\rm U}(1))$, are recovered by appropriately twisting the boundary terms akin as in the lower-dimensional setting~\cite{Garre24, Wang18}. This is most easily facilitated for by the triangular lattice.}

\bigskip\noindent As a final comment we consider an analogous construction of a three-dimensional state with a corresponding stabilizer Hamiltonian, which is constructed by starting from a state $|\hat\psi\rangle$ invariant under a $1$-form $\widehat G$ symmetry generated by the Wilson loop operators~(\ref{eq:defUell}):
\begin{equation}
    |\hat\Psi[\hat\psi]\rangle := \cdots \circ\G^{1}_{\msf 2}\circ\G^{0}_{\msf{\frac{3}{2}}}\circ\G^{1}_{\msf 1} |\hat\psi\rangle.
\end{equation}
The boundary terms in that case are given by truncating  $\mbb a^{\msf v}_ g$ and $\mbb b^{\msf p}_\chi$ as follows:
\begin{align}
    \mbb e^{\msf v}_ g &:=\CSSHam{bdryv} \,\,\,, \\[5pt]
    \mbb f^{\msf e_{y}}_\chi &:= \CSSHam{bdryey}, \\[10pt]
    \mbb f^{\msf e_{x}}_\chi &:= \CSSHam{bdryex}\,.
\end{align}

Given a $g \in G$, $\mbb e^{\msf v}_ g$ is a symmetry of $|\hat\Psi[\hat\psi]\rangle$ if and only if
\begin{equation}
    \CSSHam{bdryv2}\ \ |\hat\psi\rangle = |\hat\psi\rangle\,.
\end{equation}
Similarly as in the previous case one could define $|\hat\psi\rangle_{\widehat H}$ hosting a (partial) symmetry breaking phase of the 1-form symmetry with unbroken symmetry group $\widehat H \le \widehat G$. Commuting boundary stabilizers for such a state are given by
\begin{align}
    \mbb R_H^\msf v &:= \frac{1}{|H|}\sum_{h\in H} \mbb e^\msf v_h , \\
    \mbb R_H^\msf e &:= \frac{1}{|G/H|}\sum_{\chi\in\widehat{G/H}} \mbb f^\msf e_\chi.
\end{align}
Notice thus in particular that $|\hat\psi\rangle$ can host non-trivial 2D topological order. It is understood that besides these boundaries and the $G$ symmetry breaking ones considered above, it is also possible to define additional gapped boundaries which amount to (trivially) stacking a 2D topological order or stacking a 2D topological order while condensing composites of boundary anyons of the emergent state and (bulk) anyons from the stacked topological order \cite{Luo22}.

%% file: _general_framework.tex
\section{General framework}\label{sec:GeneralFramework}
What are the main takeaways from this introductory example? From iteratively gauging a global abelian 0-form symmetry and its dual 1-form symmetry we obtain stabilizer codes (local commuting projector Hamiltonians) in one dimension higher. Moreover, the obtained codes can be readily extended to include gapped boundary terms, and we have written down the gauging maps and states explicitly in terms of tensor networks.

The goal of this section is to provide a general framework for iteratively gauging a state $\ket{\psi}$ which is symmetric under a collection of abelian symmetries defined via certain \emph{check operators} which commute with them. Hereby we extend the formalism of~\cite{Williamson16} to arbitrary abelian groups.  We argue that in this framework the concatenated gauging maps generically exhibit \emph{local symmetries} originating from the Gauss constraints of the symmetries under scrutiny. These local symmetries can in turn be organized in an \emph{emergent} local commuting projector Hamiltonian that naturally lives in one dimension higher than the initial state $\ket{\psi}$ which is being gauged. Towards the end of this section we comment on the inclusion of the boundary terms in the emergent Hamiltonian which can be identified with certain spontaneous symmetry breaking patterns of the boundary symmetry realized by $\ket{\psi}$. Since these gauging maps can be defined in any dimension, we recover the results from~\cite{Garre24} as a specific case.

\subsection{Gauging abelian symmetries}
To set the scene, we assume a lattice $\Lambda$ in which each vertex $\msf v\in\msf V(\Lambda)$ supports a copy of $\mbb C[\widehat G]$. The tensor product Hilbert space $\bigotimes_{\msf v\in\msf V(\Lambda)}\mbb C[\widehat G]$ will be written as $\mc H := \bigotimes_{\msf v\in\msf V(\Lambda)}\mbb C[\widehat G]$ henceforth. Notice that we don't make any assumption about the underlying topology or dimension of the spatial manifold on which $\Lambda$ is defined.

Throughout this section we deal with symmetry operators that are fully specified by a \emph{complete collection} of \emph{check operators}, labeled $c\in\mc C$, that commute with all of them. The formal gauging procedure of Williamson~\cite{Williamson16} then happens via the introduction of a gauge degree of freedom for every such check. Formally, consider for any $g\in G$ the operators
\begin{equation}\label{eq:GeneralSymGen}
    \mbb U_g[\Omega] := \bigotimes_{\msf v\in\Omega} Z_g^\msf v, \quad g\in G,
\end{equation}
where $\Omega$ is a subset of the vertices, i.e. $\Omega\subseteq\msf V(\Omega)$. Borrowing the notation of \cite{Williamson16}, we will denote by $\eta$ a function 
\begin{equation}
\begin{split}
	\eta:\, &\mc C \rightarrow P(\msf V(\Lambda)) \\
			&c\rightarrow \eta(c)\subseteq \msf V(\Lambda)
\end{split}\,\,,
\end{equation}
from $\mc C$ to the power set $P(\msf V(\Lambda))$. Physically, $\eta$ maps a given constraint $c$ from the collection $\mc C$, that will below be identified with the gauge degrees of freedom, to a corresponding set of vertices $\eta(c)\subseteq \msf V(\Lambda)$. Furthermore we introduce the involution $o^c_\msf v(\chi)$ which evaluates to $\chi$ or $\bar\chi$ depending on the \emph{relative orientation} of the constraint $c$ and the vertex $\msf v$. An operator $\mbb U_g[\Omega]$ is then a \emph{valid symmetry generator} if it commutes with the checks
\begin{equation}
    X_\chi^c := \bigotimes_{\msf v\in\eta(c)} X^\msf v_{o^c_\msf v (\chi)}
\end{equation}
for all $c\in\mc C$ and all $\chi\in\widehat G$. Below, we will sometimes refer to the symmetry structure generated by these valid symmetry generators as the \emph{initial} symmetry.

To connect with the previous section, let us remark that the global 0-form $G$ symmetry corresponds to the case $\Omega=\msf V(\Lambda)$, and thus $\mbb U_g[\msf V(\Lambda)]\equiv\mbb U_g$. For these symmetry generators the checks can be chosen so that there is exactly one check per edge of the lattice, i.e. $\mc C \equiv \msf E(\Lambda)$. Explicitly, the checks then read $X^\msf e_\chi = X^{\msf v(\rightarrow \msf e)}_\chi \otimes X^{\msf v(\leftarrow \msf e)}_{\bar \chi}$, where we refer to the previous section for a detailed explanation of the notation.

An important observation is that the local constraints can be interpreted as local Hamiltonian terms that commute with the symmetry \cite{Williamson16}. As such, they can be used to construct \emph{generalized Ising models} and when perturbed, for example with single-vertex $Z_g$ terms, they host symmetry breaking phases of such symmetries.

\bigskip\noindent We now proceed by gauging the symmetry generated by the operators $\mbb U_g[\Omega]$, essentially mimicking the procedure used in the previous section based on~\cite{Haegeman14, Williamson16} adapted to this more general setting. To this end we first enlarge the `matter' Hilbert space $\mc H$ with a $\mbb C[G]$-valued \emph{gauge degree of freedom} for every check in $\mc C$. We will denote the physical support of the gauge degrees of freedom by $\msf c\in\msf C(\Lambda)$ where it has to be understood that there is bijection between $\msf C(\Lambda)$ and $\mc C$. Typically, $\msf C(\Lambda)$ can be identified with a (subset) of k-cells of the lattice $\Lambda$ rendering the aforementioned bijection canonical.

Having again the example from the previous section in mind, $\msf C(\Lambda)$ can thus be identified with $\msf E(\Lambda)$ as anticipated above.

Writing $\widehat{\mc H}:=\bigotimes_{\msf c\in\mc C}\mbb C[G]$, the first step of the gauging procedure thus entails enlarging $\mc H$ according to $\mc H\rightarrow\mc H\otimes\widehat{\mc H}$. Given the map $\eta$ defined above, we define $\eta^{-1}(\msf v)$ as the preimage of $\msf v$ under $\eta$. In other words, $\eta^{-1}$ maps the matter degree of freedom localized on $\msf v$ to its associated gauge degrees of freedom. In light of the above identification of gauge degrees of freedom and constraints, we will also interpret the set of constraints $\eta^{-1}(\msf v)$ as its corresponding subset of $\msf C(\Lambda)$. For every $g\in G$ we then define the \emph{gauge transformation} $\mc G_g^\msf v$ on the enlarged Hilbert space $\mc H\otimes\widehat{\mc H}$ as:
\begin{equation}\label{eq:GeneralGaugeTrans}
    \mc G_g^\msf v := Z_g^\msf v \bigotimes_{\msf c\in\eta^{-1}(\msf v)} X^\msf c_{o^c_\msf v(g)}.
\end{equation}
Out of these gauge transformations, a \emph{local projector} is constructed by taking the $G$-average:
\begin{equation}
    \mc P^\msf v := \frac{1}{|G|} \sum_{g\in G}\mc G_g^\msf v,
\end{equation}
$\forall \msf v\in \msf V(\Lambda)$. Indeed remark that from a redefinition of summation variables it follows that $\mc P^\msf v\cdot \mc P^\msf v = \mc P^\msf v$ for all $\msf v\in\msf V(\Lambda)$.

The \emph{gauging map} is then finally defined as the operator obtained by acting with the local projectors on the trivial gauge field $\bigotimes_{\msf c\in\msf C(\Lambda)}\ket{1}^\msf c$:
\begin{equation}
\begin{gathered}
    \mc G := \bigg(\prod_{\msf v\in\msf V(\Lambda)}\mc P^\msf v\bigg)\bigg(\bigotimes_{\msf c\in\msf C(\Lambda)}\ket{1}^\msf c\bigg) \, : \\
    \mc H \rightarrow \mc H \otimes \widehat{\mc H}.
\end{gathered}
\end{equation}
Let us summarize the most important properties of $\mc G$ before proceeding. We refer to~\cite{Williamson16} for a more detailed digression. For a given $\msf v\in\msf V(\Lambda)$, the gauge transformations $\mc G_g^\msf v$ form by construction a $G$-representation supported on $\eta^{-1}(\msf v)$. It then follows from the definitions that $\mc G^\msf v_g\cdot\mc G = \mc G$. By a relabeling of summation variable in the definition of $\mc P^\msf v$ it also follows immediately that $\mc G \cdot \mbb U_g[\Omega]=\mc G$ for all allowed symmetry operators. As illustrated in \cref{sec:TopoOrder} this property implies that only states $\ket{\psi}$ transforming trivially under the symmetry, i.e. $\mbb U_g[\Omega]\ket{\psi}=\ket{\psi}$, $\forall g\in G$ and for all allowed $\Omega$, are \emph{gaugeable}. Given such a state, we coin $\mc G\ket{\psi}$ the \emph{gauged state}. From $\mc G^\msf v_g\cdot\mc G = \mc G$ it readily follows that the gauged state is gauge invariant as expected. Even though we only deal with gauging \emph{states} throughout this manuscript, it should be noted that also operators can be gauged using this prescription. In particular, \emph{local} and \emph{symmetric} operators in the initial matter theory are mapped to \emph{dual} local gauge invariant operators under the gauging procedure in such a way that the matrix elements of these symmetric operators w.r.t. a symmetric state are preserved.

As an example, consider once more the case of a 0-form symmetry. Notice that with the choices of relative orientations $o^\msf e_\msf v$ implicitly defined above, we recover the Gauss constraints defined above in~\cref{eq:0formGaugeTrans}, from which eventually the gauging map given in~\cref{eq:G0} of~\cite{Haegeman14} is recovered. 

\bigskip\noindent Crucially, the gauge theory obtained via the aforementioned proposal exhibits global \emph{emergent} symmetries that, by construction, are anomaly-free and hence gaugeable~\cite{Wen18}. Let us continue by identifying these emergent symmetries. First notice that even though the constraints $X_\chi^c$ do not commute with the gauging map, the \emph{dressed constraints} $Z_\chi^\msf c\otimes X^c_\chi$ do commute with all Gauss operators $\mc G^\msf v_g$ and, a fortiori, with the gauging map $\mc G$. Indeed, this follows from a direct computation. For fixed $\msf v\in\msf V(\Lambda)$ and $c\in\mc C$ let us assume without loss of generality that $\msf v\in\eta(c)$, which necessarily implies that $\msf c\in\eta^{-1}(\msf v)$:
\begin{align}
    &\bigg( Z_\chi^\msf c \otimes X^c_\chi \bigg) \cdot
    \mc G_g^\msf v\nonumber\\
    &=\,\, \bigg( Z^\msf c_\chi \bigotimes_{\msf w\in\eta(c)}X^\msf w_{o^c_\msf w(\chi)} \bigg) \cdot
    \bigg( Z_g^\msf v \bigotimes_{\msf d\in\eta^{-1}(\msf v)} X^\msf d_{o^d_\msf v(g)} \bigg)\nonumber\\
    &=\,\, 
    o^c_\msf v(\bar{\chi})(g) \,\cdot\, \chi(o^c_\msf v(g)) \cdot
    \mc G_g^\msf v \cdot \bigg( Z_\chi^\msf c \otimes X^c_\chi \bigg) \nonumber\\
    &= \,\, \mc G_g^\msf v \cdot \bigg( Z_\chi^\msf c \otimes X^c_\chi \bigg).\label{eq:DressedConstComm}
\end{align}
In the second line the definition of $X^c_\chi$ and $\mc G_g^\msf v$ was employed. In the third line we used the commutation relations \cref{eq:ZchiXg} and \cref{eq:ZgXchi}. Combined with the fact that $Z^\msf c_\chi$ evaluates to $1$ on the trivial gauge field $\bigotimes_{\msf c\in\msf C(\Lambda)}\ket{1}^\msf c$, this implies that the constraint $X^c_\chi$ for every $c$ and $\chi\in\widehat G$ is the preimage of $Z^\msf c_\chi\otimes X^c_\chi$ under the gauging map $\mc G$. In particular, it follows that for a choice of constraints $\Gamma\subseteq\mc C$ and choice of involution $\hat o^c(\chi)$, $c\in\Gamma$ such that
\begin{equation}\label{eq:prodXid}
    \prod_{c\in\Gamma} X_{\hat o^c(\chi)}^c = {\rm id},
\end{equation}
there exists a corresponding operator in the gauge theory
\begin{equation}\label{eq:GeneralEmergSym}
    \mbb U[\Gamma,\hat{o}^c] =\prod_{\msf c\in\Gamma} Z^\msf c_{\hat{o}^c(\chi)}
\end{equation}
which is an emergent symmetry of $\mc G$ in the sense that
\begin{equation}\label{eq:GeneralUG=G}
    \mbb U[\Gamma,\hat{o}^c]\cdot\mc G = \mc G.
\end{equation}

Applying the above to the example of $\mc G^0$ (\ref{eq:G0}) considered in \cref{sec:TopoOrder}, we note that \cref{eq:prodXid} is exactly satisfied for any choice $\Gamma$ which coincides with a closed loop $\ell$ supported on the edges $\msf E(\Lambda)$ of the lattice. With the above choice of $o^c_\msf v$, $\hat{o}^c$ exactly takes the relative orientation of $\ell$ with respect to the orientation of the edges into account.

\bigskip\noindent In order to gauge the emergent symmetry operators we adapt a similar approach as for the initial symmetry (\ref{eq:GeneralSymGen}). Notice that the symmetry generators defined in \cref{eq:GeneralEmergSym} commute with the local checks
\begin{equation}
    X^v_g := \bigotimes_{\msf c\in\eta^{-1}(\msf v)} X^\msf c_{o^c_\msf v(g)}
\end{equation}
which are naturally localized around the vertices $\msf V(\Lambda)$. Therefore we will identify the set of these checks with the set of vertices $\msf V(\Lambda)$. To prove that these checks indeed commute with the emergent symmetry generators, consider some vertex $\msf v\in\eta(\Gamma)$. The identity (\ref{eq:prodXid}) guarantees the existence of constraints $c,d\in\eta^{-1}(\msf v)$ such that $\hat o^c(o^c_\msf v(\chi))=\overline{\hat{o}^d(o^d_\msf v(\chi))}$. It follows immediately that $Z^\msf c_{\hat{o}^\msf c(\chi)}\otimes Z^\msf d_{\hat{o}^d (\chi)}$ commutes with $X^\msf c_{o^c_\msf v(g)}\otimes X^\msf d_{o^d_\msf v(g)}$, from which the claim follows. In order to define the corresponding dual gauging map $\widehat{\mc G}$, we once more enlarge the Hilbert space according to $\widehat{\mc H}\rightarrow\widehat{\mc H} \otimes \mc H$, on which we define the Gauss operators
\begin{equation}\label{eq:dualGaugeTrans}
    \G^\msf c_\chi := Z^\msf c_\chi \bigotimes_{\msf v\in\eta(c)} X^\msf v_{\bar o^c_\msf v(\chi)} .
\end{equation}
Notice the complex conjugate appearing in $\bar o^c_\msf v$. Corresponding local projectors localized around $\msf c$ follow from averaging over $\widehat{G}$:
\begin{equation}
    \mc P^\msf c := \frac{1}{|G|}\sum_{\chi\in\widehat G}\G^\msf c_\chi 
\end{equation}
Ultimately the dual gauging map reads
\begin{equation}
    \widehat{\mc G} := \bigg(\prod_{\msf c\in\msf C(\Lambda)} \mc P^\msf c\bigg)\bigg(\bigotimes_{\msf v\in\msf V(\Lambda)}\ket{\underline 1}^\msf v\bigg).
\end{equation}
It follows that after gauging these emergent symmetries using the gauging map $\widehat\G$ the original $G$ symmetry is recovered, as expected. This is quickly demonstrated by noticing that all symmetry generators $\mbb U_g[\Omega]$ commute by definition with the checks $X^c_\chi$, and by extension with the Gauss constraints $\mc G^\msf c_\chi$, so that $\mbb U_g[\Omega]\cdot\widehat\G=\widehat\G$.

Finalizing the example of the previous section, one verifies immediately that the 1-form Gauss constraints \cref{eq:1formGaugeTrans} coincide precisely with the definition provided in  \cref{eq:dualGaugeTrans} applied to the case at hand.
\subsection{Concatenation of gauging maps}
Let us now consider the composition of the gauging operators defined above. With slight abuse of notation we consider in this section thus the composed operators
\begin{equation*}
\begin{gathered}
    \widehat{\mc G}\circ\mc G: \\
    \mbb C[\widehat G]^{\otimes |\msf V|} \rightarrow
    \mbb C[\widehat G]^{\otimes |\msf V|} \otimes \mbb C[G]^{\otimes |\msf C|} \otimes \mbb C[\widehat G]^{\otimes |\msf V|}
\end{gathered}  
\ ,
\end{equation*}
and
\begin{equation*}
\begin{gathered}
    \G\circ\widehat\G: \\
    \mbb C[G]^{\otimes |\msf C|} \rightarrow
    \mbb C[G]^{\otimes |\msf C|} \otimes \mbb C[\widehat G]^{\otimes |\msf V|} \otimes \mbb C[G]^{\otimes |\msf C|}
\end{gathered} 
\ .
\end{equation*}
Herein, $|\msf V|$ stands for the number of vertices of the lattice $\Lambda$ and by definition $|\msf C|$ is equal to the number of constraints defining the initial symmetry (\ref{eq:GeneralSymGen}), i.e. $|\msf C|=|\mc C|$. As anticipated above, these concatenated gauging maps exhibit local symmetries that can be traced back to the Gauss constraints enforced by $\mc G$ and $\widehat{\mc G}$. Crucial in the derivation of the local symmetries are following identities which can be derived in a similar fashion as \cref{eq:DressedConstComm}:
\begin{equation}
\begin{multlined}
    \bigg( Z_{\bar g}^\msf v \,\bigotimes_{c \in \eta^{-1}(\msf v)} X_{o^c _ \msf v(g)}^{\msf c} \bigg) \cdot \widehat \G \\
     =  \widehat \G \cdot \bigg( \bigotimes_{c \in \eta^{-1}(\msf v)}\, X_{o^c _ \msf v(g)}^{\msf c} \bigg) ,
\end{multlined}
\end{equation}
\begin{equation}\label{eq:GeneralZXG=GX}
\begin{multlined}
    \bigg( Z_{\bar\chi}^{\msf c} \,\bigotimes_{\msf v \in \eta(c)} X_{\bar o^c _ \msf v(\chi)}^\msf v \bigg) \cdot {\G} \\
    =  {\G} \cdot \bigg( \bigotimes_{\msf v \in \eta(c)}\, X_{\bar o^c _ \msf v(\chi)}^\msf v \bigg) .
\end{multlined}
\end{equation}
Given these identities, it is now straightforward to argue that the mutually commuting operators
\begin{equation}
\begin{gathered}
    \mbb a^\msf v_g := Z_{\bar g}^\msf v \otimes \bigg( \bigotimes_{c\in\eta^{-1}(\msf v)} X^\msf c_{o^c_\msf v(g)} \bigg)\otimes Z_g^\msf v \\
    \in {\rm End} \big(
    \mbb C[\widehat G]^{\otimes |\msf V|} \otimes \mbb C[G]^{\otimes |\msf C|} \otimes \mbb C[\widehat G]^{\otimes |\msf V|}
    \big)
\end{gathered} \, ,
\end{equation}
and
\begin{equation}
\begin{gathered}
    \mbb b^\msf c_\chi := Z^\msf c_{\bar\chi} \otimes \bigg(\bigotimes_{\msf v\in\eta(c)} X^\msf v_{\bar o^c_\msf v(\chi)}\bigg)\otimes Z^\msf c_\chi \\
    \in {\rm End} \big(
    \mbb C[G]^{\otimes |\msf C|} \otimes \mbb C[\widehat G]^{\otimes |\msf V|} \otimes \mbb C[G]^{\otimes |\msf C|}
    \big)
\end{gathered}
\end{equation}
are symmetries of the concatenated gauging maps $\widehat{\mc G}\circ\mc G$ and $\mc G\circ\widehat{\mc G}$ respectively. Indeed:
\begin{align}
    \mbb a^\msf v_g \cdot
    \big( \widehat{\mc G}\circ\mc G \big) &= \widehat{\mc G}\circ\mc G, \\
    \mbb b^\msf c_\chi \cdot
    \big(\mc G\circ\widehat{\mc G}\big) &= \mc G\circ\widehat{\mc G},
\end{align}
for all $\msf v\in\msf V(\Lambda)$, $\msf c\in\msf C(\Lambda)$, $g\in G$, $\chi\in\widehat G$. The iterated gauging procedure put forward in~\cite{Garre24} now dictates that given any state $\ket{\psi}\in\mc H$ stabilized by the initial $G$ symmetry, that is
\begin{equation}
    \mbb U_g[\Omega] \ket{\psi} = \ket{\psi}
\end{equation}
for every \emph{valid} symmetry operator $\mbb U_g[\Omega]$, one can define the \emph{emergent} state $\ket{\Psi[\psi]}$ as
\begin{equation}
\begin{gathered}
    \label{eq:GeneralEmergState}
    \ket{\Psi[\psi]} := \cdots \mc G_{\msf{\frac{3}{2}}} \circ\widehat{\mc G}_\msf 1\circ\mc G_{\msf{\frac{1}{2}}} \ket{\psi} \\
    \in \bigotimes_{\msf{\frac{i}{2}}} \mbb C[\widehat G]^{\otimes|\msf V|} \bigotimes_{\msf i}\mbb C[G]^{\otimes|\msf C|}
\end{gathered}\, ,
\end{equation}
where the big tensor products run over all appearances of the gauging maps in the definition.

This state is then by construction stabilized by the operators $\mbb a^\msf v_g$ and $\mbb b^\msf c_\chi$ defined above. Given these stabilizers one can subsequently construct a stabilizer Hamiltonian for which $\ket{\Psi[\psi]}$ is a ground state. Remark that additional stabilizers are given by the emergent symmetries of every gauging map. For the examples that follow we will refrain from including these stabilizers as they are in those cases non-local operators. However, note that this is by no means always the case. Recall that in the example of \cref{sec:TopoOrder}, the emergent 1-form symmetries around individual plaquettes gave rise to the plaquette terms defined in \cref{eq:TopoBpxy}. In fact, we rather would like to think of these terms as kinematical constraints on the Hilbert space on which the emergent state is defined and as such we are allowed to include them in the Hamiltonian.

Given the foliated way in which the emergent state (\ref{eq:GeneralEmergState}) is constructed it is now a sensible thing to consider this foliation as an additional dimension so that $\ket{\Psi[\psi]}$ is a state that lives in exactly one dimension higher than the initial state $\ket{\psi}$. In this interpretation $\ket{\psi}$ behaves as a boundary condition for the emergent state. The lattice on which $\ket{\Psi[\psi]}$ is supported can then also be thought of as \emph{emerging} from $\Lambda$ on which $\ket{\psi}$ is defined. More precisely, the lattice is made up of layers which each carry a copy of $\Lambda$. In every other layer $\mbb C[\widehat G]$-valued degrees of freedom are placed on $\msf V(\Lambda)$ and $\mbb C[G]$-valued degrees of freedom supported on $\msf C(\Lambda)$. This is most easily demonstrated by concrete examples.

\subsection{Gapped boundary conditions}
In the interpretation of $\ket{\Psi[\psi]}$ as a state living in one dimension higher than $\ket{\psi}$, the stabilizers $\mbb a^\msf v_g$ and $\mbb b^\msf c_\chi$ are interpreted as acting on the \emph{bulk} of the state $\ket{\Psi[\psi]}$. The question of how to include boundary terms that stabilize $\ket{\Psi[\psi]}$ poses itself. As it turns out, the answer to this question is intimately interwoven with the exact way in which the initial $G$ symmetry is realized on the boundary state $\ket{\psi}$.

Let us begin by considering the simplest local candidate stabilizer. It can be verified readily that for a given $g\in G$ and $\msf v\in\msf V(\Lambda)$
\begin{equation}
\begin{gathered}
    Z_g^\msf v \otimes {\rm id} \\
    \in {\rm End} \big( \mbb C[\widehat G]^{\otimes|\msf V|}\otimes \mbb C[G]^{\otimes|\msf C|} \big)
\end{gathered}
\end{equation}
commutes with the gauging map $\G$ since it obviously commutes with the local gauge transformations $\G^\msf v_g$. Hence, it is a stabilizer of the emergent state exactly when it is one of the boundary condition, i.e. $Z_g^\msf v\ket{\psi}=\ket{\psi}$. Note that one could think of this operator as a `truncated' vertex term $\mbb a^\msf v_g$.

Consider likewise the truncated $\mbb b^\msf c_\chi$ term
\begin{equation}
\begin{gathered}
    \label{eq:GeneralTruncatedCube}
    {\rm id} \otimes Z_{\bar\chi}^\msf c \bigotimes_{\msf v\in\eta(c)} X^\msf v_{\bar o^c_\msf v(\chi)} \\
    \in {\rm End} \big(\mbb C[\widehat G]^{\otimes |\msf V|} \otimes \mbb C[G]^{\otimes |\msf C|} \otimes \mbb C[\widehat G]^{\otimes |\msf V|} \big)
\end{gathered}\, ,
\end{equation}
Recall that the preimage of this operator under $\widehat\G\circ\G$ was shown to be
\begin{equation}
    \bigotimes_{\msf v \in \eta(c)}\, X_{\bar o^c _ \msf v(\chi)}^\msf v,
\end{equation}
in \cref{eq:GeneralZXG=GX}. This observation thus implies that the truncated term (\ref{eq:GeneralTruncatedCube}) is a stabilizer of the emergent state, whenever $\ket{\psi}$ is stabilized by the check $X^c_{\bar \chi}$. Since these local constraints $X^c_\chi$ can be understood as local symmetric order parameters for the global $G$ symmetry (\ref{eq:GeneralSymGen}), inequivalent boundary conditions realizing different symmetry breaking patterns of this symmetry can be devised. More formally, consider for any subgroup $H\leq G$ the state $\ket{+}_H$ which we recall was defined in \cref{sec:TopoOrderBdry} as $\ket{+}_H := \frac{1}{\sqrt{|G/H|}}\sum_{[\chi]\in\widehat{G/H}} \ket{\chi}$. Constructing then the tensor product state
\begin{equation}
    \bigotimes_{\msf v\in\msf V(\Lambda)} \ket{+}_H^\msf v,
\end{equation}
it is clear that this reference state is stabilized by $Z_h^\msf v$ and $X^c_\chi$ provided that $h\in H$, and $\chi\in\widehat{G/H}$. Remember indeed from \cref{sec:TopoOrderBdry} that $\widehat{G/H}$ stands for the subgroup of characters in $\widehat G$ which are trivial when restricted to $H$. From this it also follows that the corresponding boundary stabilizers of the emergent state $\ket{\Psi[\psi_H]}$ commute. More generally, we could equally well consider boundary states of the form
\begin{equation}
    \bigg(\sum_{[g]\in G/H}\mbb U_g[\Omega]\bigg)
    \bigg( \bigotimes_{\msf v\in\msf V(\Lambda)} \ket{+}_H^\msf v \bigg)
\end{equation}
for any allowed $\Omega$, which enjoy the same stabilizers.

Also remark that, depending on the specifics of the $G$ symmetry generators, even more generic boundary conditions can be defined, for example by also (partially) breaking the translational invariance of the boundary state. An exhaustive study of gapped boundary conditions for the models under scrutiny will be the topic of future work.

%% file: _fracton_order_LSS.tex
\section{Foliated type-I fracton order from 2D linear subsystem symmetries}\label{sec:FractonLSS}
{\it In this section we showcase the gauging of abelian linear subsystem symmetries and their dual emergent symmetries, which are themselves linear subsystem symmetries on the dual lattice, following the prescription of \cref{sec:GeneralFramework}. Concatenation of these gauging maps results in foliated type-I fracton codes, generalizing the anisotropic models studied in~\cite{Shirley19} defined for cyclic groups to arbitrary finite abelian groups.}

\subsection{Gauging 2D abelian linear subsystem symmetries and their emergent symmetries}
Throughout this section we consider again the two-dimensional torus, now endowed with a square lattice denoted by $\square$. The initial $\mbb C[\widehat G]$-valued `matter' degrees of freedom are located on the vertices $\msf v\in\msf V(\square)$ as customary. It will be useful to express the vertices in terms of their Cartesian coordinates as $\msf v \equiv (\msf i,\msf j)\in\mbb Z_{L_x}\times\mbb Z_{L_y}$. Naturally, $L_x$ and $L_y$ stand for the linear system size in the $x$ and $y$ direction respectively. The linear subsystem symmetry (LSS) generators are then represented on the matter Hilbert space $\mc H := \bigotimes_{\msf v\in\msf V(\square)}\mbb C[\widehat G]$ as
\begin{align}
    Z^{\hat x, \msf j}_g := \bigotimes_{\msf i\in\mbb Z_{\! L_x}} Z_g^{(\msf i,\,\msf j)}, \label{eq:LSSx}\\
    Z^{\hat y, \msf i}_g := \bigotimes_{\msf j\in\mbb Z_{\! L_y}} Z_g^{(\msf i,\,\msf j)}\label{eq:LSSy}.
\end{align}
There thus exists an extensive number of symmetry generators, one for every $\msf j$ in the $\hat x$ direction of the square lattice, $Z^{\hat x, \msf j}_g$, and one for every $\msf i$ in the $\hat y$ direction, $Z^{\hat y, \msf i}_g$. Notice however that not all of the symmetries generators are independent since clearly $\prod_{\msf j\in\mbb Z_{L_y}} Z^{\hat x, \msf j}_g = \prod_{\msf i\in\mbb Z_{L_x}} Z^{\hat y, \msf i}_g$, $\forall g\in G$.

Graphically, an example of symmetry operators $\color{Tan} Z^{\hat x, \msf j}_g$ and $\color{black!50} Z^{\hat y, \msf i}_{g'}$ on a portion of the square lattice looks like:
\begin{equation}
    \hspace{-8pt}\LSS{0}\hspace{-16pt}.
\end{equation}
In order to employ the recipe of the general framework laid out in \cref{sec:GeneralFramework} we need to consider the checks that fully specify these symmetry generators. An adequate choice is given by
\begin{equation}
    X^c_\chi = X_\chi^{(\msf i,\,\msf j)} \otimes X_{\bar \chi}^{(\msf i+1,\,\msf j)} \otimes X_{\bar \chi}^{(\msf i,\,\msf j+1)} \otimes X_\chi^{(\msf i +1,\,\msf j +1)},
\end{equation}
$\forall \chi\in \widehat G $, $\forall \msf i,\msf j$. These operators are naturally associated with the plaquettes $\msf P(\square)$, and we will therefore identify the constraints with the plaquettes, i.e. $\mc C\ni c\equiv\msf p\in\msf P(\square)$. Indeed, graphically, one such operator can be depicted as
\begin{equation}
    \LSSCheck{0} \,\, .
\end{equation}

\bigskip\noindent Let us continue by gauging this symmetry. To this end we introduce thus $\mbb C[G]$-valued gauge fields on the plaquettes $\msf p\in\msf P(\square)$ and thereby extend the Hilbert space according to $\mc H \rightarrow \mc H \otimes \widehat{\mc H}$, where $\widehat{\mc H}:=\bigotimes_{\msf p\in\msf P(\square)} \mbb C[\widehat G]$. The local gauge transformations in the neighborhood of the vertex $\msf v\in\msf V(\square)$ read then
\begin{equation}
    \mc G_g^{\msf v} := Z_g^\msf v \,\bigotimes_{\msf p\rightarrow\msf v}\, X_{\bar g}^\msf p \,\bigotimes_{\msf p\leftarrow\msf v}\, X_g^\msf p\, ,
\end{equation}
for every $g\in G$. The tensor products appearing in this definition are over the sets of plaquettes $\msf p\rightarrow\msf v$, respectively $\msf p\leftarrow\msf v$, which point into, respectively point away from, the vertex $\msf v$ according to the following orientations:
\begin{equation}\label{eq:FractonGauss}
    \LSSGauss{0}\,\,.
\end{equation}
As such, the definition of $\mc G_g^\msf v$ is in accordance with \cref{eq:GeneralGaugeTrans}. Notice that here the same {\color{\colorb} blue} and {\color{\colora} red} color code of the previous sections for depicting degrees of freedom in respectively $\mbb C[\widehat G]$ and $\mbb C[G]$ is recycled.

As prescribed by our general gauging procedure, we define the following local projector:
\begin{equation}
    \mc P^{\msf v} := \frac{1}{|G|}\sum_{g\in G} \mc G_g^{\msf v}\,,
\end{equation}
for every vertex $\msf v\in\msf V(\Lambda)$, out of which we construct the projector onto the gauge invariant subspace as:
\begin{equation}\label{eq:fracproj}
    \mc P := \prod_{\msf v\in\msf V(\square)} \mc P^{\msf v}.
\end{equation}
Gauging of the LSS then happens by initializing the gauge fields in the trivial tensor product configuration $\bigotimes_{\msf p\in\msf P(\square)}\ket{1}^\msf p$ and applying the above projector $\mc P$ so as to obtain
\begin{equation}\label{eq:G01}
    \G^{(0,1)} :=  \mc P \bigg(\bigotimes_{\msf p\in\msf P(\square)} \ket{1}^{\msf p}\bigg).
\end{equation}
The notation $\G^{(0,1)}$, following \cite{Rayhaun23}, is motivated by the fact that there is a sense in which linear subsystem symmetries can be thought of as `$(0,1)$-form' symmetries: $0$-form symmetries acting on a codimension $1$ foliation of the spatial manifold. From the above definition of the gauging map it follows that it satisfies:
\begin{align}
    \G^{(0,1)}  \cdot  Z^{\hat x, \msf j}_g &= \G^{(0,1)},\label{eq:G01Z=G01a}\\
    \G^{(0,1)} \cdot  Z^{\hat y, \msf i}_g &= \G^{(0,1)},\label{eq:G01Z=G01b}
\end{align}
$\forall g\in G$, $\msf i\in\mbb Z_{L_x}$, $\msf j\in\mbb Z_{L_y}$. Hence, in line with the expectation from the general framework, states transforming non-trivially under one or more of the global linear subsystem symmetry generators are annihilated by the gauging map $\G^{(0,1)}$.

\bigskip\noindent Let us now present the emergent symmetries. These happen to be themselves linear subsystem symmetries supported on the dual square lattice $\square^\vee$ and are labeled by characters $\chi\in\widehat G$. Similarly as in the case of the initial LSS it proves useful to define the vertices of the dual square lattice, $\msf v^\vee\in\msf V(\square^\vee)$, in terms of their Cartesian coordinates, i.e. $\msf v^\vee\equiv(\msf i^\vee,\msf j^\vee)\in\mbb Z_{L_y}\times\mbb Z_{L_x}$. Also notice that the dual vertices can be identified under Poincar\'e duality with the plaquettes of the primal square lattice, formally: $\msf V(\square^\vee)\ni\msf v^\vee\equiv\msf p\in\msf P(\square)$. We then define for any character $\chi\in\widehat G$ the \emph{dual} LSS generators:
\begin{align}
    Z^{\hat x^\vee\!,\, \msf j^\vee}_\chi := \bigotimes_{\msf i^\vee\in\mbb Z_{L_y}} Z_\chi^{(\msf i^\vee\!,\, \msf j^\vee)}, \label{eq:DualLSSx}\\
    Z^{\hat y^\vee\!,\, \msf i^\vee}_\chi := \bigotimes_{\msf j^\vee\in\mbb Z_{L_x}} Z_\chi^{(\msf i^\vee\!,\, \msf j^\vee)}. \label{eq:DualLSSy}
\end{align}
Examples of $\color{Tan} Z^{\hat y^\vee\!, \msf j^\vee}_\chi$ and $\color{black!50} Z^{\hat x^\vee\!, \msf i^\vee}_{\chi'}$ then look like
\begin{equation*}
    \LSS{1}\hspace{-20pt},
\end{equation*}
wherein the dual lattice $\square^\vee$ is drawn in gray on top of the primal lattice $\square$ drawn in black. 

The emergent symmetries are apparent from
\begin{align}
    Z^{\hat x^\vee\!,\, \msf j^\vee}_\chi\cdot\G^{(0,1)}  = \G^{(0,1)}, \\
    Z^{\hat y^\vee\!,\, \msf i^\vee}_\chi \cdot  \G^{(0,1)}  = \G^{(0,1)},
\end{align}
for all $\chi\in\widehat G,\, \msf i^\vee\in\mbb Z_{L_y},\,\msf j^\vee\in\mbb Z_{L_x}$.

Despite these identities being a mere application of \cref{eq:GeneralUG=G} to the case at hand, let us nevertheless derive these emergent symmetries explicitly for completeness. To this end, note that the operators $Z^{\hat x^\vee\!,\, \msf j^\vee}_\chi, Z^{\hat y^\vee\!,\, \msf i^\vee}_\chi$ commute with the projector in \cref{eq:fracproj} and leave invariant the initial trivial gauge field on the plaquettes $\bigotimes_{\msf p\in\msf P(\square)} \ket{1}^{\msf p}$. The former is satisfied because along dual horizontal and vertical directions $\hat x^\vee/\hat y^\vee$, the operators $Z^{\hat x^\vee\!,\, \msf j^\vee}_\chi, Z^{\hat y^\vee\!,\, \msf i^\vee}_\chi$ overlap in two faces with the projectors $\mc P_g^{\msf v}$ appearing in the definition of $\mc G^{(0,1)}$ that act as $X_g\otimes X^\dagger_g$ on them.

To conclude the construction of the $\mc G^{(0,1)}$ gauging map, let us provide its PEPO representation. To this end we use tensors which slightly generalize those introduced in \cref{sec:TopoOrder}. Their symmetry properties are therefore also completely analogous and are omitted here. Again restricting to one unit cell of the PEPO, which in this case is made up of one Z-type tensor on the vertex $\msf v\in\msf V(\square)$ of the unit cell and one $+$-type tensor on an adjacent plaquette $\msf p\in\msf P(\square)$, we claim that the PEPO is constructed from:
\begin{equation}\label{eq:LSSPEPOUnit0}
    \LSSPEPOUnit{0}\,\,\,.
\end{equation}
Indeed, let us verify its relevant properties. In the first place it follows from \cref{eq:ZTensorSym} combined with \cref{eq:+TensorInv} that the gauging map absorbs the initial LSSs supported on the vertices of the primal lattice, i.e. (\ref{eq:G01Z=G01a}-\ref{eq:G01Z=G01b}). Analogously, the gauge invariance of the gauging map, i.e. $\G_g^\msf v\cdot\G^{(0,1)}$, $\forall\msf v\!\in\!\msf V(\square)$, $g\!\in\! G$, is proven by making use of \cref{eq:ZTensorSym} and (\ref{eq:+TensorSym}). Finally, the existence of the emergent symmetries is a direct consequence of the flatness of the gauge field enforced by the $+$-tensors, combined with the observation that the Z-type tensors are diagonal in its virtual indices.

\subsection{Iterative gauging of linear subsystem symmetries}
Due to the structural similarity between the dual LSS generators (\ref{eq:DualLSSx}, \ref{eq:DualLSSy}) and the initial ones (\ref{eq:LSSx}, \ref{eq:LSSy}), the gauging procedure of the latter can be readily adopted to gauge the emergent LSS symmetries. To this end, we introduce $\mbb C[\widehat G]$-valued gauge fields on the plaquettes $\msf p^\vee$ of the dual lattice $\square^\vee$, which, as we mentioned, can equivalently be thought of as the vertices $\msf v\in\msf V(\square)$. As is custom by now, we initiate these gauge fields as $\bigotimes_{\msf v\in\msf V(\square)}\ket{\underline 1}^{\msf v}$. On this enlarged Hilbert space we define the local gauge transformation
\begin{equation}
    \mc G_\chi^{\msf p} := Z_\chi^\msf p \,\bigotimes_{\msf v\rightarrow\msf p}\, X_{\bar \chi}^{\msf v} \,\bigotimes_{\msf v\leftarrow\msf p}\, X_\chi^{\msf v}.
\end{equation}
The orientations determining the sets of `incoming' and `outgoing' plaquettes on the dual square lattice, respectively $\msf v\rightarrow\msf p$ and $\msf v\leftarrow\msf p$, are dual to those on the primal lattice as per the convention in (\ref{eq:FractonGauss}). From the perspective of the primal lattice this gauge transformation can hence be depicted as
\begin{equation}
    \LSSGauss{1}
\end{equation}
Using the local projectors
\begin{equation}
    \mc P^{\msf p} := \frac{1}{|G|}\sum_{\chi\in\widehat G} \mc G_\chi^{\msf p}\,\, ,
\end{equation}
the following gauging map is then constructed:
\begin{equation}\label{eq:G01dual}
    \widehat{\G}^{(0,1)} :=  \bigg(\prod_{\msf p\in\msf P(\square)} \mc P^{\msf p}\bigg)
    \bigg(\bigotimes_{\msf v\in\msf V(\square)} \ket{\underline 1}^{\msf v}\bigg).
\end{equation}
After the application of $\widehat{\G}^{(0,1)}$ LSSs on the primal lattice represented by (\ref{eq:LSSx}-\ref{eq:LSSy}) are reattained.

\bigskip\noindent The PEPO representation of $\widehat{\G}^{(0,1)}$ in turn is constructed from the unit cell
\begin{equation}\label{eq:LSSPEPOUnit1}
    \LSSPEPOUnit{1}\,\,\,.
\end{equation}
In this depiction it has to be understood that the dual Z-type tensor is placed on a plaquette $\msf p\in\msf P(\square)$ of the lattice whereas the dual +-type tensor is located on a vertex $\msf v\in\msf V(\square)$. The symmetries of this PEPO are immediately found from dualizing those of the above PEPO defined in \cref{eq:LSSPEPOUnit0}; from them the properties of $\widehat{\G}^{(0,1)}$ are recovered immediately.

\bigskip\noindent An initial state $\ket{\psi}$ that is in the even sector of the LSSs (\ref{eq:LSSx}-\ref{eq:LSSy}) can now be iteratively gauged by using the usual prescription as follows:
\begin{equation}\label{eq:ConcGau}
     \ket{\Psi[\psi]} := \cdots  \circ \G_{\msf{\frac{3}{2}}}^{(0,1)} \circ \widehat \G_{\msf 1}^{(0,1)} \circ \G_{\msf{\frac{1}{2}}}^{(0,1)}\ket{\psi}.
\end{equation}
Following our general paradigm we then interpret $\ket{\Psi[\psi]}$ as a three-dimensional state living on a cubic lattice $\smallcube$ endowed with a copy of $\mbb C[\widehat G]$ to every edge in the $\hat z$ direction ({\color{\colorb} blue}) and a local Hilbert space $\mbb C[G]$ ({\color{\colora} red}) associated to every plaquette in the $xy$-plane. Putting the PEPO representations \cref{eq:LSSPEPOUnit0,eq:LSSPEPOUnit1} together, it then follows that the emergent state away from the boundary can be represented by the PEPO unit cell
\begin{equation}
    \LSSPEPOState.
\end{equation}
From the general procedure outlined in \cref{sec:GeneralFramework}, it then follows that the bulk stabilizers of the state $\ket{\Psi[\psi]}$ can be pictorially represented as
\begin{align}
   \mbb a_g^\msf v &= \FractonHam{v}, \\
   \mbb b_\chi^\msf c &= \FractonHam{c},
\end{align}
for every vertex $\msf v\in\msf V(\smallcube)$ and every cube $\msf c\in\msf C(\smallcube)$ respectively. The state \eqref{eq:ConcGau} is thus a ground state of the commuting projector Hamiltonian
\begin{equation}\label{eq:LSSBulkHam}
    \mbb H^{\rm blk} := -\sum_{\msf v\in\msf V(\smallcube)} \mbb P^\msf v - \sum_{\msf c\in\msf C(\smallcube)} \mbb P^\msf c,
\end{equation}
in which
\begin{align}
    \mbb P^\msf v &:= \frac{1}{|G|} \sum_{g\in G} \mbb a_g^\msf v, \\
    \mbb P^\msf c &:= \frac{1}{|G|} \sum_{\chi\in\widehat G} \mbb b_\chi^\msf c.
\end{align}
This generalizes the anisotropic model introduced in \cite{Shirley19}, which exhibits foliated type-I fracton order, to arbitrary finite groups. Our representation can be brought in the form showcased in \cite{Shirley19} by applying the local basis transformation defined in \cref{eq:UBasisTrans} to the red plaquettes. It is expected that our approach will facilitate the classification and construction of these fracton orders.

\subsection{Gapped boundary conditions}
Let us now also extend the bulk Hamiltonian to the boundary of the three-dimensional cubic lattice. To this end we consider following operators near the boundary:
\begin{align}
    \mbb d^\msf c_\chi &:= \FractonHam{d}\,\, , \label{eq:FractonBdryd}\\[10pt]
    \mbb c^{\msf e_z}_ g &:= \qquad\CSSHam{bdrysinglez}\,\,, \label{eq:FractonBdryc}
\end{align}
for all $\chi\in\widehat G$ and $g \in G$.

Being truncated versions of the bulk stabilizers $\mbb b^\msf c_\chi$ and $\mbb a^{\msf v}_ g$, they naturally commute with all bulk terms appearing in the bulk Hamiltonian $\mbb H^{\rm blk}$. Note that, similar as to the boundary terms defined for the introductory example in \cref{sec:TopoOrderBdry}, $\mbb d^\msf c_\chi$ and $\mbb c^{\msf e_z}_g$ only commute for particular choices of $\chi$ and $g$ such that $\chi(g)=1$. This statement will be refined below.

From the general framework it follows that $\mbb d^\msf c_\chi$ stabilizes the three-dimensional state $\ket{\Psi[\psi]}$ if the boundary condition $\ket{\psi}$ is symmetric under the operator
\begin{equation}\label{bdry-op-LSS}
    \LSSCheck{1}\,\,\,,
\end{equation}
for the boundary plaquette $\msf p\in\msf P(\square)$ corresponding to $c$. One recognizes these operators as the checks $X^c_\chi$ defined above. For illustrative purposes, let us construct one particular reference boundary before proceeding to the more general case.

A suitable choice for $\ket{\psi}$ is any of the extensive number of ground states of the two-dimensional \emph{classical $\widehat G$ Ising plaquette model}. Indeed, recall that the Hamiltonian of this model is exactly given by $\mbb H^{\rm Ising} := -\sum_{c}\sum_{\{\chi_c\}_c} X^c_{\chi_c}$. One such ground state is for example constructed by starting from the product state
\begin{equation}
    \bigotimes_{\msf v\in\msf V(\square)} \ket{+}_{\{1\}}^\msf v,
\end{equation}
where the notation introduced in eq. (\ref{eq:+Hstate}) was used, and acting on this reference state with all LSS generators according to
\begin{equation}
\begin{split}
    \ket{\psi}_{\{1\}} := \bigg(\sum_{\msf j=0}^{L_y-1}\sum_{\substack{\{g_\msf j\}_\msf j\\ \in G}} Z^{\hat x,\msf j}_{g_\msf j}\bigg) 
    \cdot \bigg(\sum_{\msf i=0}^{L_x-1}\sum_{\substack{\{g_\msf i\}_\msf i\\ \in G}} Z^{\hat y,\msf i}_{g_\msf i}\bigg) \\ 
    \bigg(\bigotimes_{\msf v\in\msf V(\square)}
    \ket{+}_{\{1\}}^\msf v\bigg).
\end{split}
\end{equation}
We stress that in this definition every summation over group elements is over the entire group $G$. From the observation that $X^c_\chi\big(\bigotimes_{\msf v\in\msf V(\square)} \ket{+}_{\{1\}}^\msf v\big) = \bigotimes_{\msf v\in\msf V(\square)} \ket{+}_{\{1\}}^\msf v$ for every $\chi\in\widehat G$, together with the fact that the checks commute by definition with all LSS generators, it indeed follows that $\ket{\psi}_{\{1\}}$is a good boundary condition for the emergent state $\ket{\Psi[\psi_{\{1\}}]}$ for which additional boundary stabilizers are given by
\begin{equation}
    \mbb Q^\msf c_{\{1\}} := \sum_{\chi\in\widehat G} \mbb d^\msf c_\chi, \quad \forall \msf c\in\msf C(\dCube).
\end{equation}

Similar as to the introductory example in \cref{sec:TopoOrder}, and as per the general paradigm in \cref{sec:GeneralFramework}, we can also define more general boundary conditions for different symmetry breaking patterns of the LSS on the boundary. Indeed, given again the state $\ket{+}_H$ defined in \cref{eq:+Hstate}, it is natural to define
\begin{equation}\label{eq:FractonBdryState}
\begin{multlined}
    \ket{\psi}_H := \bigg(\sum_{\msf j=0}^{L_y-1}\sum_{\substack{\{[g_\msf j]\}_\msf j\\ \in G/H}} Z^{\hat x,\msf j}_{g_\msf j}\bigg) \\
    \cdot \bigg(\sum_{\msf i=0}^{L_x-1}\sum_{\substack{\{[g_\msf i]\}_\msf i\\ \in G/H}} Z^{\hat y,\msf i}_{g_\msf i}\bigg)
    \bigg(\bigotimes_{\msf v\in\msf V(\square)}
    \ket{+}_H^\msf v\bigg).
\end{multlined}
\end{equation}
Given this choice of boundary condition, appropriate commuting boundary stabilizers are given by:
\begin{align}
    \mbb Q^\msf c_H &:= \frac{1}{|G/H|}\sum_{\chi\in\widehat{G/H}} \mbb d^\msf c_\chi, \\
    \mbb Q^{\msf e_z}_H &:= \frac{1}{|H|}\sum_{h\in H} \mbb c^{\msf e_z}_h.
\end{align}
Indeed given that $h\in H$, and thus $\chi(h)=1$ for any $\chi\in\widehat{G/H}$, the boundary operators $\mbb d^\msf c_\chi$ and $\mbb c^{\msf e_z}_h$ defined in (\ref{eq:FractonBdryd},\ref{eq:FractonBdryc}) commute. These boundary terms are then added to the bulk Hamiltonian (\ref{eq:LSSBulkHam}) to obtain the Hamiltonian
\begin{equation}
    \mbb H_H := \mbb H^{\rm blk} + \sum_{\msf c\in\msf C(\dCube)} \mbb Q^\msf c_H + \sum_{\msf e_z\in\msf E(\dCube)} \mbb Q^{\msf e_z}_H.
\end{equation}
Before proceeding we note that the boundary conditions presented here are by no means exhaustive. In particular it should be pointed out that the boundary states defined in \cref{eq:FractonBdryState} are manifestly translationally invariant, which is a condition which can be relaxed~\cite{Schuster23}. A complete study of all (gapped) boundary conditions for these models will be presented elsewhere.

%% file: _fracton_order_fractal.tex
\section{Type-I fracton order from 2D Sierpinski fractal symmetry}\label{sec:FractonFractal}
{\it In this section we consider $\mbb Z_2$ Sierpinski fractal symmetry as initial symmetry to our iterative gauging procedure. It is shown that the original Castelnovo-Chamon fractal model proposed in \cite{Castelnovo12} emerges.}

\subsection{Gauging Sierpinski fractal symmetry and its emergent dual fractal symmetry}
Let us begin by defining a set of generators for the fractal Sierpinski symmetry. Therefore we consider the two-dimensional torus endowed with a triangular lattice that we will denote by $\triangle$. Given a vertex $\msf v\in \msf V(\triangle)$, we will denote its rectangular coordinates in the basis $\{\hat x,\hat y\}$, indicated in the figure below, by $\msf v\equiv (\msf i, \msf j)$, $\msf i, \msf j\in\mbb Z_L$, $L$ being the \emph{linear system size}. Every vertex $\msf v$ carries a copy of the local Hilbert space $\mbb C[\mbb Z_2]\simeq\mbb C^2={\rm Span}_\mbb C\{\ket{0},\ket{1}\}$ on which the Pauli operators $X$ and $Z$ are defined.\footnote{In fact, in the remainder of this section we will identify $\mbb C[\mbb Z_2]$ and $\mbb C[\widehat{\mbb Z_2}]$ by making use of the obvious isomorphism $\mbb Z_2\simeq\widehat{\mbb Z}_2$.} We will consider a subset $\msf P(\triangle)^+\!\subset\!\msf P(\triangle)$ of the plaquettes whose member plaquettes are shaded in red in the figure below, foreshadowing that we will put gauge degrees of freedom on these plaquettes when gauging the fractal symmetry. Note that the number of such plaquettes equals the number of vertices, i.e. $|\msf P(\triangle)^+|=|\msf V(\triangle)|$.\footnote{Indeed, this follows from the fact that the Euler characteristic of the torus is zero, which can be expressed as $|\msf V(\triangle)|-|\msf E(\triangle)|+|\msf P(\triangle)|=0$, combined with the facts that $|\msf P(\triangle)|=2|\msf P(\triangle)^+|$ and $|\msf E(\triangle)|=3|\msf V(\triangle)|$.} This allows us to assign to every $\msf p^+\in\msf P(\triangle)^+$ a \emph{base point} ${\rm bp}(\msf p^+)$ which we conventionally choose as dictated by the dotted lines in the figure:
\begin{equation}
    \SierpinskiLattice{0} \hspace{-10pt}.
\end{equation}

As pointed out in \cite{Devakul19} the number of symmetry generators depends in a non-trivial way on the (linear) system size $L$. In this section we will for demonstrative purposes restrict to the case of $L=2^m-1$ for an integer $m$, in which case the number of symmetry generators is exactly equal to $L-1$. The Sierpinski symmetry generators are then of the form
\begin{equation}\label{eq:SierpinskiGen}
    \mbb U[\mc S_q] := \prod_{\msf i,\msf j} \big(Z^{(\msf i,\msf j)}\big)^{\mc S_q(\msf i,\msf j)},
\end{equation}
where $\mc S_q$ is a function $\mc S_q:\mbb Z_L\times \mbb Z_L \rightarrow\mbb F_2$ defined such that $\mbb U[\mc S_q]$ commutes with the checks $X^{\msf p^+}$ which are given by
\begin{equation}
\begin{gathered}
    X^{\msf p^+} := X^{(\msf i,\msf j)} \otimes X^{(\msf i+1,\msf j)} \otimes X^{(\msf i,\msf j+1)}, \\
    (\msf i,\msf j) = {\rm bp}(\msf p^+),
\end{gathered}
\end{equation}
for all $\msf p^+\in\msf P(\triangle)^+$. A check thus consists of three Pauli $X$ operators acting on the three vertices surrounding a plaquette $\msf p^+\in\msf P(\triangle)^+$.

As follows from a direct computation, imposing commutation of the generators (\ref{eq:SierpinskiGen}) with all these checks requires $\mc S_q$ to satisfy the recursive identity
\begin{equation}
    \mc S_q(\msf i,\msf j+1) := \mc S_q(\msf i +1,\msf j) + \mc S_q(\msf i,\msf j)\label{eq:SierpinskiRule},
\end{equation}
$\forall\ \msf i,\msf j\in\mbb Z_L$. A \emph{valid symmetry generator} is hence fully specified by a \emph{valid} choice of `initial condition' at $\msf j=0$:
\begin{equation}
    \mc S_q(\msf i, 0) := q(\msf i) \in \mbb F_2^L.
\end{equation}
The function $\mc S_q(\msf i,\msf j)$, for $\msf j\neq 0$ follows then from applying the recursion relation to this choice of $q$, and as such the corresponding generator $\mbb U[\mc S_q]$ commutes by construction with all checks. The $L-1$ independent choices for $q$ which give rise to the $L-1$ independent Sierpinski symmetry generators that we expect can be chosen as
\begin{equation}
    q^{(\alpha)}(\msf i) = \delta_{\msf i, \alpha} + \delta_{\msf i, \alpha+1}, \quad \alpha=1,2,...,L-1.
\end{equation}

Let us mention as a side remark that it turns out that one could think of $\mc S_q$ as a \emph{valid history} of a \emph{Sierpinski cellular automaton} in the following sense. For fixed $\msf j$, $\mc S(\msf i,\msf j)$ corresponds to a state valued in $\mbb F_2^L$. Given this state, the state at $\msf j+1$, i.e. $\mc S_q(\msf i,\msf j+1)$ for fixed $\msf j+1$, is fully specified by the above recursion relation from $\mc S_q(\msf i, \msf j)$. Therefore $\msf j$ can be thought of as labeling discrete `time' steps, where this notion of time thus flows in the direction of increasing $\msf j$. In this picture $q(\msf i)$ serves as an initial condition for the automaton, a posteriori justifying our jargon introduced above.

As a concrete example take $L=3$, and $q^{(1)}(\msf i) = [1\ 1\ 0]_\msf i$. The corresponding history $\mc S_q$ is then computed to be
\begin{equation}
    \mc S_q(\msf i,\msf j) =
    \begin{bmatrix}
        1 & 1 & 0 \\
        0 & 1 & 1 \\
        1 & 0 & 1
    \end{bmatrix}_{\msf j,\msf i},
\end{equation}
whose matching lattice operator $\mbb U[\mc S_q]$ can be depicted as:
\begin{equation}
    \SierpinskiLattice{1}.
\end{equation}

Let us now gauge the Sierpinski symmetry using the general paradigm of \cref{sec:GeneralFramework}. To this end we introduce qubit gauge degrees of freedom on the plaquettes in $\msf P(\triangle)^+$. A local gauge transformation for every $\msf v\in\msf V(\triangle)$ reads then 
\begin{equation}\label{eq:GTransfomSierp}
    \mc G^\msf v_1 := Z^\msf v\!\!\!\! \bigotimes_{\msf p^+ \in \msf P(\triangle)^+ |\msf v} X^{\msf p^+},
\end{equation}
wherein the tensor product is understood to be over plaquettes $\msf p^+ \in \msf P(\triangle)^+ |\msf v$ that contain $\msf v$. Graphically:
\begin{equation}
    \SierpinskiGauss{0}.
\end{equation}
From the local projector
\begin{equation}
    \mc P^\msf v := \frac{1}{2} \left({\rm id} + \mc G^\msf v_1\right),
\end{equation}
one constructs, as is by now routine, the global projector $\mc P = \prod_{\msf v\in\msf V(\triangle)} \mc P^\msf v$ from which one obtains the gauging map by trivially initializing the gauge field:
\begin{equation}
    \mc G := \mc P\bigg(\bigotimes_{\msf p\in\msf P(\triangle)^+}\ket{0}^\msf p\bigg).
\end{equation}
Evidently, one has $\mc G^\msf v_1\cdot \mc G=\mc G$. Let us also verify explicitly that this gauging map indeed projects out states transforming non-trivially under the Sierpinski symmetry, i.e. $\mc G\cdot \mbb U[\mc S_q] = \mc G$, for all \emph{valid} symmetry generators defined in \cref{eq:SierpinskiGen}. To this end, we could expand the local projectors appearing in the definition of the gauging map $\G$:
\begin{equation}\label{eq:SierpGaugingMapExpansion}
    \prod_{\msf v\in\msf V(\triangle)} \prod_{\{g_\msf v\}_\msf v} (\mc G_1^\msf v)^{g_\msf v}.
\end{equation}
From the definition of $\mc G_1^\msf v$ in \cref{eq:GTransfomSierp} it then follows that three $\mc G_1^\msf v$'s in the expansion (\ref{eq:SierpGaugingMapExpansion}) on vertices $(\msf i+1, \msf j)$, $(\msf i, \msf j)$ and $(\msf i, \msf j+1)$ overlap in exactly one plaquette $\msf p^+\in\msf P(\triangle)^+$, namely the one such that $(\msf i,\msf j)={\rm bp}(\msf p^+)$. Locally, this means that the property $\mc G\cdot \mbb U[\mc S_q] = \mc G$ is equivalent to $\mc S_q$ satisfying (\ref{eq:SierpinskiRule}), which it has to in order to define a genuine Sierpinski symmetry generator.

\bigskip\noindent We now turn to the emergent symmetries. As a matter of fact, after gauging we obtain a \emph{dual} Sierpinski fractal symmetry supported on the gauge degrees of freedom. Given a plaquette $\msf p^+\in\msf P(\triangle)^+$, we define its \emph{rectangular coordinates} as $(\msf i^\vee,\msf j^\vee):={\rm bp}(\msf p^+)$. Mimicking the definition of $\mc S_q$, let us define $\mc S_q^\vee:\mbb Z_L\times\mbb Z_L\rightarrow\mbb F_2$ by means of the recursive identity
\begin{equation}
\begin{multlined}
    \mc S_q^\vee(\msf i^\vee, \msf j^\vee) := \mc S_q^\vee(\msf i^\vee, \msf j^\vee+1) \\ + \mc S_q^\vee(\msf i^\vee-1,j^\vee+1)
\end{multlined}\, ,
\end{equation}
\begin{equation}
    \mc S_q^\vee(\msf i^\vee, 0) := q(\msf i^\vee),
\end{equation}
for all $\msf i^\vee, \msf j^\vee\in\mbb Z_L$. By comparison with the definition of $\mc S_q$ in (\ref{eq:SierpinskiRule}) we can infer that $\mc S_q^\vee$ defines a Sierpinski fractal symmetry in which the `time steps' labeled by $\msf j$ flow in the opposite direction compared to the one in (\ref{eq:SierpinskiRule}) and in which the `spatial direction' $\msf i$ is reflected as well.
It can then be shown that $\mc S_q^\vee$ exactly specifies the emergent fractal symmetries after gauging. Namely, defining
\begin{equation}
    \mbb U^\vee[\mc S_q^\vee] := \prod_{\msf i^\vee,\msf j^\vee} (Z^{(\msf i^\vee,\msf j^\vee)})^{\mc S^\vee_q(\msf i^\vee,\msf j^\vee)},
\end{equation}
it follows that
\begin{equation}
    \mbb U^\vee[\mc S_q^\vee]\cdot \mc G = \mc G.
\end{equation}
Indeed, this follows from the fact that exactly for this choice of $\mc S^\vee_q$ the operator $\mbb U^\vee[\mc S_q^\vee]$ commutes with every $\mc G_1^\msf v$, $\mbb U^\vee[\mc S_q^\vee]\cdot \mc G_1^\msf v = \mc G_1^\msf v\cdot\mbb U^\vee[\mc S_q^\vee]$, together with the trivial observation that the initial flat gauge configuration $\bigotimes_{\msf p^+\in\msf P(\triangle)^+}\ket{0}^{\msf p^+}$ is a $+1$ eigenstate of $\mbb U^\vee[\mc S_q^\vee]$.

\bigskip\noindent Since we are dealing with the symmetry group $\mbb Z_2$ here, we can in fact significantly simplify the tensor network calculus introduced in \cref{sec:TopoOrder} by removing the arrows and using a {\color{\colora}red} color for all tensor legs since we have identified $\mbb C[\mbb Z_2]$ and $\mbb C[\widehat{\mbb Z_2}]$ above. By virtue of these simplifications, we will also only need two type of tensors for both gauging maps which are given by:
 \begin{equation}
     \ZtwoPEPOTensor{0}{}{}{}{}{}{},
 \end{equation}
 which in the construction of the $\G$-PEPO will be placed on every vertex $\msf v\in\msf V(\triangle)$ and in the case of the $\widehat\G$-PEPO on every plaquette in $\msf P(\triangle)^+$, and
 \begin{equation}
     \ZtwoPEPOTensor{1}{}{}{}{}{}{},
 \end{equation}
 placed on the plaquettes $\msf P(\triangle)^+$ when it appears in the PEPO representation of $\G$ and on the vertices $\msf v\in\msf V(\triangle)$ when it appears in the $\widehat\G$-PEPO.

 For the PEPO representations of both gauging maps we make a choice of unit cell such that a plaquette tensor on a given $\msf p^+\in\msf P(\triangle)^+$ is combined with a vertex tensor placed on the corresponding base point ${\rm bp}(\msf p^+)$. For the gauging map $\G$, the PEPO unit cell is then explicitly given by:
\begin{equation}
 \ZtwoPEPOUnit{+}{Z}\ \ \ ,
\end{equation}
where, as mentioned above, the +-type tensor is placed on a certain plaquette $\msf p^+\in\msf P(\triangle)^+$ and the Z-type tensor on the corresponding base point ${\rm bp}(\msf p^+)$.
\subsection{Iterative gauging of Sierpinski fractal symmetry}
Gauging of the dual Sierpinski symmetry happens via the introduction of qubits on the vertices $\msf V(\triangle)$.
In the neighborhood of a plaquette $\msf p^+\in\msf P(\triangle)^+$, following gauge transformation is defined:
\begin{equation}
    \mc G_1^{\msf p^+} := Z^{\msf p^+}\!\!\!\! \bigotimes_{\msf v\in\msf V(\triangle)|\msf p^+}X^\msf v,
\end{equation}
which forms a local $\mbb Z_2$-representation. In this, $\msf V(\triangle)|\msf p^+$ stands for the set of three vertices surrounding $\msf p^+$. Concretely:
\begin{equation}
    \SierpinskiGauss{1}.
\end{equation}

Based on the local projector, defined as
\begin{equation}
    \mc P^{\msf p^+} := \frac{1}{2}\left({\rm id} + \mc G_1^{\msf p^+}\right),
\end{equation}
the corresponding gauging map reads:
\begin{equation}
    \widehat\G := \mc P\bigg(\bigotimes_{\msf v\in\msf V(\triangle)}\ket{0}^\msf v\bigg),
\end{equation}
in which $\mc P := \prod_{\msf p^+\in\msf P(\triangle)^+}\mc P^{\msf p^+}$. The expected properties $\mbb U[\mc S_q]\cdot\widehat\G=\widehat\G$ and $\widehat\G\cdot\mbb U^\vee[\mc S^\vee_q]=\widehat\G$, for all valid symmetry generators $\mbb U[\mc S_q]$ and $\mbb U^\vee[\mc S_q^\vee]$, hold by construction.

Under the aforementioned conventions, the unit cell of the PEPO representation of $\widehat\G$ can be depicted as
\begin{equation}
    \ZtwoPEPOUnit{Z}{+}\ \ \ ,
\end{equation}
where now the Z-type tensor is placed on a plaquette $\msf p^+\in\msf P(\triangle)^+$ and the second tensor on the corresponding base point.

\bigskip\noindent What is the emergent state obtained from these gauging procedures? It turns out that the emergent state can be interpreted as living on the vertices of a \emph{hexagonal close-packed} (HCP) lattice, which consists of layers of triangular lattices whose vertices are slightly displaced with respect to each other~\cite{Casasola24}. A slightly different realization of this lattice which is closer to our setup is the following. We consider also layers of two-dimensional triangular lattices $\triangle$, but in such a way that the vertices of all layers align along the direction perpendicular to the stack of layers. In every other layer we then place qubits on the vertices of the triangular lattice, i.e. on $\msf V(\triangle)$, and on the plaquettes $\msf P(\triangle)^+$.
The PEPO representation of the emergent state for one unit cell is then given by:
\begin{equation}
    \FractonPEPOState.
\end{equation}

Using a {\color{\colorb} blue} color for the degrees of freedom on $\msf V(\triangle)$ and a {\color{\colora} red} color for those on the plaquettes $\msf P(\triangle)^+$, the bulk terms of the emergent Hamiltonian are then the following:
\begin{align}\label{fracbulk}
    \mbb a^\msf v &:= \SierpinskiHam{0}\ , \\
    \mbb b^{\msf p^+} &:= \SierpinskiHam{1}.
\end{align}
With slight abuse of notation, $\msf v$ refers to the vertex of the emergent lattice on which the bottom {\color{\colorb} blue} qubit appearing in the $\mbb a$ operator is supported. On the other hand, $\msf p^+$ refers to the support of the {\color{\colora} red} qubit on the bottom of the depiction of the $\mbb b^{\msf p^+}$ operator. It is clear that these operators all mutually commute. We can construct projectors out of these operators as
\begin{align}
    \mbb P^\msf v := \frac{1}{2} \big({\rm id} + \mbb a^\msf v\big), \\
    \mbb P^{\msf p^+} := \frac{1}{2} \big({\rm id} + \mbb b^{\msf p^+}\big).
\end{align}
The emergent bulk Hamiltonian then reads
\begin{equation}
    \mbb H^{\rm blk} := -\sum_{\msf v\in\msf V} \mbb P^\msf v - \sum_{\msf p^+\in\msf P^+} \mbb P^{\msf p^+},
\end{equation}
where the sums are over all (bulk) vertices and $\msf p^+$-plaquettes of the HCP lattice. This model was proposed by Caselnovo and Chamon in \cite{Castelnovo12}, is covered in the family of models by Yoshida \cite{Yoshida13} and was recently studied in detail in \cite{Casasola24}. 

If we start the iterative gauging by $\G$, the possible boundary terms correspond to the truncation of (\ref{fracbulk}) near the boundary or individual $Z$ terms.
Explicitly the truncated (\ref{fracbulk}) terms read:
\begin{equation}
    \mbb d^{\msf p^+} = \SierpinskiHam{3}.
\end{equation}
which also have been identified and studied in \cite{Casasola24}. The term $\mbb d^{\msf p^+}$ acts on the boundary state $\ket{\psi}$, which is symmetric under the fractal Sierpinski symmetry, as:
\begin{equation}
    \SierpinskiCheck,
\end{equation}
which can be recognized as the local checks $X^{\msf p^+}$ and serve as a local order parameter for the initial fractal symmetry. As such, there are two possible boundaries depending on whether the fractal symmetry is spontaneously broken by $\ket{\psi}$ or not. In the first case, the boundary stabilizers are given by
\begin{equation}
    \mbb Q^{\msf p^+} := \frac{1}{2} \bigg( {\rm id} + X^{\msf p^+} \bigg),
\end{equation}
for all $\msf p^+\in\msf P(\triangle)^+$ on the boundary $\triangle$, whereas in the symmetry breaking case the stabilizers are given by
\begin{equation}
    \mbb Q^{\msf p^+} := \frac{1}{2} \bigg( {\rm id} + Z^\msf v \bigg).
\end{equation}

%% file: _conclusions.tex
\section{Conclusion and outlook}
In this work we have put the iterative gauging paradigm of~\cite{Garre24} for abelian symmetries on a systematic footing and extended it to arbitrary dimensions and generic types of abelian symmetries. We provided explicit examples where the emergent states are three-dimensional. Concretely we have shown the emergence of surface codes, foliated type-I fracton codes and non-foliated type-I fracton codes from the iterative gauging of 0-form, 1-form, linear subsystem and Sierpinski fractal symmetry respectively.

The method proposed in this work resembles the \emph{sandwich} construction of SymTFT. In particular, we constructed states of the following form:
\begin{equation*}
	\langle \hat{\psi} |\mc G \circ\widehat{\mc G} \circ \cdots \circ \mc G \circ\widehat{\mc G}\circ\mc G \ket{\psi},
\end{equation*}
made up of the gauging operator $\mc G$ of a $G$ symmetry, its dual $\widehat{\mc G}$ and states $\ket{\psi}, |\hat{\psi}\rangle$ symmetric under the initial abelian symmetry and its dual. For the case where the boundary supports an abelian 1-form symmetry, our method exactly recovers the quantum double in correspondence with the SymTFT proposal. The symmetries of this sandwich state are considered as stabilizer local Hamiltonian terms giving rise to the Hamiltonian:
\begin{equation*}
	\mbb H^{\rm blk} + \mbb H^{\rm bdry},
\end{equation*}
where the bulk terms only depend on the symmetry $G$ and the boundary terms depend on the quantum phase of $\ket{\psi}$.

We predict a number of ways in which this framework can be further refined and propose the following open problems. From the case of the emergent surface codes, it is clear that further refinement is required to accommodate boundary conditions hosting SPT phases on top of the partial symmetry breaking. It is plausible that akin to the lower-dimensional case~\cite{Garre24} this requires modification of the boundary stabilizers. A similar question arises regarding the boundary conditions of the emergent anisotropic models studied in \cref{sec:FractonLSS}. In particular, it would be revealing to study boundaries that (potentially partially) break translation symmetry. It is plausible to expect that by including boundary states which realize different symmetry breaking patterns in distinct regions, we could realize domain walls in the emergent orders such as the electromagnetic domain wall in the three-dimensional toric code~\cite{Delcamp21}. One could also wonder to what extent our framework is exhaustive in producing classes stabilizer codes. 

The main assumption of this work is the symmetry group being abelian, which results in abelian higher-dimensional codes. As such we leave the study of non-abelian, and more general categorical, symmetries in the iterative gauging framework for future work. As a first step, we will generalize the one-dimensional gauging map of~\cite{Haegeman14} to generalized symmetries described by unitary fusion categories in forthcoming work~\cite{BVDC25b}.

An obvious family of models not treated in this work are type-II fracton models, which differ from type-I fracton models in the mobility of their excitations. Adapting our approach to accommodate these fracton orders could prove to be a splendid challenge.

We further foresee that the explicit tensor network constructions of the states provided here will enhance their numerical study. Having direct access to the boundary states in this framework could facilitate the study of boundary phase transitions of these emergent models.